\documentclass[letterpaper,twocolumn,10pt]{article}
\usepackage{usenix}

\usepackage{graphicx}
\usepackage{tikz}
\usepackage{amsmath}
\usepackage{amssymb}
\usepackage{pifont}
\usepackage{xspace}
\usepackage{mathbbol}
\usepackage{dsfont}
\usepackage{subcaption}
\usepackage{caption}
\usepackage{listings}
\usepackage{paralist}
\usepackage[breakable]{tcolorbox}
\usepackage{multirow}
\usepackage{booktabs}
\usepackage{colortbl}
\usepackage{algorithm}
\usepackage{algorithmicx}
\usepackage{algpseudocode}
\algrenewcommand{\algorithmiccomment}[1]{\hfill$\triangleright$ {\scriptsize #1}}
\usepackage{tabularx}
\usepackage{placeins}

\definecolor{findingborder}{HTML}{4A90A4}
\definecolor{findingbg}{HTML}{EDF4F7}
\newtcolorbox{findingbox}{
  colback=findingbg,
  colframe=findingborder,
  leftrule=2.5pt,
  rightrule=0pt,
  toprule=0pt,
  bottomrule=0pt,
  left=8pt,
  right=8pt,
  top=4pt,
  bottom=4pt,
  before skip=\medskipamount,
  after skip=\medskipamount,
  sharp corners,
  breakable
}

\definecolor{sawin}{HTML}{D6EAF8}
\definecolor{sarunner}{HTML}{FADBD8}
\newcommand{\hlwin}[1]{{\setlength{\fboxsep}{1.5pt}\colorbox{sawin}{\strut\textbf{#1}}}}
\newcommand{\hlrun}[1]{{\setlength{\fboxsep}{1.5pt}\colorbox{sarunner}{\strut #1}}}
\definecolor{deltagain}{HTML}{D5F5E3}
\newcommand{\hldelta}[1]{{\setlength{\fboxsep}{1.5pt}\colorbox{deltagain}{\strut\textbf{#1}}}}

\newcommand{\system}{{\textsf{PhantomCall}}}

\newcommand*\wcircle[1]{\scalebox{1.15}{\ding{\the\numexpr171+#1\relax}}}
\newcommand*\bcircle[1]{\scalebox{1.15}{\ding{\the\numexpr181+#1\relax}}}
\newcommand*\scircle[1]{{\small\ding{\the\numexpr181+#1\relax}}}

\begin{document}



\title{ \Large \bf PhantomCall: Evading ML Malware Detectors via 
\\ Function Call Graph Perturbation}


\author{
\rm Md Ajwad Akil$^{*}$, Adrian Shuai Li$^{*}$, Imtiaz Karim$^{**}$, Arun Iyengar$^{\ddagger}$, \\
\rm Ashish Kundu$^{\dagger}$, Elisa Bertino$^{*}$ \\
$^{*}$Purdue University, $^{**}$The University of Texas at Dallas, $^{\ddagger}$Intelligent 
\\ Data Management and Analytics (USA), $^{\dagger}$Cisco Research,  \\
$^{*}$\{makil, li3944, bertino\}@purdue.edu, \\
$^{**}$imtiaz.karim@utdallas.edu, \\
aki@akiyengar.com, $^{\dagger}$ashkundu@cisco.com,
}

\maketitle
\begin{abstract}
Prior adversarial attacks on Windows PE malware detectors
target raw bytes, PE headers, or intra-function
control-flow graphs, leaving the function call graph (FCG)
unexplored as an attack surface. Yet the FCG structure is
an important feature in graph-based malware detectors. We
present \system{}, a black-box attack that perturbs the FCG
of Windows PE malware by injecting fully executable dummy functions at targeted call sites, adding new nodes and edges to both the CFG
and FCG while preserving program semantics. We pair this structural perturbation with classifier-guided search and tunable injection parameters, effective across three architecturally distinct classifiers.
Evaluated on a 2025-collected Windows malware corpus against
MalConv (raw-byte CNN), MalGraph (graph-based GNN), and
SAFE+GNN (pure FCG GNN trained from scratch on a 2024
corpus) at two FPR thresholds, the best \system{} variant achieves
85-100\% attack success rate across all configurations,
exceeding prior state-of-the-art by up to 14.78 percentage
points on MalGraph and 95.5 percentage points on SAFE+GNN,
and generating evasive variants up to 2.9$\times$ faster on
average across all targets. For MalConv and MalGraph, the
majority of evasions require only a single call site
modification, and 86-97\% of evaluated evasive variants preserve
the original malicious behavior in sandbox-based semantic
testing across all configurations.
\end{abstract}

\section{Introduction}\label{sec:introduction}

Despite decades of defensive research, malware remains among the most persistent and economically devastating threats in modern computing. In 2025 alone, global cybercrime damages reached an estimated \$10.5~trillion~\cite{cybersecurityventures2026}, with threat intelligence systems flagging over 450,000 new malicious files every day~\cite{avtest2025stats}, a rate that has increased year-over-year~\cite{intelligentciso2026kaspersky}. Microsoft Windows still remains the primary attack platform, accounting for approximately 87\% of all desktop malware detections in 2025, which is nearly seven times the rate seen on macOS~\cite{surfshark2025malware}. The financial toll is severe as the median ransom payment in 2025 reached \$1~million per incident~\cite{sophos2025ransomware}, while the average cost of an extortion or ransomware breach stands at \$5.08~million~\cite{ibm2025cost}. 

The defensive landscape has evolved considerably in response to these evolving threats. Signature-based antivirus, which emerged in the 1990s, blacklists files against a database of known malware patterns, is trivially defeated by slight variant generation or code transformation that is absent in the database~\cite{ling2023adversarial}. Over the past decade, machine learning (ML)-based detectors have therefore become a central defensive layer. These include raw-byte convolutional neural network-based models~\cite{raff2018malware, raff2021classifying}, static PE (Portable Executable) feature ensembles trained on header and section metadata~\cite{anderson2018ember}, and graph neural networks (GNNs) that operate on control-flow graphs (CFG) and function-call graph (FCG) structure using node and edge features~\cite{ling2022malgraph,yan2019classifying,wu2021mcbg,someya2023fcgat,8752028,https://doi.org/10.1049/ise2.12082, someya2023graph}. Modern commercial products, including Microsoft Defender, now integrate ML-based components alongside traditional signatures~\cite{microsoft2024mddr}, enabling continuous retraining on fresh telemetry from billions of endpoints. ML-based static malware detection is fast, scalable, requires no sandbox and has become a vital deployed layer of real-world malware detection.

However, recent advances in adversarial machine learning have demonstrated that ML systems are inherently vulnerable to carefully crafted perturbations, whether in image classifiers~\cite{zhu2024learning,goodfellow2015explaining}, language models~\cite{zou2023universal}, or vision-language models~\cite{xie2025chain}. These attacks exploit structural properties of learned models, including high-dimensional feature spaces and brittle decision boundaries, enabling adversaries to induce misclassification with minimal, targeted changes.
ML-based malware detectors are particularly exposed to such vulnerabilities. Unlike other domains, adversaries in this setting have strong control over the input generation process and can modify malware samples while preserving their functionality and evasiveness~\cite{pierazzi2020intriguing,ling2023adversarial}. This creates a natural and powerful adversarial loop, making robustness a central challenge. Consequently, studying adversarial malware generation against these systems serves two complementary purposes: \wcircle{1} Realistically characterizing the attack surface of production systems under query-only black-box access, and \wcircle{2} Driving the development of more robust detectors through adversarial training and red-teaming. This is particularly important because ML detectors already largely resist classical obfuscation techniques~\cite{ling2023adversarial}, making structured adversarial perturbations a novel and practically significant threat vector.
While prior adversarial attacks on Windows PE malware detectors have targeted raw bytes~\cite{kolosnjaji2018adversarial,kreuk2018deceiving,suciu2019exploring}, PE headers and overlays~\cite{demetrio2021functionality,li2025minimal}, or intra-function control-flow graphs~\cite{ling2024wolf,zhang2022semantics,zapzalka2024semantics}, graph-based detectors increasingly rely on the function call graph structure as a discriminative feature~\cite{ling2022malgraph,someya2023fcgat,kargarnovin2024mal2gcn, someya2023graph}, making its adversarial robustness directly relevant to the security of these systems. FCG-level perturbation has been
studied in Android
malware~\cite{li2023black,li2025efficient,zhao2021structural,song2025fcghunter}
and WebAssembly~\cite{kim2025does}, but these approaches
inject either provably dead calls or trivially
short-circuiting stubs, and none treat the perturbation
configurations such as injected function body content, injection volume,
or injected NOP content search strategy as variables that can be tuned
against the target classifier. Whether fully executable FCG
perturbations with configurable injection strategies can
evade architecturally diverse detectors under black-box
access, and how the attack mechanism differs across
classifier architectures, remains an open and understudied problem.


\vspace{2pt}
\noindent\textbf{Problem.} Given the aforementioned gaps, we investigate the following 
question: \emph{Given a Windows PE malware binary detected by an ML-based classifier, can we efficiently generate a functionally-equivalent, executable adversarial variant that evades the classifier by perturbing its function call graph, under a black-box setting?}

\vspace{2pt}
\noindent\textbf{Our Approach.} \system{} perturbs the function call graph of a malware binary by injecting fully executable dummy functions at targeted call sites, creating new nodes and edges in both the CFG and FCG (the latter being our primary contribution, since CFG-level displacement has been explored in prior work) while preserving execution semantics. A classifier-guided search iteratively explores injection configurations at each site, using the detector's maliciousness score as feedback. The same attack framework achieves high evasion rates across three architecturally distinct classifiers by adjusting only its key hyperparameters — the number of injected dummy functions, the number of call sites to perturb, and the search iteration budget, while generating evasive variants at practical throughput.




\textbf{Evaluation.} We evaluate \system{} on a 2025-collected Windows malware corpus against MalConv (raw-byte CNN)\cite{raff2018malware}, MalGraph (CFG+FCG-based GNN)\cite{ling2022malgraph}, and SAFE+GNN (pure FCG GNN)~\cite{someya2023graph} at 0.1\% and 1\% FPR thresholds. The best \system{} variant achieves 85-100\% attack success rate (ASR) across all configurations, exceeding prior state-of-the-art by up to 14.78 percentage points on MalGraph and 95.5 percentage points on SAFE+GNN (where both prior baselines fail to exceed 2\% ASR), while generating evasive variants up to 2.9$\times$ faster on average. ASR remains 84.5-100\% when restricted to high-confidence detections (P(malware) $\geq$ 0.95), and 86-97\% of evasive samples retain their original malicious functionality. Adversarial finetuning reduces ASR to 26-83\% depending on the classifier, and structural heuristic defenses achieve 100\% detection but at 9.5-17.4\% false positive rates, and a simple attack modification (linking injected functions via call edges) further weakens the lowest-FPR heuristic. Neither defense strategy alone neutralizes the attack.


\vspace{2pt}
\noindent\textbf{Contributions.} To summarize, our contributions are:
\begin{compactitem}
   \item To our knowledge, we present the first adversarial attack on Windows PE malware that perturbs the function call graph by injecting fully executable, reachable dummy functions.
    
    \item We design a configurable attack framework with classifier-guided search (greedy and adaptive simulated annealing) and tunable injection parameters, effective across three architecturally distinct classifiers through hyperparameter adjustment alone.
    
    \item We conduct a rigorous evaluation on a 2025-collected malware corpus measuring both evasion rate and generation throughput, with comprehensive ablation studies, semantic preservation analysis, and defense evaluation including adversarial finetuning and structural heuristic detection.
\end{compactitem}

All of our experimental code and dataset hashes will be released soon.

\section{Background}\label{sec:background}
\noindent This section introduces the relevant background on ML-based malware detection and adversarial malware generation.

\subsection{Malware Detection}
\noindent
An ML-based malware classifier is defined as $y=f(m)$ where $f$ is the classifier and $m$ is a malware executable in the problem space $\mathcal{M}$ (the domain of Windows PE executables). The output $y \in \{0,1\}$ indicates benign ($y=0$) or malicious ($y=1$). Since ML models operate on vectorized numerical features~\cite{bishop2006pattern}, we introduce a feature computing function $\psi(m)=x$ that maps $m$ to a feature vector $x$ in the feature space $\mathcal{X}$. The detector $f$ yields a maliciousness probability $\mathcal{P}(m) \in [0,1]$, and we define $f(m) = \mathbf{1}[\mathcal{P}(m) \geq \tau^{clsf}]$, where $\tau^{clsf}$ is a deployment-dependent classification threshold. Samples with $\mathcal{P}(m) \geq \tau^{clsf}$ are flagged as malware.


\subsection{Adversarial Malware Generation}
Given a learning-based detector $f$ and a malware executable $m \in \mathcal{M}$, adversarial malware generation generates a modified executable $m^{*} \in \mathcal{M}$ such that $f$ misclassifies $m^{*}$ as goodware i.e., $f(m^{*}) = 0$ while $m^{*} \sim m$, where $\sim$ denotes \emph{semantic equivalence}: $m^{*}$ preserves the malicious payload and runtime behavior of $m$~\cite{pierazzi2020intriguing,ling2023adversarial}. Attacks can be carried out in two settings. In \emph{feature-space} attacks, the adversary perturbs the feature representation $\psi(m) = x \in \mathcal{X}$ directly, which may not produce a valid executable~\cite{pierazzi2020intriguing}. \emph{Problem-space} attacks instead modify the raw executable, ensuring every adversarial sample remains a runnable PE file.

\section{Overview}\label{sec:overview}
\noindent In this section, we introduce our 
threat model, formal problem definition, and discuss the challenges and their solutions.

\subsection{Threat Model}
We assume that the primary goal of the adversary is to induce misclassification of malware as goodware by generating an adversarial sample $m^*$ from $m \in \mathcal{M}$ that preserves program semantics while evading an ML-based classifier $f$~\cite{pierazzi2020intriguing}.

We consider a classical black-box threat model in which the adversary has no access to the target model's internals. The attacker can only query the classifier with a malware sample $m$ and observe the predicted label along with the associated maliciousness probability $\mathcal{P}(m)$. The adversary can apply semantics-preserving transformations to Windows PE binaries, such as injecting new code sections and redirecting internal call instructions, while maintaining PE format compliance~\cite{pespecifications}. We scope the attack to classifiers whose feature extraction recovers the FCG from internal direct call instructions resolving to user-defined functions, as in MalGraph~\cite{ling2022malgraph}, SAFE+GNN~\cite{someya2023graph}, and similar GNN-based detectors. FCG representations based on external API calls, obfuscated \texttt{push+return} sequences, or computed indirect calls are outside scope, as \system{} targets only internal direct calls for displacement. 

\subsection{Problem Formulation}
Following the \emph{problem-space} attack setting~\cite{pierazzi2020intriguing}, we directly modify the executable rather than perturbing features.  Let $\Sigma$ denote the set of semantics-preserving atomic transformations, and let $\boldsymbol{\sigma} = (\sigma_1, \dots, \sigma_L) \in
\Sigma^L$ be a transformation sequence whose composition $\Phi_{\boldsymbol{\sigma}} = \sigma_L \circ \cdots \circ \sigma_1$ produces the
adversarial sample $m^* = \Phi_{\boldsymbol{\sigma}}(m)$. The classifier flags a sample as malicious when $\mathcal{P}(m) \ge \tau^{clsf}$. The objective is to find $\boldsymbol{\sigma}$ minimizing $\mathcal{P}$ subject to evasion and semantic equivalence:
\begin{equation}\small
\label{eq:prob_threshold}
\begin{array}{cl}
\displaystyle\min_{\boldsymbol{\sigma}} & \mathcal{P}(m^*) \\[6pt]
\text{s.t.} & \mathcal{P}(m^*) < \tau^{clsf},\quad
m^* = \Phi_{\boldsymbol{\sigma}}(m),\quad
m^* \sim m
\end{array}
\end{equation}

\subsection{Challenges \&  Solution Outline}
\noindent We now discuss the main challenges and approaches to their solution that guide our design of \system{}.

\noindent \textbf{(C1) FCG Perturbation on Compiled Windows PE Binaries:}
Prior FCG-level adversarial malware generation attacks target Android~\cite{li2023black,li2025efficient,zhao2021structural,song2025fcghunter}, IoT~\cite{mwangi2024adversarial}, and WebAssembly~\cite{kim2025does} malware, but inject either provably dead calls or trivially simple functions under managed runtimes that guarantee stack safety and calling convention compliance. No prior attack produces fully executable injected functions with semantically non-trivial bodies on compiled PE binaries. In the PE setting, semantic preservation of malware functionality requires that every injected function adhere to x86 calling conventions, correctly preserve register and stack state, use proper relative addressing, and ensure runtime reachability.

\noindent \textbf{(A1)} We design an FCG perturbation mechanism that injects fully executable dummy functions into Windows PE malware via call-site displacement through a trampoline that preserves the original call's stack semantics, simultaneously altering both the CFG and FCG while preserving program semantics (Section~\ref{sec:detailed_design}). The perturbation mechanism is identical across all target classifiers; only the function body content and dummy function node injection volume (number of dummy functions) differ, as hyperparameters.

\noindent \textbf{(C2) Classifier-Guided Search over the
Injection Configuration Space:} The perturbation mechanism
from~(C1) provides the structural attack surface, but the
attacker must still decide how to allocate the search budget
across two dimensions: which NOP sequences to place in the
dummy function bodies and trampoline padding, and how many
call sites to perturb at what injection volume per displacement. The only feedback is a scalar
$\mathcal{P}(m)$ per query with no gradient signal, and the
score landscape is non-monotonic. Which dimension dominates
depends on the target: classifiers that process
instruction content are more sensitive to NOP variation,
while classifiers whose representations aggregate over
function nodes benefit more from larger dummy node injection volumes
and are less sensitive to NOP content. Prior approaches rely
on random insertion~\cite{rigaki2023power}, benign opcode
transplantation~\cite{peng2025evading}, gradient
access in white or grey-box settings~\cite{kolosnjaji2018adversarial,kreuk2018deceiving,lucas2021malware},
or MCTS-guided
search~\cite{ling2024wolf}. No prior approach treats NOP content and injection volume as jointly tunable search variables or evaluates how the effective search dimension shifts across classifier architectures.

\noindent \textbf{(A2)} We design a family of
classifier-guided search algorithms that explore the
injection configuration space from~(C2), spanning greedy
exploitation to Simulated Annealing-based exploration
(Section~\ref{sec:sa-sem-nop-search}). The algorithms
expose externally tunable hyperparameters (NOP search
iteration budget, node injection
volume, function body content, NOP pool diversity) that allow the attacker to adjust the balance
between per-site NOP exploration and call-site coverage
based solely on observed changes in $\mathcal{P}(m)$, with
the same framework effective across architecturally
distinct classifiers.

\begin{figure*}[t]
\centering
\includegraphics[width=0.9\textwidth]{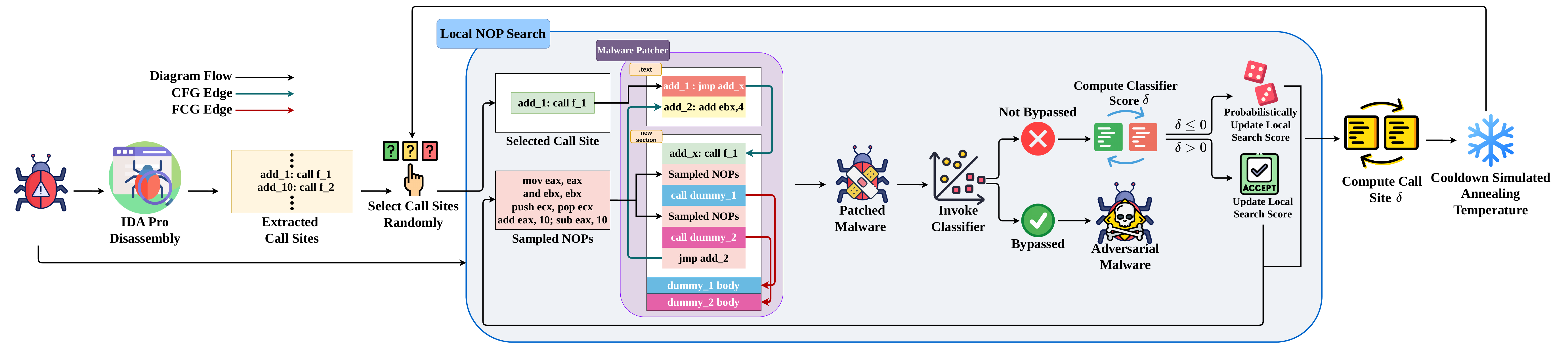}
\caption{Overview of \system{}. Call sites (internal \texttt{call} instructions) are extracted via IDA and sampled uniformly at random for perturbation. At each outer SA iteration, the selected call site is passed to the \emph{Local NOP Search} (\textsc{LocalSA}), which samples candidate semantic NOP sequences, applies the FCG perturbation via \emph{Malware Patcher} (\textsc{PatchAtCall}) to inject a new section and $k$ dummy leaf functions, and queries the classifier for a score $\mathcal{P}(\hat{m})$. Candidate moves are accepted on improvement ($\delta > 0$) or probabilistically otherwise, allowing the search to escape local minima. The outer loop updates the SA temperature via a classic or adaptive cooling schedule.}
\label{fig:main_diagram}

\end{figure*}

\section{Detailed Design}\label{sec:detailed_design}

In this section, we present the detailed design of \system{}, which consists of two modules: (1)~the \textbf{Function Call Graph Perturbation Module}, which injects dummy leaf functions to perturb the FCG topology and (2)~the \textbf{Classifier-Guided Injection Search Module}, which iteratively selects call sites and explores semantic NOP sequences at each site to minimize the classifier score. Each call-site perturbation instantiates one semantics-preserving transformation $\sigma_i \in \Sigma$, and the search algorithms below realize the composition $\Phi_{\boldsymbol{\sigma}}$ that produces the final adversarial binary $m^*$. The high-level overview is presented in Figure~\ref{fig:main_diagram}.

\subsection{Function Call Graph Perturbation Module}
We inject fully executable dummy functions into the FCG of compiled Windows PE malware via a call-site hijacking mechanism. At each targeted call site, the original \texttt{call} instruction is overwritten with an unconditional \texttt{jmp} into a newly injected code region or trampoline. The trampoline is carefully structured to \emph{first} perform the original \texttt{call} to its intended target, \emph{then} execute explicit \texttt{call} instructions to $k$ injected \emph{dummy leaf functions}, and \emph{finally} transfer control back to the instruction immediately following the original call. Here $k$ is sampled uniformly from $[k_{\min}, k_{\max}]$ independently per call-site injection (in our experiments, $k_{\min}/k_{\max}=1/3$ for MalConv and MalGraph, and $2{,}500/3{,}500$ for SAFE+GNN), so the number of new FCG nodes and edges introduced varies across sites. 

\noindent \textbf{Distinction from prior displacement-based approaches.}
The instruction displacement primitive was introduced
by Koo et al.~\cite{koo2016juggling}, who displace arbitrary
instruction sequences to a dedicated \texttt{.ropf} section as a
defense against return-oriented programming, preserving basic
block boundaries by design and introducing no new functions or
call edges. MalGuise~\cite{ling2024wolf} adapts this to
\texttt{call} instructions by splitting the enclosing basic block
and filling the resulting gap with semantic NOPs, which perturbs
CFG basic block structure but does not inject new functions or
modify the FCG. \system{} utilizes the same displacement
primitive but layers FCG perturbation on top of the existing
CFG modification: each displaced call site retains semantic NOP padding in the trampoline region while additionally anchoring
$k$ fully executable dummy functions that introduce new FCG
nodes and edges. The dummy functions can optionally be connected via inter-function call edges to form richer graph topologies
(evaluated as an ablation in Section~\ref{par:fcg_complexity}). This simultaneous perturbation of both the CFG and FCG is what
distinguishes \system{} from prior related displacement-based
approaches.

The trampoline's instruction ordering is semantically critical. Under stack-based x86 calling conventions~\cite{microsoftcallingconventions, fogcallingconventions}, the caller pushes arguments onto the stack before the \texttt{call} instruction. Since the original \texttt{call} is replaced by a \texttt{jmp} (which, unlike \texttt{call}, does not push a return address), the trampoline is entered with those arguments already live on the stack in the state the original callee expects, making it essential to issue the original \texttt{call} first. The subsequent dummy function calls are safe under both \texttt{stdcall} and \texttt{cdecl} conventions as the injected functions are parameterless leaf functions and consist of semantic NOP sequences, whose bodies preserve all registers and memory by construction, so they never dereference below their own return address and leave any residual caller arguments on the stack entirely untouched. The final \texttt{jmp}~$a_{\text{next}}$ returns control to the original execution flow, ensuring any caller-side stack cleanup present in the original code executes as normal and preserving original semantics unconditionally. The site-selection decision is fully decoupled from the injection step (Section~\ref{sec:call_site_selection}).

We target direct \texttt{call} instructions whose operand resolves to an internal user-defined function. Windows API calls and library imports are excluded, as their IAT-indirect encoding carries base-relocation entries that would be corrupted by an overwrite. Internal direct calls use PC-relative offsets with no relocation entry, making them safe to overwrite, and their uniform 5-byte encoding allows a same-sized unconditional \texttt{jmp} to the trampoline as a clean replacement.

\subsubsection{Dummy Leaf Functions}
\label{subsec:dummy-leaf}
Each dummy function is a self-contained leaf function placed sequentially in the injected section immediately after the trampoline. The function bodies are register and memory-neutral by construction, so execution leaves the program state unchanged despite the explicit call and return. We designed two variants of the dummy leaf functions depending on the target classifier. For MalConv and MalGraph targets, each dummy function starts with a standard callee prologue (\texttt{push ebp; mov ebp, esp}), followed by a body of self-canceling semantic NOP sequences, and ending with a matching epilogue (\texttt{pop ebp; ret}). These sequences are effective because these classifiers use bytes (raw bytes for MalConv, basic-block instruction counts for MalGraph) in their feature sets. For SAFE+GNN, which embeds entire function bodies as token sequences via a learned Word2Vec model, the semantic NOP body content has limited effect on evasion as the classifier's global mean pooling aggregates over function node embeddings rather than instruction content (see Section~\ref{sec:rq2}). We instead use \emph{minimal bodies} of the form \texttt{push reg; op reg, reg; pop reg; ret} ($\sim$5 bytes each, no prologue or epilogue), sampled from a weighted pool of five patterns selected through preliminary experiments (Appendix~\ref{app:hyperparameters}). Both designs are parameterless with no API calls, branches, or loops, prioritizing runtime reachability and stack safety. Enriching function bodies to better resemble benign code is discussed in Section~\ref{section:conclusion}.


\subsubsection{Graph Perturbation Effects}
The injection perturbs the sample's graph topology at three levels: \wcircle{1} \textbf{FCG nodes}: Each dummy function is a distinct function entry, adding $k$ new nodes per displacement; \wcircle{2} \textbf{FCG edges}: The trampoline's $k$ explicit \texttt{call} instructions add $k$ directed edges from the hijacked caller to each new dummy callee, altering its degree and neighborhood structure and \wcircle{3} \textbf{CFG basic blocks}: Each dummy body forms a single new basic block ending in \texttt{ret}, the trampoline introduces another, and the hijacked block now terminates with a \texttt{jmp}.

The complete construction is summarized in Algorithm~\ref{alg:patch-at-call}, which corresponds to the \textsc{PatchAtCall} method invoked by the search algorithms described in Section~\ref{sec:sa-sem-nop-search}. A conceptual illustration is also shown in Figure~\ref{fig:fcg_perturbation}.  We describe the implementation details of \textsc{PatchAtCall} in Appendix~\ref{app:patch-at-call}.

\begin{algorithm}[t]
\fontsize{7}{8}\selectfont
\caption{\textsc{PatchAtCall}: FCG Perturbation via Trampoline Injection}
\label{alg:patch-at-call}
\begin{algorithmic}[1]
\Require Malware $m$, call site $c$ (target $a_{\text{tgt}}$, next $a_{\text{next}}$), $k \sim \mathrm{Uniform}(k_{\min}, k_{\max})$, selected NOP subset $n$ (sampled by caller from $\mathcal{N}$), max repetitions $r$.
\Ensure Patched binary $\hat{m}$ with $k$ new FCG nodes and edges.

\For{$i = 1$ to $k$} \Comment{\textbf{Build $k$ dummy leaf functions}}
    \State $n_i \sim \textsc{Pick}(n)$ \Comment{Select NOPs from caller-provided subset}
    \State $\texttt{body}_i \gets [\texttt{push ebp};\ \texttt{mov ebp, esp}] \mathbin\Vert n_i \mathbin\Vert [\texttt{pop ebp};\ \texttt{ret}]$\footnotemark
\EndFor

\State Allocate contiguous region in new PE section; compute $\texttt{va}_{\text{tramp}}$
\State $\texttt{tramp} \gets \textsc{Asm}(\texttt{call } a_{\text{tgt}})$ \Comment{Original call \emph{first}}
\For{$i = 1$ to $k$}
    \State $r_i \sim \mathrm{Uniform}(0, r)$ \Comment{Repetition count for gap}
    \State $n_i^{\text{tramp}} \gets \textsc{Repeat}(\textsc{Pick}(n),\, r_i)$ \Comment{Inter-call NOP gap}
    \State $\texttt{tramp} \gets \texttt{tramp} \mathbin\Vert n_i^{\text{tramp}} \mathbin\Vert \textsc{Asm}(\texttt{call } \texttt{va}_{\texttt{body}_i})$
\EndFor
\State $n_{\text{final}} \gets \textsc{Repeat}(\textsc{Pick}(n),\, \mathrm{Uniform}(0, r))$
\Comment{Final NOP gap}
\State $\texttt{tramp} \gets \texttt{tramp} \mathbin\Vert n_{\text{final}} \mathbin\Vert \textsc{Asm}(\texttt{jmp } a_{\text{next}})$ \Comment{Return to original flow}

\State Write $\texttt{tramp} \mathbin\Vert \texttt{body}_1 \mathbin\Vert \cdots \mathbin\Vert \texttt{body}_{k}$ into injected PE section
\State Overwrite $c$ with $\textsc{Asm}(\texttt{jmp } \texttt{va}_{\text{tramp}})$ \Comment{Hijack call site}
\State \Return patched binary $\hat{m}$
\end{algorithmic}
\end{algorithm}
\footnotetext{For SAFE+GNN, $\texttt{body}_i$ is instead sampled from a pool of self-contained $\sim$5-byte patterns of the form \texttt{push reg; op reg, reg; pop reg; ret} with no prologue or epilogue (Appendix~\ref{app:hyperparameters}).}

\subsection{Call Site Selection}\label{sec:call_site_selection}
Given the full set of candidate call sites $\mathcal{C}$ for a sample, the default strategy in \system{} is a \textbf{random selection}. At each outer SA iteration, a call site is drawn uniformly at random from  $\mathcal{C}$.
We also evaluate structural ranking strategies as an ablation, ranking $\mathcal{C}$ by node-centrality measures (raw degree, betweenness, eigenvector) and local assembly instruction density surrounding the displaced call. No strategy consistently outperforms random selection across classifiers, so random remains the default. Full methodology and results are in Appendix~\ref{app:structural-filtering} and~\ref{app:structural-filtering-results}.

\subsection{Classifier-Guided Injection Search Module}\label{sec:sa-sem-nop-search}
Given a set of candidate call sites $\mathcal{C}$ for a malware sample, the search module determines which call sites to perturb and which semantic NOP sequences to use at each site to minimize the maliciousness probability $\mathcal{P}(m)$. The search space and semantic NOP combinations across multiple call sites are discrete, combinatorial, and expensive to explore exhaustively. Hence, we experiment with two search strategies of increasing exploration capability: \wcircle{1}~a \textbf{Greedy} search that purely exploits classifier feedback to select the locally best-performing semantic NOPs and \wcircle{2}~a \textbf{Simulated Annealing (SA)–inspired} search that augments greedy exploitation with probabilistic acceptance of worse moves (NOP selection) to escape local minima. The per-site search depth and call-site coverage are externally configurable, enabling the same framework across classifiers with different NOP sensitivities.

\begin{figure}[!htbp]
\centering
\includegraphics[width=0.78\columnwidth]{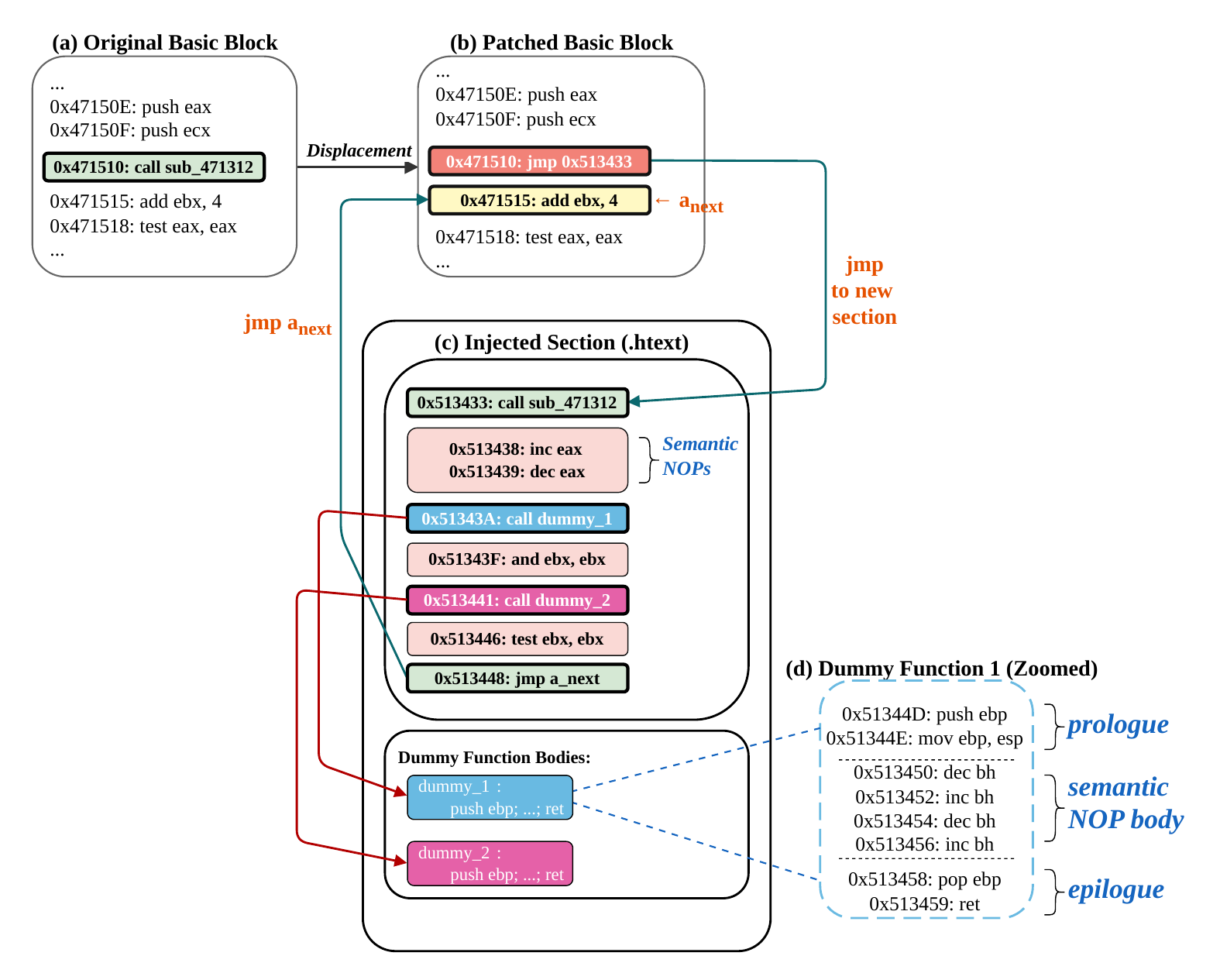}
\caption{Illustration of the FCG perturbation method \textsc{PatchAtCall}. 
(a)~A \texttt{call} instruction targeting an internal function is selected for 
perturbation. (b)~The original \texttt{call} is overwritten with a \texttt{jmp} 
into the injected section (\texttt{.htext}). (c)~The trampoline first issues the 
original \texttt{call} to preserve stack semantics, then makes explicit 
calls to $k$ dummy functions, and finally jumps back to $a_{\text{next}}$. 
(d)~Dummy function body structure for MalConv/MalGraph targets; SAFE+GNN uses minimal bodies.}
\label{fig:fcg_perturbation}
\end{figure}

\subsubsection{\textbf{Greedy Search}}
Greedy search evaluates candidate NOP injections and deterministically selects those yielding the best immediate reduction in classifier score, prioritizing local exploitation without stochastic exploration.
Concretely, at each call site the inner loop samples candidate semantic NOPs, patches the sample via \textsc{PatchAtCall} (Algorithm~\ref{alg:patch-at-call}), and queries the classifier $\mathcal{P}(\cdot)$. The candidate yielding the lowest score is accepted only if it strictly improves the site-level best $p_{\text{site}}$. It returns $(m^*, p^*, \texttt{bypassed})$, where $m^*$ is the adversarial binary on bypass or the best non-bypassing sample otherwise. Because greedy search never accepts worse intermediate states, it performs no exploration and may stall in local minima. The full procedure is given in Algorithms~\ref{alg:greedy-sem-nop} and~\ref{alg:local-greedy} (Appendix~\ref{app:greedy-outer}).

\subsubsection{\textbf{SA-Inspired Search Algorithm}}\label{subsec:sa-sem-nop}
Simulated Annealing (SA)~\cite{kirkpatrick1983optimization} is a metaheuristic for discrete optimization that escapes local minima by accepting worse moves with probability $\exp(-\Delta E / T)$, where $T$ is a temperature decreased over time via a cooling schedule~\cite{aarts1988quantitative,sechen2003timberwolf}. It has been applied in the security domain for fuzzing, robustness evaluation, and hardware security~\cite{bohme2017directed,eykholt2023uret,wang2025optilock,10.1145/1982185.1982509}. We adapt SA to semantic NOP selection by treating the maliciousness probability $\mathcal{P}(m) \in [0,1]$ as the energy to minimize below the classifier threshold $\tau^{clsf}$, which decouples the search from the classifier's input representation so the same framework applies whether the target operates on raw bytes or graphs.

For a given call site $c \in \mathcal{C}$, semantic NOPs are sampled uniformly at random from $\mathcal{N}$ (set of semantic NOPs). The inner SA loop (\textsc{LocalSA}, Algorithm~\ref{alg:local-sa}) explores semantic NOP combinations for that fixed site, patching the sample at $c$ via \textsc{PatchAtCall} (Algorithm~\ref{alg:patch-at-call}) and scoring each candidate with $\mathcal{P}(\cdot)$. Rather than modifying the binary in-place on every SA acceptance, we maintain a scalar reference score $p_{\text{local}}$ that drifts under SA transitions, where improvements ($\delta > 0$) are always accepted and worse moves are accepted with probability $\exp(\delta/T)$. Only the best-scoring candidate $m_{\text{site}}$ found during the inner loop is propagated to the global state $m^*$. After each outer iteration, $\textsc{UpdateCallSiteList}(\mathcal{C}, c, m^*)$ appends any new \texttt{call} instructions formed by displacing $c$.

\begin{algorithm}[t]
\fontsize{7}{8}\selectfont
\caption{SA-Driven Semantic NOP Search over Call Sites}
\label{alg:sa_sem_nop}
\begin{algorithmic}[1]
\Require Malware $m_0$, classifier $\mathcal{P}(\cdot)$ with threshold $\tau^{clsf}$, call sites $\mathcal{C}$, NOP library $\mathcal{N}$, size budget $B$, temperatures $T_0, T_{\min}$, max iterations $I_{\max}$.
\Ensure $(m^*, p^*, \texttt{bypassed})$: best binary, its score, and evasion flag.
\State $m^* \gets m_0$,\ $p^* \gets \mathcal{P}(m_0)$,\ $T \gets T_0$
\State $\texttt{no\_improve} \gets 0$,\ $\texttt{bypassed} \gets \texttt{false}$,\ $i \gets 0$
\While{$i < I_{\max}$ \textbf{and} $T \ge T_{\min}$ \textbf{and} $\mathcal{C} \neq \emptyset$ \textbf{and} $\texttt{no\_improve} < \texttt{patience}$}
    \If{$p^* < \tau^{clsf}$} $\texttt{bypassed} \gets \texttt{true}$;\ \textbf{break} \EndIf
    \State $p^*_{\text{(prev)}} \gets p^*$
    \State Select call site $c \in \mathcal{C}$
    \State $(m_{\text{site}}, p_{\text{site}}, b_{\text{in}}) \gets \textsc{LocalSA}(c, m_0, m^*, p^*, T, \mathcal{N}, B, J_{\max}, \tau^{clsf})$
    \If{$b_{\text{in}}$} $m^* \gets m_{\text{site}},\ p^* \gets p_{\text{site}},\ \texttt{bypassed} \gets \texttt{true}$;\ \textbf{break} \EndIf \Comment{Evasion; exit}
    \If{$p_{\text{site}} < p^*$} \Comment{Update best}
        \State $m^* \gets m_{\text{site}}$,\ $p^* \gets p_{\text{site}}$,\ $\texttt{no\_improve} \gets 0$
    \Else
        \State $\texttt{no\_improve} \gets \texttt{no\_improve} + 1$
    \EndIf
    \State \textsc{UpdateCallSiteList}$(\mathcal{C}, c, m^*)$ \Comment{Append new calls from displacement}
    \State $\delta_{\text{site}} \gets p^*_{\text{(prev)}} - p_{\text{site}}$ \Comment{For cooling schedule}
    \State $T \gets \textsc{UpdateTemp}(T, \delta_{\text{site}})$ \Comment{Cool or reanneal}
    \State $i \gets i + 1$
\EndWhile
\State \Return $(m^*, p^*, \texttt{bypassed})$
\end{algorithmic}
\vspace{-4pt}
\end{algorithm}

We update the temperature according to one of two cooling schedules, corresponding to our two SA variants. \textbf{Classic exponential cooling} applies a fixed multiplicative decay $T \leftarrow \alpha T$ with $\alpha \in (0,1)$. \textbf{Adaptive cooling with reannealing} tracks site-level score changes $\delta_{\text{site}}$ over a short history window, cooling more aggressively on consistent progress and raising the temperature up to a capped fraction of $T_0$ when scores stall. Setting $T=0$ reduces Algorithm~\ref{alg:local-sa} to the greedy procedure (Appendix~\ref{app:greedy-outer}). 

\noindent \textbf{Classifier-specific configuration.} For MalConv and MalGraph, we use the semantic NOP templates from MalGuise~\cite{ling2024wolf}, sampling $s$ distinct NOP types uniformly at random at each step and repeating each up to $r$ times. We use diverse NOP sequences rather than single dead-code instructions because instruction diversity significantly affects evasion for graph-based classifiers (Appendix~\ref{app:nop-sensitivity}). For SAFE+GNN, inner NOP search iterations are reduced, and the search budget is allocated toward outer call-site iterations (Algorithm~\ref{alg:sa_sem_nop}), with minimal bodies (Section~\ref{subsec:dummy-leaf}), as this classifier responds primarily to injection volume rather than body content. Default hyperparameters are in Appendix~\ref{app:hyperparameters}.




\begin{algorithm}[t]
\fontsize{7}{8}\selectfont
\caption{\textsc{LocalSA}: Inner SA Loop at a Fixed Call Site}
\label{alg:local-sa}
\begin{algorithmic}[1]
\Require Call site $c$, original malware $m_0$, global state $m^*$ with score $p^*$, temperature $T$, NOP library $\mathcal{N}$, size budget $B$, max inner iterations $J_{\max}$, classifier threshold $\tau^{clsf}$.
\Ensure $(m_{\text{site}}, p_{\text{site}}, \texttt{bypassed})$: best state for this call site, its score, and evasion flag.
\State $p_{\text{local}} \gets p^*$,\ $m_{\text{site}} \gets m^*$,\ $p_{\text{site}} \gets p_{\text{local}}$ \Comment{SA reference score}
\State $\texttt{bypassed} \gets \texttt{false}$
\For{$j = 1$ to $J_{\max}$}
    \State $n_j \sim \textsc{SelectSemNops}(\mathcal{N})$
    \State $\hat{m} \gets \textsc{PatchAtCall}(m^*, c, n_j)$ \Comment{From global $m^*$}
    \If{\textsc{SizeIncrease}$(\hat{m}, m_0) > B$} \textbf{continue} \EndIf
    \State $p_{\text{cand}} \gets \mathcal{P}(\hat{m})$
    \If{$p_{\text{cand}} < \tau^{clsf}$} $m_{\text{site}} \gets \hat{m}$,\ $p_{\text{site}} \gets p_{\text{cand}}$,\ $\texttt{bypassed} \gets \texttt{true}$
    \State \textbf{return} $(m_{\text{site}}, p_{\text{site}}, \texttt{bypassed})$
    \EndIf
    \State $\delta \gets p_{\text{local}} - p_{\text{cand}}$ 
    
    
    \If{$\delta > 0$} $p_{\text{local}} \gets p_{\text{cand}}$ \Comment{Accept improvement} 
        \If{$p_{\text{cand}} < p_{\text{site}}$} $m_{\text{site}} \gets \hat{m}$,\ $p_{\text{site}} \gets p_{\text{cand}}$ \Comment{Update site best}
        \EndIf
    \ElsIf{$\textsc{Uniform}(0,1) \le \exp(\delta / T)$} $p_{\text{local}} \gets p_{\text{cand}}$\Comment{SA Accept}
    \EndIf
\EndFor
\State \Return $(m_{\text{site}}, p_{\text{site}}, \texttt{bypassed})$
\end{algorithmic}
\end{algorithm}

\subsubsection{Binary Patching}
\label{subsec:binary-patching}
For reconstructing adversarial binaries, we first allocate space for all relocated call instructions, semantic NOP blocks, and trampoline jumps in a \emph{new} code section appended to the end of the PE file (the trampoline). In our implementation, we use the new section of the binary and align its size to a multiple of the architecture page size (4~KB), as required by Windows PE specifications~\cite{pespecifications}. To ensure a fair comparison with the MalGuise baseline, we also evaluate their attack under the same new-section configuration.

\section{Evaluation}\label{sec:evaluation}
This section 
conducts an extensive set of evaluations aiming to answer
the following research questions:
\begin{compactitem}
    \item \textbf{RQ1} -  How effectively does \system{} evade detection by state-of-the-art ML-based malware detectors?
    \item \textbf{RQ2} - What is the contribution of each attack component to evasion effectiveness, and what is the resulting structural footprint of the generated adversarial samples?
    \item \textbf{RQ3} - How sensitive is attack effectiveness to injection topology and search hyperparameters?
    \item \textbf{RQ4} - Do the generated adversarial samples preserve the semantics and functionality of the unperturbed originals?
    \item \textbf{RQ5} - How robust is \system{} against defensive countermeasures?
\end{compactitem}

\subsection{Evaluation Setup}

\subsubsection{Target Systems Details}
We evaluate \system{} against three ML-based static Windows malware classifiers representing distinct detection paradigms:

\wcircle{1} \textbf{MalConv}~\cite{raff2018malware}, a CNN-based classifier that operates directly on raw malware bytes, widely used in both white and black-box adversarial malware generation ~\cite{lucas2021malware, ling2024wolf, li2025minimal, tian2024functionality, wang2020mdea}.

\wcircle{2} \textbf{MalGraph}~\cite{ling2022malgraph}, a hierarchical GNN that combines CFG-level basic-block features with FCG-level function aggregation. For both MalConv and MalGraph, we use the pretrained models and FPR thresholds from~\cite{ling2024wolf}, trained on 210{,}251 Windows executables (108,610 goodware and 101,641 malware across 848 malware families)~\cite{ling2022malgraph}.

\wcircle{3} \textbf{SAFE+GNN}, a three-layer GraphSAGE~\cite{graphsage} classifier that operates exclusively on the FCG, with each function node represented by a 100-dimensional Word2Vec~\cite{mikolov2013efficient} embedding of its tokenized disassembly and the graph-level representation obtained via global mean pooling. We trained this model on a 2024-collected corpus (Section~\ref{sec:dataset}) as a binary classifier, achieving 95.25\% test accuracy. FPR thresholds were calibrated on a 7886 held-out benign samples. Architecture, training and dataset details are in Appendix~\ref{app:SAFE+GNN} and~\ref{app:SAFE+GNN_dataset}.

We compare \system{} with two state of the art adversarial malware generation baselines. For fairness, we use the same evaluation settings specified in their original implementations.

\wcircle{1} \textbf{MalGuise}~\cite{ling2024wolf} is a problem-space black-box adversarial attack that manipulates the CFG via call-based redividing, using Monte-Carlo Tree Search to select call sites and choose semantic NOPs to inject. It is our primary baseline because it operates in the same threat model and attack surface as \system{}. For fairness, all shared parameters between \system{} and Malguise (size budget, timeout, search iterations) are kept identical within each classifier and search configuration.

\wcircle{2} \textbf{SRL}~\cite{zhang2022semantics} is a black-box, feature-space attack that uses reinforcement learning to iteratively modify the malware's CFG by injecting semantic NOPs. We repeatedly contacted the author but received no response. Hence, we implemented a faithful re-implementation of the work and used it for evaluation. The maximum number of iterations is set to 100, and the number of modified blocks to 200. Because it operates in feature space and produces perturbed adversarial CFGs rather than executables, it is compatible only with the graph-based detector and cannot be evaluated against MalConv, which operates on raw bytes.

\subsubsection{Target and Calibration Datasets}
\label{sec:dataset}
We collected x86 Windows PE malware samples from MalwareBazaar~\cite{MalwareBazaar} (2025 submissions) and partitioned them into classifier-specific pools after removing packed binaries using DIE~\cite{die} and filtering for samples with at least one identifiable \texttt{call} instruction. MalConv requires $\leq$1.9\, MB to remain within its 2\, MB input limit after section injection, MalGraph uses an ${\sim}$8.4\, MB outlier cutoff, and SAFE+GNN uses a 10\, KB-10\, MB range to accommodate the larger dummy function node injection volumes. Each pool was scored at 0.1\% and 1\% FPR, retaining only samples above the classifier threshold. The final pools are 559/3{,}662 (MalConv), 2{,}605/3{,}234 (MalGraph), and 1{,}000/1{,}000 (SAFE+GNN) at 0.1\%/1\% FPR. SAFE+GNN pools are smaller because higher per-sample dummy-node injection increases sample processing time.

SAFE+GNN was trained from scratch on a separate 2024-collected corpus consisting of 14{,}000 malware from the MalwareBazaar corpus, 17{,}886 benign from the Practical Security Analytics dataset~\cite{lesterpsadataset} (details in Appendix~\ref{app:SAFE+GNN_dataset}). To assess the impact of temporal separation between the 2024 training and 2025 test corpora, we verify that 72-74\% of attack-pool families appear in the training set, accounting for 95-97\% of test samples by count (Appendix~\ref{app:family_overlap}). For the SRL baseline, we trained on 2{,}000 MalwareBazaar samples (March-August 2024). For defense evaluation (Section~\ref{sec:rq5}), we finetuned on 300 adversarial samples with 300 benign samples from 2024, with separate calibration sets. All pools are disjoint from the 2025 test set. Additional details are in Appendix~\ref{app:dataset-details}.

\subsection{Evaluation Metrics}
\label{sec:metrics}
We evaluate \system{} along three dimensions: evasion success, semantic preservation, and generation throughput.

Let $f(m) = \mathbf{1}[\mathcal{P}(m) \geq \tau^{clsf}]$ denote the hard classification decision of the target classifier, where $\mathcal{P}(m)$ is the maliciousness probability and $\tau^{clsf}$ is the classifier threshold. Let $\mathcal{M}_{\text{orig}} = \{m \in \mathcal{M} \mid f(m)=1\}$ be the set of samples  classified as malicious, and $\mathcal{M}_{\text{adv}} = \{m^* \mid \exists\, m \in \mathcal{M}_{\text{orig}}: m^* \sim m \land f(m^*) = 0\}$ the set of successful adversarial samples.

\noindent\textbf{Attack Success Rate (ASR)}:
ASR is the most commonly used metric for assessing adversarial attacks against malware classifiers~\cite{lucas2021malware, ling2024wolf, zhang2022semantics, ling2019deepsec}. It measures the fraction of adversarial samples, generated from originally correctly classified malicious inputs, that successfully evade the target classifier. Formally, it is defined as: 
\begin{equation}
    ASR = \frac{|\mathcal{M}_{adv}|}{|\mathcal{M}_{orig}|}
\end{equation}

\noindent\textbf{Semantics Preservation Rate (SPR)}:
To assess whether evasive adversarial samples preserve their original malicious behavior, we measure the fraction of successful evasions that remain semantically equivalent to the original malware. 
Following prior work~\cite{ling2024wolf}, we define:
\begin{equation}
    \text{SPR} = \frac{|\{m^* \in \mathcal{M}_{\text{adv}} \mid \text{Sem}(m, m^*) = 1\}|}{|\mathcal{M}_{\text{adv}}|}
\end{equation}

where $m$ denotes the original sample from which $m^*$ was derived, and $\text{Sem}(\cdot,\cdot)$ is a semantic equivalence function that determines whether $m^*$ preserves the original malicious behavior. The details of this function is provided in Section~\ref{sec:rq4}.


\noindent \textbf{Attack Throughput ($\Theta$)}:
ASR alone does not characterize practical threat, as a high-ASR attack requiring hours of computation poses a different risk than one achieving the same rate in minutes. We therefore measure the rate at which successful adversarial samples are produced: 
\begin{equation}
    \Theta = \frac{|\mathcal{M}_{\text{adv}}|}{\sum_{m^* \in \mathcal{M}_{\text{adv}}} t(m^*)}
\end{equation}
where $t(m^*)$ is the wall-clock generation time for adversarial sample $m^*$. We report throughput in successful bypassed samples generated per hour, rounded to the nearest integer.

\subsubsection{Implementation Details}
\system{} is implemented in Python and evaluated on a server with 3 RTX 3090 GPUs, 252 GB RAM, and 48 processors. We use IDA Home~\cite{idapro} for call site extraction, disassembly, and CFG feature computation. Binary patching uses LIEF\footnote{\url{https://lief.re/}} for PE section injection and pefile\footnote{\url{https://github.com/erocarrera/pefile}} for call-site overwriting. No instruction relocation, reference fixup, or disassembly-reassembly is required. We adapt the section injection and call-site displacement primitives from~\cite{ling2024wolf}. We implement the trampoline construction, dummy function assembly, address calculations, FCG manipulation pipeline, and all search algorithms. E9patch\footnote{\url{https://github.com/GJDuck/e9patch}} was considered but supports only x86\_64, whereas our corpus is 32-bit x86.


\begin{table}[t]
\centering
\caption{ASR (\%) and Throughput $\Theta$ (successful
variants/hr) against three target classifiers.
\textsuperscript{\dag}Ceiling: 100\,\% ASR at MalConv
0.1\,\% FPR (559 samples).
\colorbox{sawin}{\strut Best} and
\colorbox{sarunner}{\strut Second-best} values per column.}
\label{tab:full_comparison}
\setlength{\tabcolsep}{2.2pt}
\renewcommand{\arraystretch}{1.1}
\resizebox{\columnwidth}{!}{%
\scriptsize
\begin{tabular}{@{}lcccccccccccc@{}}
\toprule
 & \multicolumn{4}{c}{\textbf{MalConv}} & \multicolumn{4}{c}{\textbf{MalGraph}} & \multicolumn{4}{c}{\textbf{SAFE+GNN}} \\
\cmidrule(lr){2-5} \cmidrule(lr){6-9} \cmidrule(lr){10-13}
\multirow{2}{*}{\textbf{Strategy}} & \multicolumn{2}{c}{0.1\,\%} & \multicolumn{2}{c}{1\,\%} & \multicolumn{2}{c}{0.1\,\%} & \multicolumn{2}{c}{1\,\%} & \multicolumn{2}{c}{0.1\,\%} & \multicolumn{2}{c}{1\,\%} \\
\cmidrule(lr){2-3} \cmidrule(lr){4-5} \cmidrule(lr){6-7} \cmidrule(lr){8-9} \cmidrule(lr){10-11} \cmidrule(lr){12-13}
 & ASR & $\Theta$ & ASR & $\Theta$ & ASR & $\Theta$ & ASR & $\Theta$ & ASR & $\Theta$ & ASR & $\Theta$ \\
\midrule
MalGuise & 100.0\textsuperscript{\dag} & 649 & 97.05 & 70 & 91.09 & 22 & 82.16 & 14 & 1.90 & 6 & 0.9 & 7 \\
SRL & -- & -- & -- & -- & 51.82 & \hlwin{1352} & 30.12 & \hlwin{587} & 0.10 & -- & 0.0 & -- \\
\midrule
\system{} (Greedy) & 100.0\textsuperscript{\dag} & \hlwin{1234} & 98.31 & \hlwin{245} & \hlwin{96.81} & \hlrun{36} & \hlwin{96.94} & \hlrun{26} & \hlwin{97.40} & \hlwin{42} & \hlwin{85.00} & \hlwin{9} \\
\system{} (SA) & 100.0\textsuperscript{\dag} & 930 & \hlwin{99.45} & 148 & 95.28 & 28 & \hlrun{94.53} & 20 & 88.70 & 16 & 64.50 & 5 \\
\system{} (Ad.\ SA) & 100.0\textsuperscript{\dag} & \hlrun{945} & \hlrun{99.40} & \hlrun{150} & \hlrun{96.01} & 30 & 94.28 & 20 & \hlrun{88.60} & \hlrun{16} & \hlrun{64.50} & \hlrun{5} \\
\bottomrule
\end{tabular}%
}
\end{table}

\subsection{Evaluation Results}

\subsubsection{\textbf{RQ1: Attack Effectiveness against Target Classifiers}}
\label{sec:rq1}

Table~\ref{tab:full_comparison} compares \system{} with all baselines across the target classifiers at both FPRs.
We report both the ASR and $\Theta$.


SRL achieves substantially lower ASR than \system{} across all graph-based classifiers (Table~\ref{tab:full_comparison}), and produces near-zero ASR against SAFE+GNN as its feature-space CFG mutations (instruction count additions) do not alter FCG topology. SRL's throughput advantage is not practically meaningful because it is a feature-space attack that doesn't produce exeuctable malware.



MalGuise is the most directly comparable baseline, operating in the same threat model and attack surface (see Section ~\ref{sec:detailed_design} for the detailed mechanism comparison). The key difference in results stems from \system{}'s dual CFG+FCG perturbation, which directly targets graph-based classifiers' structural features.

In our experiments, we consider the Graph-based classifiers (MalGraph and SAFE+GNN) as the primary evaluation targets, whereas MalConv evaluation serves a complementary purpose to test whether the structural perturbations transfer to a classifier that does not observe graph topology. The largest gains over MalGuise appear on graph-based classifiers, where dual CFG+FCG perturbation directly targets the classifier's structural features. \system{} exceeds MalGuise by up to +14.78,pp ASR and 1.85$\times$ throughput on MalGraph. For SAFE+GNN, MalGuise achieves $<$2\% ASR while \system{} reaches 85-97\% while being 7$\times$ faster at 0.1\%FPR, demonstrating that CFG-only perturbation is insufficient for classifiers operating on FCG topology. Against MalConv, both methods achieve comparable ASR and the distinction is throughput, where \system{} is 1.43-3.5$\times$ faster.


The attack succeeds on graph-based classifiers because their pooling layers aggregate over all function nodes without distinguishing injected from original, so each injected dummy function directly shifts the learned representation. This holds under a fully black-box threat model as the attacker does not need knowledge of classifiers internal feature extraction pipeline. For MalConv, the injected section and displaced call sites perturb the raw byte content directly. More details on how the attack affects each classifier's internal representation and feature space visualization is provided in Appendix~\ref{app:feature-space}.

All samples in Table~\ref{tab:full_comparison} were correctly classified as malware above the respective FPR thresholds before the attack. To further verify that the observed ASR is not driven by borderline classifications, we restrict the evaluation to samples where the classifier assigned P(malware) $\geq 0.95$ before any perturbation, and across all classifiers and thresholds, the ASR remains 84.5-100\% with full confidence sweep in Appendix~\ref{app:high_confidence}. The family overlap analysis (See Appendix~\ref{app:family_overlap}) confirms training-test corpus consistency. 


Among \system{}'s three search variants, Greedy achieves the highest ASR and throughput in nearly all configurations. However, as we analyze in
Section~\ref{sec:ablation}, the SA variants achieve comparable evasion rates with a smaller perturbation footprint for MalConv and MalGraph. For SAFE+GNN, Greedy leads SA by +20\,pp at 1\% FPR, and the factors behind this gap are examined in Section~\ref{sec:ablation}. We also evaluate transferability to commercial antivirus engines (Appendix~\ref{app:av_analysis}) and show that traditional obfuscation tools (UPX, Enigma) achieve negligible ASR ($\leq$3.76\%) across MalConv and MalGraph, confirming that packing-based approaches are ineffective against learning-based classifiers (Appendix~\ref{app:upx_enigma}).

\begin{findingbox}
\textbf{Answer to RQ1:} \system{} consistently matches or exceeds MalGuise in both ASR and throughput. The largest gains appear on graph-based classifiers, where dual CFG+FCG perturbation exploits structural features that CFG-only perturbation cannot reach. Feature-space attacks (SRL) are ineffective, confirming FCG topology perturbation as the essential attack surface.
\end{findingbox}
%
\subsubsection{\textbf{RQ2: Ablation Studies and Attack Footprint}}\label{sec:rq2}\mbox{}\par
\noindent\textbf{Ablation Studies.}
\label{sec:ablation}
We identify two key components of \system{} for ablation:
\wcircle{1}~FCG perturbation via dummy function injection, and
\wcircle{2}~the semantic NOP body content and associated search
algorithm. For each component, we compare \system{} against a
variant with that component disabled or replaced, while keeping
all other components constant. For MalConv and MalGraph, ablations use the full test pool.
For SAFE+GNN, ablations use 209 samples unless
noted otherwise, as the larger injection volume increases
per-sample IDA processing time and attack cost.


\textbf{\wcircle{1} Effect of FCG Perturbation.}
FCG perturbation adds new function nodes and call edges to the recovered
graph. To isolate its effect, we construct a CFG-only variant that still
displaces the call through a trampoline, executes the original call, and
returns to the original control flow, but injects no dummy function nodes.
MalConv and MalGraph use Adaptive SA with semantic-NOP padding in both
variants and SAFE+GNN uses Greedy without semantic-NOP padding because those
patterns are ineffective for that classifier. The search strategy is held
fixed within each comparison.


Table~\ref{tab:ablation_fcg} shows only a marginal ASR benefit
($\leq$1\,pp) for byte-level MalConv, whereas MalGraph gains
7.37-10.15\,pp ASR and approximately 1.43$\times$ throughput. The
effect is most pronounced for SAFE+GNN where removing FCG edges while retaining
the same call-site displacement reduces ASR from 97.4\% to 0.9\% at
0.1\% FPR and from 85.0\% to 0.1\% at 1\% FPR. For this FCG-only
classifier, perturbing graph topology is therefore the effective evasion mechanism.

\begin{table}[tb]
\centering
\caption{Ablation: FCG perturbation. MalConv and MalGraph use Adaptive\,SA and SAFE+GNN uses Greedy. Highlights: \colorbox{sawin}{\strut Better} value per column and \colorbox{deltagain}{\strut Notable $\Delta$.}}
\label{tab:ablation_fcg}
\renewcommand{\arraystretch}{1.05}
\resizebox{\columnwidth}{!}{%
\begin{tabular}{@{}l*{12}{c}@{}}
\toprule
 & \multicolumn{4}{c}{\textbf{MalConv}} & \multicolumn{4}{c}{\textbf{MalGraph}} & \multicolumn{4}{c}{\textbf{SAFE+GNN}} \\
\cmidrule(lr){2-5} \cmidrule(lr){6-9} \cmidrule(lr){10-13}
\multirow{2}{*}{\textbf{Variant}} & \multicolumn{2}{c}{0.1\,\%} & \multicolumn{2}{c}{1\,\%} & \multicolumn{2}{c}{0.1\,\%} & \multicolumn{2}{c}{1\,\%} & \multicolumn{2}{c}{0.1\,\%} & \multicolumn{2}{c}{1\,\%} \\
\cmidrule(lr){2-3} \cmidrule(lr){4-5} \cmidrule(lr){6-7} \cmidrule(lr){8-9} \cmidrule(lr){10-11} \cmidrule(lr){12-13}
 & ASR & $\Theta$ & ASR & $\Theta$ & ASR & $\Theta$ & ASR & $\Theta$ & ASR & $\Theta$ & ASR & $\Theta$ \\
\midrule
w/ FCG & \hlwin{100.0} & 945 & \hlwin{99.40} & 150 & \hlwin{96.01} & \hlwin{30} & \hlwin{94.28} & \hlwin{20} & \hlwin{97.40} & 42 & \hlwin{85.00} & 9 \\
w/o FCG & 99.46 & 946 & 98.36 & \hlwin{212} & 88.64 & 21 & 84.13 & 14 & 0.90 & \hlwin{55} & 0.10 & \hlwin{25} \\
\midrule
$\Delta$ & $+$0.54 & $\approx$ & $+$1.04 & $-$1.4$\times$ & \hldelta{$+$7.37} & \hldelta{$+$1.4$\times$} & \hldelta{$+$10.15} & \hldelta{$+$1.4$\times$} & \hldelta{$+$96.50} & $-$1.3$\times$ & \hldelta{$+$84.90} & $-$2.8$\times$ \\
\bottomrule
\end{tabular}%
}
\end{table}

\textbf{\wcircle{2} Effect of Search Algorithm and Function Body
Content.}
The second component we ablate is the semantic NOP body content
and the iterative search over it. For MalConv and MalGraph, we
compare Adaptive SA (which iterates over NOP candidates at each
call site) against a single-pass configuration that randomly samples
one NOP configuration per site without explicit search.
Table~\ref{tab:search_ablation} shows that iterative search is critical, improving ASR by $\sim$26-62 pp across configurations. Without it, fewer than half of samples evade at 1\% FPR.

\begin{figure}[tb]
\centering
\includegraphics[width=0.5\columnwidth]{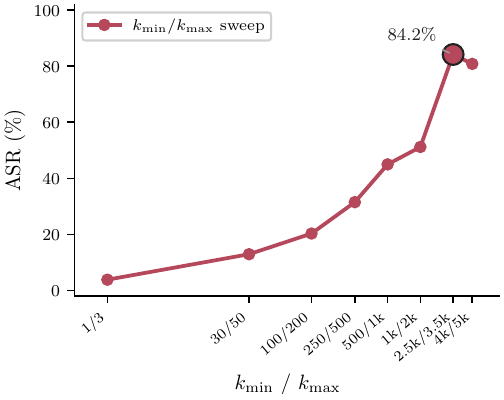}
\caption{ASR vs.\ $k_{\min}$/$k_{\max}$ for SAFE+GNN
at 1\% FPR (Greedy, 209 samples per configuration, x-axis log scaled).
The highlighted point marks the
Table~\ref{tab:full_comparison} configuration.}
\label{fig:k_sensitivity}
\end{figure}

\textbf{SAFE+GNN Function Body Ablation.}
For SAFE+GNN, we evaluate whether function body content affects evasion by replacing the minimal bodies ($\sim$5 bytes, $k_{\min}/k_{\max}$=2{,}500/3{,}500) with regular dummy function bodies ($\sim$11-13 bytes, prologue/epilogue + semantic NOPs), evaluated on 209 samples at 1\% FPR with Greedy search. With minimal bodies at $k$=2{,}500/3{,}500, ASR is 84.21\%. Replacing them with regular bodies drops ASR to 17.2\% ($k$=1{,}500/2{,}000), and adding NOP padding further reduces it to 8.2\% ($k$=500/1{,}000). Because regular bodies are larger, they reduce the maximum $k$ within the size budget. To verify that this is not an artifact of reduced $k$, we cross-check against the $k$-sensitivity results (Figure~\ref{fig:k_sensitivity}) at matching $k$ ranges, which confirms that NOP content itself degrades evasion, not the reduction in $k$ alone. Hence, the useful parameters to tune for SAFE+GNN are the number of injected nodes via outer call-site iterations and the choice of function-body design.  

\begin{figure}[tb]
\centering
\includegraphics[width=\columnwidth]{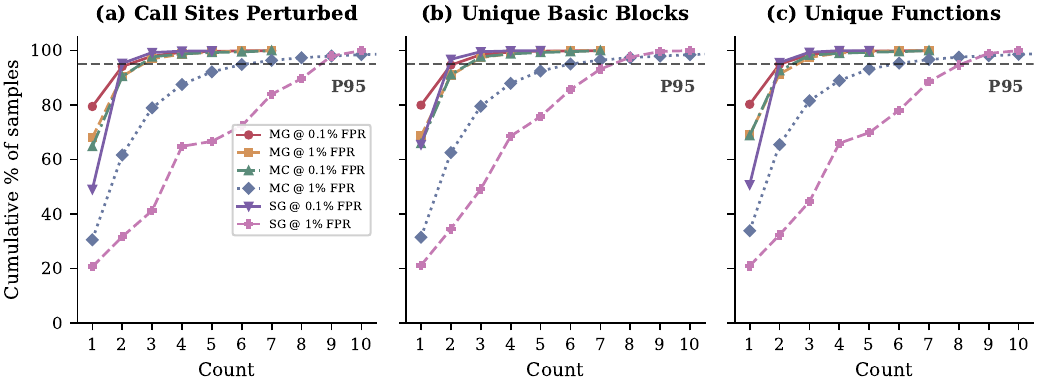}
\caption{CDF of structural perturbations per bypassed sample, aggregated across all three search algorithms. Each subplot overlays the call sites, unique basic blocks, and perturbed functions. The dashed line marks the 95th percentile.}
\label{fig:footprint_cdf}
\end{figure}

\textbf{Attack Footprint.}
\label{sec:rq2_footprint}
To assess the footprint of our proposed attack, we measure the number of call sites, unique basic blocks and functions perturbed for the evading samples. Figure~\ref{fig:footprint_cdf} presents the cumulative distribution across all six configurations, aggregated across all three search algorithms.

\begin{table}[tb]
\centering
\caption{Ablation: Iterative NOP search. Both variants use FCG perturbation with Random call site selection. 
Highlights: \colorbox{sawin}{\strut Better} value per column, \colorbox{deltagain}{\strut Notable $\Delta$}.}
\label{tab:search_ablation}
\renewcommand{\arraystretch}{1.05}
\resizebox{\columnwidth}{!}{%
\scriptsize
\begin{tabular}{@{}l*{8}{c}@{}}
\toprule
 & \multicolumn{4}{c}{\textbf{MalConv}} & \multicolumn{4}{c}{\textbf{MalGraph}} \\
\cmidrule(lr){2-5} \cmidrule(lr){6-9}
\multirow{2}{*}{\textbf{Variant}} & \multicolumn{2}{c}{0.1\,\% FPR} & \multicolumn{2}{c}{1\,\% FPR} & \multicolumn{2}{c}{0.1\,\% FPR} & \multicolumn{2}{c}{1\,\% FPR} \\
\cmidrule(lr){2-3} \cmidrule(lr){4-5} \cmidrule(lr){6-7} \cmidrule(lr){8-9}
 & ASR & $\Theta$ & ASR & $\Theta$ & ASR & $\Theta$ & ASR & $\Theta$ \\
\midrule
Ad.\ SA & \hlwin{100.0} & 945 & \hlwin{99.40} & 150 & \hlwin{96.01} & 30 & \hlwin{94.28} & 20 \\
1-Pass & 73.52 & \hlwin{2571} & 37.90 & \hlwin{1140} & 48.37 & \hlwin{96} & 41.84 & \hlwin{94} \\
\midrule
$\Delta$ & \hldelta{$+$26.48} & $-$2.7$\times$ & \hldelta{$+$61.50} & $-$7.6$\times$ & \hldelta{$+$47.64} & $-$3.2$\times$ & \hldelta{$+$52.44} & $-$4.7$\times$ \\
\bottomrule
\end{tabular}%
}
\end{table}
Across MalConv and MalGraph, the median number of perturbed call sites is~1, with 65-80\% of samples achieving evasion through a single call site modification (Figure~\ref{fig:footprint_cdf}). Even the hardest configuration (MalConv at 1\% FPR) requires a median of only~2 sites, with the 95th percentile at~7. The median size overhead of injected shellcode is 0.21-0.41\% of the original binary and median basic block size at perturbed call sits is $\sim$31-37 bytes. For SAFE+GNN, the call site count is comparable (median 2 at 0.1\% FPR, 4 at 1\% FPR), but the per-site injection volume is higher ($k$=2{,}500-3{,}500 vs.\ 1-3), resulting in median byte overhead of 5.0\% at 0.1\% FPR and 16.5\% at 1\% FPR (Greedy).

\textbf{Search Convergence Efficiency.}
We also evaluate how search strategy affects convergence efficiency (Figure~\ref{fig:search_efficiency} in  Appendix~\ref{app:search_efficiency}). SA variants achieve single call-site perturbation evasion for 10-18\,pp more samples than Greedy across MalConv and MalGraph, with the gap most pronounced at MalConv 1\% FPR (23.3\% Greedy vs 34.6\% SA). For SAFE+GNN, the per-site pattern holds (24.2\% SA vs 15.9\% Greedy at 1\% FPR), but SA achieves only 64.5\% overall ASR compared to Greedy's 85\% (Table~\ref{tab:full_comparison}). Because function-body content and injected semantic NOPs as padding do not meaningfully contribute to evading this classifier, SA's inner NOP exploration consumes budget better spent targeting additional call sites to inject more dummy functions, and Greedy's broader site coverage enables bypass of harder samples. Details are in Appendix~\ref{app:search_efficiency}


\textbf{Sensitivity to Injection Volume.}
SAFE+GNN requires more dummy functions per call site than MalConv and MalGraph, where $k_{\min}/k_{\max}=1/3$ per call-site suffices. We evaluate ASR across eight $k$ ranges on 209 samples at 1\% FPR (Figure~\ref{fig:k_sensitivity}). ASR generally increases with $k$, from 3.8\% at $k$=1/3 (2.7\% injected byte overhead) to 84.2\% at 2{,}500/3{,}500 (15.3\% injected overhead). SAFE+GNN uses global mean pooling over all function node embeddings, so the classifier averages every node without distinguishing injected from original. Shifting this averaged representation past the decision boundary requires enough injected nodes with embeddings distinct from the original malware nodes. This presents a stealth-evasion tradeoff: lower $k$ yields a smaller footprint but lower ASR. The $k$=2{,}500/3{,}500 range used for the runs in Table~\ref{tab:full_comparison} sits at the point of diminishing returns. This volume requirement is a property of SAFE+GNN's aggregation architecture, not a limitation of the attack mechanism.

\begin{findingbox}
\textbf{Answer to RQ2:}
FCG perturbation is essential for graph-based classifiers (removing it collapses ASR), and iterative NOP search is critical for MalConv and MalGraph. Most samples evade with a single call site modification and minimal byte overhead for MalConv/MalGraph, while SAFE+GNN requires higher injection volume.
\end{findingbox}

\subsubsection{\textbf{RQ3: Injection Topology and Hyperparameter Sensitivity}}
\label{sec:rq3}
We explore whether increasing the structural complexity of injected dummy functions (via inter-function calls) enhances evasion, and discuss how sensitive the attack is to search hyperparameters.

\wcircle{1} \textbf{Effect of FCG Complexity (Leaf Functions vs.\
DAG-Constrained Injection).}
\label{par:fcg_complexity}
We explore whether connecting the $k$ dummy functions via
randomly sampled call edges (DAG-constrained injection)
improves or hinders evasion compared to injecting them as
isolated leaf nodes (the default). Acyclicity is required so
that chained calls terminate and return control to the
trampoline (Appendix~\ref{app:dag-construction}). For MalConv
and MalGraph ($k_{min}$/$k_{max}$ = 1/3), we use a fixed edge
probability $p_{edge} = 0.8$. For SAFE+GNN since
($k_{min}$/$k_{max}$ = 2{,}500/3{,}500), a fixed probability
would produce a quadratic number of inter-fake edges, so we
use $p_{edge} = c/k$ with $c = 2$, yielding approximately
$k$ expected edges per call site (linear in $k$). SAFE+GNN uses Greedy search, while the other two classifiers use our Adaptive SA algorithm.

\begin{table}[tb]
\centering
\caption{Isolated leaf vs.\ DAG-constrained injection. MalConv/MalGraph
use $p_{edge} = 0.8$ with Adaptive\,SA; SAFE+GNN uses
$p_{edge} = 2/k$ with Greedy. Highlights:
\colorbox{sawin}{\strut Better} value per column,
\colorbox{deltagain}{\strut Notable $\Delta$}.}
\label{tab:fcg_chaining}
\renewcommand{\arraystretch}{1.05}
\resizebox{\columnwidth}{!}{%
\footnotesize
\begin{tabular}{@{}l*{12}{c}@{}}
\toprule
 & \multicolumn{4}{c}{\textbf{MalConv}} & \multicolumn{4}{c}{\textbf{MalGraph}} & \multicolumn{4}{c}{\textbf{SAFE+GNN}} \\
\cmidrule(lr){2-5} \cmidrule(lr){6-9} \cmidrule(lr){10-13}
\multirow{2}{*}{\textbf{Topology}} & \multicolumn{2}{c}{0.1\,\%} & \multicolumn{2}{c}{1\,\%} & \multicolumn{2}{c}{0.1\,\%} & \multicolumn{2}{c}{1\,\%} & \multicolumn{2}{c}{0.1\,\%} & \multicolumn{2}{c}{1\,\%} \\
\cmidrule(lr){2-3} \cmidrule(lr){4-5} \cmidrule(lr){6-7} \cmidrule(lr){8-9} \cmidrule(lr){10-11} \cmidrule(lr){12-13}
 & ASR & $\Theta$ & ASR & $\Theta$ & ASR & $\Theta$ & ASR & $\Theta$ & ASR & $\Theta$ & ASR & $\Theta$ \\
\midrule
Leaf & \hlwin{100.0} & \hlwin{945} & \hlwin{99.40} & \hlwin{150} & \hlwin{96.01} & \hlwin{30} & \hlwin{94.28} & \hlwin{20} & \hlwin{97.40} & \hlwin{42} & \hlwin{85.00} & 9 \\
DAG & 99.64 & 584 & 96.86 & 102 & 93.59 & 28 & 88.81 & 15 & 95.50 & 33 & 50.70 & 13 \\
\midrule
$\Delta$ & $+$0.36 & \hldelta{$+$1.6$\times$} & \hldelta{$+$2.54} & \hldelta{$+$1.5$\times$} & \hldelta{$+$2.42} & $+$1.1$\times$ & \hldelta{$+$5.47} & $+$1.3$\times$ & $+$1.90 & $+$1.3$\times$ & $+$\hldelta{34.30} & $-1.44\times$ \\
\bottomrule
\end{tabular}%
}
\end{table}

As shown in Table~\ref{tab:fcg_chaining}, DAG-constrained
chaining uniformly degrades attack performance across all
three classifiers. For MalConv, the ASR loss is modest
($\leq$2.5\,pp) since MalConv does not observe FCG structure,
with the throughput penalty stemming from the larger per-site
injection footprint. For MalGraph, the degradation is more
pronounced (up to 5.5\,pp ASR) as the FCG-level GNN
propagates messages among the connected dummy functions,
forming a recognizable cluster rather than dispersing across
independent nodes in the global pool. For SAFE+GNN, DAG
chaining costs a significant degradation upto 34.30\,pp ASR, which may be due to call instructions inside dummy nodes degrade ASR or the subgraphs that are formed due to DAG chaining makes it structurally more distinctive to the model, degrading evasion.

\wcircle{2} \textbf{Hyperparameter Robustness.} We verified sensitivity
to three hyperparameters - SA initial temperature, NOP pool diversity, and dummy function count on MalConv and MalGraph at 0.1\% FPR (full results in
Appendices~\ref{app:temp-sensitivity},~\ref{app:nop-sensitivity},
and~\ref{app:k-sensitivity}). SA initial temperature has no
notable effect on ASR ($<$0.6\,pp variation). Semantic NOP
variety is classifier-dependent with MalConv favoring concentrated
single-type repetition ($s{=}1$), while MalGraph requires a
minimum diversity of NOP types and collapses to 5.3\% ASR
when the pool is restricted to a single instruction type.
For the dummy function count $k$, restricting to $k_{\min}/k_{\max}{=}1/1$
drops MalGraph ASR by 11.5\,pp, but all configurations with
$k \geq 3$ are within 1.7\,pp of each other, justifying the
default $k_{\min}/k_{\max}{=}1/3$. These ablations are not
repeated for SAFE+GNN classifier as the $k$-range sensitivity is analyzed
in Section~\ref{sec:rq2_footprint}, and semantic NOP content
does not contribute to evasion for this classifier.


\begin{findingbox}
\textbf{Answer to RQ3:} Isolated leaf injection consistently outperforms DAG-constrained injection across all three classifiers, as explicitly linking injected functions forms a recognizable subgraph that reduces evasion. Attack performance is robust to SA temperature, but sensitive to semantic NOP pool diversity for MalGraph.
\end{findingbox}

\subsubsection{\textbf{RQ4: Semantic Preservation Analysis}}
\label{sec:rq4}
We evaluate semantic preservation by comparing Windows API-call sequences collected from a sandbox execution following the protocol of Malguise~\cite{ling2024wolf}. Since malware can exhibit non-deterministic behavior across runs~\cite{kasama2012malware}, exact sequence matching would be too strict. We instead calibrate an equivalence threshold from each sample's own variability.


For each adversarial $m^* \in \mathcal{M}_{\text{adv}}$ we collect 
three sandbox runs, two independent runs of the original 
($m^{r_1}$, $m^{r_2}$) and one run of the adversarial ($m^*$). 
Let $\text{API}(\cdot)$ denote the ordered Windows API 
sequence captured during execution. We measure similarity using 
the length-normalized Levenshtein distance over API tokens: 
\begin{equation}
dist(x,y) \;=\; \frac{\text{Lev}\!\left(\text{API}(x),\,\text{API}(y)\right)}{\max\!\left(|\text{API}(x)|,\,|\text{API}(y)|\right)} \;\in\; [0,1]
\end{equation}

{\sloppy
We compute $\text{dist}(m^{r_1}, m^{r_2})$ for every sample in
the per-configuration evaluation pool and take the 99.5th percentile as the threshold:
$\text{dist}_\Delta = P_{99.5}\bigl[\{\text{dist}(m^{r_1}, m^{r_2})\}\bigr]$.
The semantic-equivalence indicator used in $\text{SPR}$
(Section~\ref{sec:metrics}) is then
$\text{Sem}(m, m^*) = \mathbf{1}\bigl[\text{dist}(m^{r_1}, m^*) < \text{dist}_\Delta\bigr]$.
}

We evaluate adversarial samples per (classifier, FPR) configuration, drawing from the bypass set with a fixed random seed. A sample is
eligible if both baseline runs of its original produce $\geq 20$ API calls, since a few differing calls in a short trace produce disproportionately large normalized distances that would exceed the threshold regardless of actual behavioral similarity. We select 300 eligible samples per configuration, except MalConv at 0.1\% FPR (211 eligible) and SAFE+GNN at 1\% FPR (295 eligible). We cap evaluation at 300 samples per configuration since each requires three separate sandbox runs.

\begin{table}[tb]
\centering
\caption{Semantic preservation rate (SPR, \%) of \system{} adversarial samples validated against a sandbox. Higher is better.}
\label{tab:semantic_preservation}
\scriptsize
\setlength{\tabcolsep}{4pt}
\renewcommand{\arraystretch}{1.05}
\begin{tabular}{@{}lcccccc@{}}
\toprule
& \multicolumn{2}{c}{\textbf{MalConv}} & \multicolumn{2}{c}{\textbf{MalGraph}} & \multicolumn{2}{c}{\textbf{SAFE+GNN}} \\
\cmidrule(lr){2-3} \cmidrule(lr){4-5} \cmidrule(lr){6-7}
& 0.1\% & 1\% & 0.1\% & 1\% & 0.1\% & 1\% \\
\midrule
$N$           & 211   & 300   & 300   & 300   & 300   & 295 \\
$\text{SPR}$ (\%) & 86.26 & 87.00 & 90.00 & 93.00 & 93.67 & 96.95 \\
\bottomrule
\end{tabular}
\end{table}


Table~\ref{tab:semantic_preservation} shows that \system{} preserves the original malicious behavior in 93-97\% of SAFE+GNN bypasses, 90-93\% of MalGraph bypasses and 86-87\% of MalConv bypasses. SAFE+GNN achieves the highest SPR across all configurations, consistent with the minimal function bodies being a structurally minimal perturbation ($\sim$5 bytes/injected function) that does not alter the original code's instruction stream.
Of the 1706 evaluated samples across all classifiers, 147 (8.6\%) fail the preservation test with 64 \emph{diverged} and 83 \emph{short-trace} (the adversarial produced fewer than 20 API calls despite both original runs exceeding this threshold). The dominant failure cause for MalConv and MalGraph is overlay-dependent malware whose absolute file offset shifts after PE section insertion. SAFE+GNN on the other hand had fewer overall failures (28 out of 595 vs. \ 119 out of 1111 for MalConv/MalGraph) and follow a different pattern, with short-trace types accounting for the majority rather than overlay. Per-configuration breakdowns, overlay analysis and detailed failures statistics are provided in Appendix~\ref{app:semantic-failures}.

\begin{findingbox}
\textbf{Answer to RQ4:} \system{} preserves the original malicious behavior in 86–97\% of evaluated bypasses. SAFE+GNN achieves the highest SPR (93-97\%), consistent with its minimal function-body perturbation.
\end{findingbox}

\FloatBarrier
%
\subsubsection{\textbf{RQ5: Defense Robustness}}
\label{sec:rq5}
\textbf{Adversarial Finetuning.} We evaluate whether lightweight adversarial finetuning can harden the target classifiers against \system{}. Using the defense pool described in Section~\ref{sec:dataset}, we finetune each classifier and recalibrate at 1\% FPR, the harder attack threshold (finetuning hyperparameters and design justifications in Appendix~\ref{app:hyperparameters} and~\ref{app:dataset-details}).

\begin{figure}[t]
\centering
\includegraphics[width=\columnwidth]{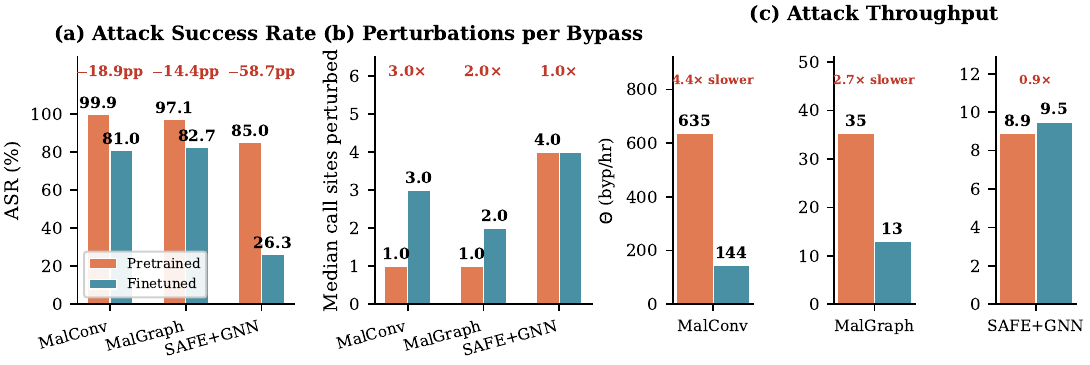}
\caption{Effect of adversarial finetuning on attack effectiveness at 1\% FPR. (a)~ASR. (b)~Median call sites perturbed per successful bypass. (c)~Attack Throughput $\Theta$. Note the independent $y$-axes per classifier in (c).}
\label{fig:defense_metrics}
\end{figure}

Figure~\ref{fig:defense_metrics} summarizes the results
across three dimensions. For MalConv and MalGraph,
adversarial finetuning reduces ASR by 18.9\,  pp
(99.9\%$\to$81.0\%) and 14.4\,pp (97.1\%$\to$82.7\%)
respectively, while the attacker must perturb
2-3$\times$ more call sites and sustain a
2.73-4.41$\times$ throughput reduction to achieve evasion.
For SAFE+GNN, finetuning produces the largest ASR
reduction of 58.7\,pp (85.0\%$\to$26.3\%), while
per-bypass perturbation cost and throughput remain
unchanged (median 4 call sites, $\sim$9\,byp/hr).
These results demonstrate that adversarial finetuning reduces but does not neutralize the attack (26-83\% residual ASR), indicating that \system{} remains effective against hardened models.

\smallskip
\noindent\textbf{Structural heuristic defenses.}
A defender aware of the injection mechanism could deploy
rule-based heuristics that target structural artifacts
\system{} leaves in the modified binary.
We evaluate three such heuristics on all adversarial
samples generated across all three classifiers (MalConv,
MalGraph, SAFE+GNN) at both operating points with both the default leaf variant (Table~\ref{tab:full_comparison}) and the DAG variant (Table~\ref{tab:fcg_chaining}):
\wcircle{1}{\textbf{H\textsubscript{sect}}}: The binary contains an executable section outside \texttt{.text}. \system{} appends a new \texttt{.htext} section to host injected code, so this artifact is present in every adversarial sample.
\wcircle{2}\textbf{H\textsubscript{leaf}}: H\textsubscript{sect} holds \emph{and} every detected function in the non \texttt{.text} section is a leaf (no outgoing \texttt{call} instructions). This targets the default injection mode, in which each dummy function body is self-contained.
\wcircle{3}\textbf{H\textsubscript{size}}: The file size exceeds
a threshold $T_{size}$ derived from the unmodified malware population. Because the attack increases file size, we test whether this overhead is detectable: we set $T_{30}$\,=\,median\,$\times$\,1.30 (median\,=\,1.11\,MB over 7342 unique originals), using the largest budget in our experiments (30\%, SAFE+GNN), which is an \emph{upper bound} on the overhead any adversarial sample actually incurred.

\noindent We additionally evaluate
H\textsubscript{all}\,=\,$\bigwedge$\{H\textsubscript{sect},
H\textsubscript{leaf}, H\textsubscript{size}\} and
H\textsubscript{any}\,=\,$\bigvee$\{H\textsubscript{sect},
H\textsubscript{leaf}, H\textsubscript{size}\}.
To test whether a simple attacker adaptation defeats these
heuristics, we include \system{}'s \emph{DAG variant}
(Section~\ref{par:fcg_complexity}), which injects inter-function
\texttt{call} instructions between dummy functions
(edge probability $p\!=\!0.8$ for MalConv/MalGraph;
$p\!=\!c/k$ with $c\!=\!2$ for SAFE+GNN), forming a
call graph within the same pre-allocated section at zero
additional size cost.

Table~\ref{tab:defense_heuristics} reports true positive
rates (TPR, fraction of adversarial samples flagged) and
false positive rates (FPR, fraction of benign samples
wrongly flagged).
All 11572 leaf-variant and 10887 DAG-variant
adversarial samples are pooled across classifiers and
operating points. The threshold is computed from the full
original malware pool.

\begin{table}[t]
\centering
\caption{Structural heuristic detection rates.
  TPR measured on adversarial samples (Leaf and DAG
  variants, pooled across all classifiers);
  FPR on 6867 benign executables.
  $T_{30}$\,=\,median\,$\times$\,1.30.
  Distribution statistics and threshold sweep results
  are in Appendix~\ref{app:defense_sweep}.}
\label{tab:defense_heuristics}
\scriptsize
\setlength{\tabcolsep}{4pt}
\begin{tabular}{lccc}
\toprule
\textbf{Heuristic} & \textbf{TPR (Leaf)} & \textbf{TPR (DAG)} & \textbf{FPR (Benign)} \\
\midrule
H\textsubscript{sect}           & 100.0\%  & 100.0\%  & 9.5\%  \\
H\textsubscript{leaf}           & 99.9\%   & 10.5\%   & 0.1\%  \\
H\textsubscript{size}($T_{30}$) & 25.0\%   & 24.1\%   & 11.3\% \\
\midrule
H\textsubscript{all}($T_{30}$)  & 25.0\%   & 1.7\%    & 0.1\%  \\
H\textsubscript{any}($T_{30}$)  & 100.0\%  & 100.0\%  & 17.4\% \\
\bottomrule
\end{tabular}
\end{table}


H\textsubscript{sect} flags every adversarial sample regardless of variant, but 9.5\% of benign executables also contain non-\texttt{.text} executable sections, making standalone deployment impractical. H\textsubscript{leaf} achieves 99.9\% TPR with only 0.1\% FPR on the default leaf injection, but the DAG variant reduces detection to 10.5\% by adding inter-function calls that break the all-leaf pattern. H\textsubscript{size} detects only 25\% of adversarial samples even under the most defender-favorable threshold, because PE file sizes exhibit extreme natural variance (Appendix~\ref{app:defense_sweep}).
The combinations show the core challenge: H\textsubscript{any} catches all adversarial samples but at 17.4\% FPR, while H\textsubscript{all} reduces FPR to 0.1\% but detects only 1.7\% of DAG-variant samples. A sweep across percentile anchors and overhead multipliers confirms that no threshold choice resolves this tradeoff (Appendix~\ref{app:defense_sweep}).
These heuristics target implementation-specific artifacts rather than the underlying classifier vulnerability. An adaptive attacker who enriches injected function bodies with realistic instruction patterns could further reduce structural distinguishability while preserving the FCG perturbation.


\begin{findingbox}
\textbf{Answer to RQ5:} Adversarial finetuning reduces but does not eliminate the attack (residual ASR 26-83\%), and structural heuristics face a fundamental detection versus false-alarm tradeoff that a simple attacker adaptation (DAG injection) further weakens. Neither defense neutralizes \system{} attack completely.
\end{findingbox}


\section{Related Works}\label{sec:related_works}



Adversarial malware generation has been extensively studied, with a broad range of techniques proposed to produce evasive malware variants. A prominent line of work focuses on API call manipulation, in which irrelevant/benign-looking API calls are injected via gradient-based optimization, generative models, greedy strategies, or graph rewiring to evade detectors~\cite{hu2017generating, 8669079, robust_malware_detection_challenge, tarallo, 11575074}. Another major direction targets the CFG representation of malware by modifying assembly-level control flow~\cite{ling2024wolf, peng2025evading}, perturbing learned graph features~\cite{zhang2022semantics}, or injecting additional nodes into the CFG~\cite{zapzalka2024semantics}. 
In contrast, fine-grained perturbation at the source-code level remains relatively underexplored, with prior work limited to approaches such as injecting assembly stubs~\cite{app12157546} and manipulating system-call graphs in intermediate representations~\cite{ming2017impeding}.

A complementary body of work explores direct binary-level manipulation, either globally or locally, through byte injection or appending while maintaining functionality~\cite{DBLP:journals/corr/abs-2012-07994, kreuk2018deceiving, QIAO2022102762, aghakhani2020malware, Yuan2020BlackBoxAA, kolosnjaji2018adversarial, suciu2019exploring, ceschin2020no, 10.1145/3545948.3545960, 10.1145/3597503.3639141}. Other studies modify different malware binary components, such as headers or overlays~\cite{li2025minimal, demetrio2021functionality}, or apply binary diversification techniques to generate semantically equivalent variants~\cite{lucas2021malware, Lucas23}. Reinforcement learning has also been leveraged to search large transformation spaces via surrogate models, policy distillation, or semantics-preserving Deep Q-Networks~\cite{rigaki2023power, tian2024functionality, zhang2022semantics}. 

A separate line of work targets the FCG where attacks have been explored on Android malware via dead call edges or trivially simple call injection at the smali level~\cite{li2023black,li2025efficient,zhao2021structural,song2025fcghunter}, on WebAssembly through function insertion and call-site redirection~\cite{kim2025does}, and on IoT ELF binaries through branch function injection~\cite{mwangi2024adversarial}. These approaches benefit from managed runtimes or intermediate representations that abstract calling conventions and stack management, making FCG modification substantially simpler than in the compiled Windows PE settings. As FCG-based detectors such as SAFE+GNN~\cite{someya2023graph} and FCGAT~\cite{someya2023fcgat} increasingly rely on function call graph structure as a primary discriminative feature, the robustness of these representations to structural perturbation is directly relevant to deployed security. \system{} differs from all of the above in that it perturbs the topological structure of compiled Windows PE malware by injecting fully executable, reachable dummy functions with configurable body content and injection volume, simultaneously altering both CFG and FCG structure while preserving problem-space validity.

\section{Conclusion, Limitations, and Future Work}\label{section:conclusion}
\system{} demonstrates that injecting fully executable dummy functions into the FCG of compiled Windows PE malware is a viable and persistent attack surface. Across three architecturally distinct classifiers, the attack achieves 85-100\% ASR while preserving malicious functionality in 86-97\% of evaluated samples. The effective injection strategy differs by architecture: NOP content diversity drives evasion for classifiers that process instruction-level features, while injection volume becomes the governing factor when the classifier's architecture is less sensitive to individual function content. Both adversarial finetuning and structural heuristic defenses reduce but fail to eliminate our proposed attack.

The injected functions are parameterless with self-canceling or minimal bodies, making them structurally simpler than benign code. Enriching function bodies with benign API calls, conditional branches, and local variables is a natural future step. Attention-based graph classifiers that weight function nodes by structural or semantic relevance may be less susceptible to volume-based injection, and designing perturbations that target learned attention weights is a promising direction. On the defensive side, developing classifiers that can distinguish semantically inert injected functions from legitimate code remains an open direction.

\bibliographystyle{plainurl}
\bibliography{references}

\begin{thebibliography}{10}

\bibitem{aarts1988quantitative}
Emile H.~L. Aarts, Jan H.~M. Korst, and Peter J.~M. van Laarhoven.
\newblock A quantitative analysis of the simulated annealing algorithm: A case study for the traveling salesman problem.
\newblock {\em Journal of Statistical Physics}, 50(1):187--206, 1988.
\newblock \href {https://doi.org/10.1007/BF01022991} {\path{doi:10.1007/BF01022991}}.

\bibitem{MalwareBazaar}
{ABUSE(CH)}.
\newblock Malwarebazaar, 2025.
\newblock [Last Accessed: Dec 3, 2025].
\newblock URL: \url{https://bazaar.abuse.ch/}.

\bibitem{10.1145/3545948.3545960}
Ahmed Abusnaina, Afsah Anwar, Sultan Alshamrani, Abdulrahman Alabduljabbar, RhongHo Jang, DaeHun Nyang, and David Mohaisen.
\newblock Systematically evaluating the robustness of ml-based iot malware detection systems.
\newblock In {\em Proceedings of the 25th International Symposium on Research in Attacks, Intrusions and Defenses}, RAID '22, page 308–320, New York, NY, USA, 2022. Association for Computing Machinery.
\newblock \href {https://doi.org/10.1145/3545948.3545960} {\path{doi:10.1145/3545948.3545960}}.

\bibitem{aghakhani2020malware}
Hojjat Aghakhani, Fabio Gritti, Francesco Mecca, Martina Lindorfer, Stefano Ortolani, Davide Balzarotti, Giovanni Vigna, and Christopher Kruegel.
\newblock When malware is packin'heat; limits of machine learning classifiers based on static analysis features.
\newblock In {\em Network and Distributed Systems Security (NDSS) Symposium 2020}, 2020.

\bibitem{8752028}
Hisham Alasmary, Aminollah Khormali, Afsah Anwar, Jeman Park, Jinchun Choi, Ahmed Abusnaina, Amro Awad, Daehun Nyang, and Aziz Mohaisen.
\newblock Analyzing and detecting emerging internet of things malware: A graph-based approach.
\newblock {\em IEEE Internet of Things Journal}, 6(5):8977--8988, 2019.
\newblock \href {https://doi.org/10.1109/JIOT.2019.2925929} {\path{doi:10.1109/JIOT.2019.2925929}}.

\bibitem{anderson2018ember}
H.~Anderson and Phil Roth.
\newblock Ember: An open dataset for training static pe malware machine learning models.
\newblock {\em ArXiv}, abs/1804.04637, 2018.
\newblock URL: \url{https://api.semanticscholar.org/CorpusID:4888440}.

\bibitem{avtest2025stats}
{AV-TEST GmbH}.
\newblock Malware statistics \& trends report.
\newblock \url{https://www.av-test.org/en/statistics/malware/}, 2025.
\newblock Accessed: April 2026. Reports $\sim$83\% of new malware targeting Windows and $\sim$500{,}000 new malicious files detected per day (Nov 2024--Oct 2025 window).

\bibitem{bishop2006pattern}
Christopher~M Bishop.
\newblock {\em Pattern recognition and machine learning}, volume~4.
\newblock Springer, 2006.

\bibitem{bohme2017directed}
Marcel B\"{o}hme, Van-Thuan Pham, Manh-Dung Nguyen, and Abhik Roychoudhury.
\newblock Directed greybox fuzzing.
\newblock In {\em Proceedings of the 2017 ACM SIGSAC Conference on Computer and Communications Security}, CCS '17, page 2329–2344, New York, NY, USA, 2017. Association for Computing Machinery.
\newblock \href {https://doi.org/10.1145/3133956.3134020} {\path{doi:10.1145/3133956.3134020}}.

\bibitem{ceschin2020no}
Fabricio Ceschin, Marcus Botacin, Gabriel L\"{u}ders, Heitor~Murilo Gomes, Luiz Oliveira, and Andre Gregio.
\newblock No need to teach new tricks to old malware: Winning an evasion challenge with xor-based adversarial samples.
\newblock In {\em Reversing and Offensive-Oriented Trends Symposium}, ROOTS'20, page 13–22, New York, NY, USA, 2021. Association for Computing Machinery.
\newblock \href {https://doi.org/10.1145/3433667.3433669} {\path{doi:10.1145/3433667.3433669}}.

\bibitem{pierazzi2020intriguing}
Jacopo Cortellazzi, Erwin Quiring, Daniel Arp, Feargus Pendlebury, Fabio Pierazzi, and Lorenzo Cavallaro.
\newblock Intriguing properties of adversarial ml attacks in the problem space [extended version].
\newblock {\em ACM Trans. Priv. Secur.}, 28(4), September 2025.
\newblock \href {https://doi.org/10.1145/3742895} {\path{doi:10.1145/3742895}}.

\bibitem{cybersecurityventures2026}
{Cybersecurity Ventures}.
\newblock Official cybercrime report 2025.
\newblock \url{https://cybersecurityventures.com/official-cybercrime-report-2025/}, 2025.
\newblock Reports cybercrime damages reached \$10.5T annually in 2025; projects \$12.2T by 2031.

\bibitem{demetrio2021functionality}
Luca Demetrio, Battista Biggio, Giovanni Lagorio, Fabio Roli, and Alessandro Armando.
\newblock Functionality-preserving black-box optimization of adversarial windows malware.
\newblock {\em IEEE Transactions on Information Forensics and Security}, 16:3469--3478, 2021.

\bibitem{tarallo}
Gabriele Digregorio, Salvatore Maccarrone, Mario D'Onghia, Luigi Gallo, Michele Carminati, Mario Polino, and Stefano Zanero.
\newblock Tarallo: Evading behavioral malware detectors in the problem space.
\newblock In Federico Maggi, Manuel Egele, Mathias Payer, and Michele Carminati, editors, {\em Detection of Intrusions and Malware, and Vulnerability Assessment}, pages 128--149, Cham, 2024. Springer Nature Switzerland.

\bibitem{downing2021deepreflect}
Evan Downing, Yisroel Mirsky, Kyuhong Park, and Wenke Lee.
\newblock {DeepReflect}: Discovering malicious functionality through binary reconstruction.
\newblock In {\em 30th USENIX Security Symposium (USENIX Security 21)}, pages 3469--3486. USENIX Association, August 2021.
\newblock URL: \url{https://www.usenix.org/conference/usenixsecurity21/presentation/downing}.

\bibitem{DBLP:journals/corr/abs-2012-07994}
Mohammadreza Ebrahimi, Ning Zhang, James~Lee Hu, Muhammad~Taqi Raza, and Hsinchun Chen.
\newblock Binary black-box evasion attacks against deep learning-based static malware detectors with adversarial byte-level language model.
\newblock {\em CoRR}, abs/2012.07994, 2020.
\newblock URL: \url{https://arxiv.org/abs/2012.07994}, \href {https://arxiv.org/abs/2012.07994} {\path{arXiv:2012.07994}}.

\bibitem{eykholt2023uret}
Kevin Eykholt, Taesung Lee, Douglas Schales, Jiyong Jang, Ian Molloy, and Masha Zorin.
\newblock Uret: universal robustness evaluation toolkit (for evasion).
\newblock In {\em Proceedings of the 32nd USENIX Conference on Security Symposium}, SEC '23, USA, 2023. USENIX Association.

\bibitem{fogcallingconventions}
Agner Fog.
\newblock Calling conventions for different c++ compilers and operating systems.
\newblock Technical report, Technical University of Denmark, 2024.
\newblock [Last Accessed: Apr 15, 2026].
\newblock URL: \url{https://www.agner.org/optimize/calling_conventions.pdf}.

\bibitem{goodfellow2015explaining}
Ian~J. Goodfellow, Jonathon Shlens, and Christian Szegedy.
\newblock Explaining and harnessing adversarial examples.
\newblock {\em CoRR}, abs/1412.6572, 2014.
\newblock URL: \url{https://api.semanticscholar.org/CorpusID:6706414}.

\bibitem{graphsage}
William~L. Hamilton, Rex Ying, and Jure Leskovec.
\newblock Inductive representation learning on large graphs.
\newblock In {\em Proceedings of the 31st International Conference on Neural Information Processing Systems}, NIPS'17, page 1025–1035, Red Hook, NY, USA, 2017. Curran Associates Inc.

\bibitem{10.1145/3597503.3639141}
Shuai He, Cai Fu, Hong Hu, Jiahe Chen, Jianqiang Lv, and Shuai Jiang.
\newblock Malwaretotal: Multi-faceted and sequence-aware bypass tactics against static malware detection.
\newblock In {\em Proceedings of the IEEE/ACM 46th International Conference on Software Engineering}, ICSE '24, New York, NY, USA, 2024. Association for Computing Machinery.
\newblock \href {https://doi.org/10.1145/3597503.3639141} {\path{doi:10.1145/3597503.3639141}}.

\bibitem{die}
Horsicq.
\newblock {DIE} detect-it-easy.
\newblock \url{https://github.com/horsicq/Detect-It-Easy}, 2024.
\newblock Open-source packer/protector detection tool.

\bibitem{hu2017generating}
Weiwei Hu and Ying Tan.
\newblock Generating adversarial malware examples for black-box attacks based on gan.
\newblock In {\em International Conference on Data Mining and Big Data}, 2017.
\newblock URL: \url{https://api.semanticscholar.org/CorpusID:4316147}.

\bibitem{huang2025malfocus}
Weihao Huang, Chaoyang Lin, Lu~Xiang, Zhiyu Zhang, Guozhu Meng, Lei Xue, Kai Chen, Lei Meng, and Zongming Zhang.
\newblock Malfocus: Locating malicious modules in malware based on hybrid deep learning.
\newblock {\em IEEE Transactions on Dependable and Secure Computing}, 2025.

\bibitem{ibm2025cost}
{IBM Security}.
\newblock Cost of a data breach report 2025.
\newblock \url{https://www.ibm.com/reports/data-breach}, 2025.
\newblock Average cost of an extortion or ransomware incident: \$5.08M. Full report available via free registration; accessible summary at \url{https://www.helpnetsecurity.com/2025/08/04/ibm-cost-data-breach-report-2025/. Last Accessed: April 23, 2026.}

\bibitem{intelligentciso2026kaspersky}
{Intelligent CISO}.
\newblock Kaspersky detected half a million malicious files daily in 2025.
\newblock \url{https://www.intelligentciso.com/2026/01/05/kaspersky-detected-half-a-million-malicious-files-daily-in-2025/}, 2026.
\newblock [Last Accessed: April 23, 2026].

\bibitem{kargarnovin2024mal2gcn}
Omid Kargarnovin, Amir~Mahdi Sadeghzadeh, and Rasool Jalili.
\newblock Mal2gcn: a robust malware detection approach using deep graph convolutional networks with non-negative weights.
\newblock {\em Journal of Computer Virology and Hacking Techniques}, 20(1):95--111, 2024.
\newblock \href {https://doi.org/10.1007/s11416-023-00498-7} {\path{doi:10.1007/s11416-023-00498-7}}.

\bibitem{kasama2012malware}
Takahiro Kasama, Katsunari Yoshioka, Daisuke Inoue, and Tsutomu Matsumoto.
\newblock Malware detection method by catching their random behavior in multiple executions.
\newblock In {\em 2012 IEEE/IPSJ 12th International Symposium on Applications and the Internet}, pages 262--266, 2012.
\newblock \href {https://doi.org/10.1109/SAINT.2012.49} {\path{doi:10.1109/SAINT.2012.49}}.

\bibitem{8669079}
Masataka Kawai, Kaoru Ota, and Mianxing Dong.
\newblock Improved malgan: Avoiding malware detector by leaning cleanware features.
\newblock In {\em 2019 International Conference on Artificial Intelligence in Information and Communication (ICAIIC)}, pages 040--045, 2019.
\newblock \href {https://doi.org/10.1109/ICAIIC.2019.8669079} {\path{doi:10.1109/ICAIIC.2019.8669079}}.

\bibitem{kim2025does}
Taeyoung Kim, Sanghak Oh, Kiho Lee, Weihang Wang, Yonghwi Kwon, Sanghyun Hong, and Hyoungshick Kim.
\newblock When does wasm malware detection fail? a systematic analysis of their robustness to evasion.
\newblock In {\em 2025 40th IEEE/ACM International Conference on Automated Software Engineering (ASE)}, pages 2957--2969, 2025.
\newblock \href {https://doi.org/10.1109/ASE63991.2025.00243} {\path{doi:10.1109/ASE63991.2025.00243}}.

\bibitem{kirkpatrick1983optimization}
S.~Kirkpatrick, C.~D. Gelatt, and M.~P. Vecchi.
\newblock Optimization by simulated annealing.
\newblock {\em Science}, 220(4598):671--680, 1983.
\newblock URL: \url{http://www.jstor.org/stable/1690046}.

\bibitem{kolosnjaji2018adversarial}
Bojan Kolosnjaji, Ambra Demontis, Battista Biggio, Davide Maiorca, Giorgio Giacinto, Claudia Eckert, and Fabio Roli.
\newblock Adversarial malware binaries: Evading deep learning for malware detection in executables.
\newblock In {\em 2018 26th European Signal Processing Conference (EUSIPCO)}, pages 533--537, 2018.
\newblock \href {https://doi.org/10.23919/EUSIPCO.2018.8553214} {\path{doi:10.23919/EUSIPCO.2018.8553214}}.

\bibitem{koo2016juggling}
Hyungjoon Koo and Michalis Polychronakis.
\newblock Juggling the gadgets: Binary-level code randomization using instruction displacement.
\newblock In {\em Proceedings of the 11th ACM on Asia Conference on Computer and Communications Security}, ASIA CCS '16, page 23–34, New York, NY, USA, 2016. Association for Computing Machinery.
\newblock \href {https://doi.org/10.1145/2897845.2897863} {\path{doi:10.1145/2897845.2897863}}.

\bibitem{10.1145/1982185.1982509}
Orestis Kostakis, Joris Kinable, Hamed Mahmoudi, and Kimmo Mustonen.
\newblock Improved call graph comparison using simulated annealing.
\newblock In {\em Proceedings of the 2011 ACM Symposium on Applied Computing}, SAC '11, page 1516–1523, New York, NY, USA, 2011. Association for Computing Machinery.
\newblock \href {https://doi.org/10.1145/1982185.1982509} {\path{doi:10.1145/1982185.1982509}}.

\bibitem{kreuk2018deceiving}
Felix Kreuk, Assi Barak, Shir Aviv-Reuven, Moran Baruch, Benny Pinkas, and Joseph Keshet.
\newblock Deceiving end-to-end deep learning malware detectors using adversarial examples.
\newblock {\em arXiv preprint arXiv:1802.04528}, 2018.

\bibitem{lesterpsadataset}
Michael Lester.
\newblock {PE} malware machine learning dataset.
\newblock \url{https://practicalsecurityanalytics.com/pe-malware-machine-learning-dataset/}, 2024.
\newblock Practical Security Analytics.

\bibitem{li2025minimal}
Chengyi Li, Zhiyuan Jiang, Yongjun Wang, Tian Xia, Yayuan Zhang, and Yuhang Mao.
\newblock Minimal: Hard-label adversarial attack against static malware detection with minimal perturbation.
\newblock In James Kwok, editor, {\em Proceedings of the Thirty-Fourth International Joint Conference on Artificial Intelligence, {IJCAI-25}}, pages 5589--5597. International Joint Conferences on Artificial Intelligence Organization, 8 2025.
\newblock Main Track.
\newblock \href {https://doi.org/10.24963/ijcai.2025/622} {\path{doi:10.24963/ijcai.2025/622}}.

\bibitem{li2023black}
Heng Li, Zhang Cheng, Bang Wu, Liheng Yuan, Cuiying Gao, Wei Yuan, and Xiapu Luo.
\newblock Black-box adversarial example attack towards fcg based android malware detection under incomplete feature information.
\newblock In {\em Proceedings of the 32nd USENIX Conference on Security Symposium}, SEC '23, USA, 2023. USENIX Association.

\bibitem{li2025efficient}
Heng Li, Bang Wu, Wei Zhou, Wei Yuan, Cuiying Gao, Xinge You, and Xiapu Luo.
\newblock An efficient adversarial attack on fcg-based android malware detection systems.
\newblock {\em IEEE Transactions on Information Forensics and Security}, 20:9413--9426, 2025.
\newblock \href {https://doi.org/10.1109/TIFS.2025.3607270} {\path{doi:10.1109/TIFS.2025.3607270}}.

\bibitem{ling2019deepsec}
Xiang Ling, Shouling Ji, Jiaxu Zou, Jiannan Wang, Chunming Wu, Bo~Li, and Ting Wang.
\newblock Deepsec: A uniform platform for security analysis of deep learning model.
\newblock In {\em 2019 IEEE Symposium on Security and Privacy (SP)}, pages 673--690, 2019.
\newblock \href {https://doi.org/10.1109/SP.2019.00023} {\path{doi:10.1109/SP.2019.00023}}.

\bibitem{ling2022malgraph}
Xiang Ling, Lingfei Wu, Wei Deng, Zhenqing Qu, Jiangyu Zhang, Sheng Zhang, Tengfei Ma, Bin Wang, Chunming Wu, and Shouling Ji.
\newblock Malgraph: Hierarchical graph neural networks for robust windows malware detection.
\newblock In {\em IEEE INFOCOM 2022 - IEEE Conference on Computer Communications}, pages 1998--2007, 2022.
\newblock \href {https://doi.org/10.1109/INFOCOM48880.2022.9796786} {\path{doi:10.1109/INFOCOM48880.2022.9796786}}.

\bibitem{ling2023adversarial}
Xiang Ling, Lingfei Wu, Jiangyu Zhang, Zhenqing Qu, Wei Deng, Xiang Chen, Yaguan Qian, Chunming Wu, Shouling Ji, Tianyue Luo, Jingzheng Wu, and Yanjun Wu.
\newblock Adversarial attacks against windows pe malware detection: A survey of the state-of-the-art.
\newblock {\em Computers \& Security}, 128:103134, 2023.
\newblock URL: \url{https://www.sciencedirect.com/science/article/pii/S0167404823000445}, \href {https://doi.org/10.1016/j.cose.2023.103134} {\path{doi:10.1016/j.cose.2023.103134}}.

\bibitem{ling2024wolf}
Xiang Ling, Zhiyu Wu, Bin Wang, Wei Deng, Jingzheng Wu, Shouling Ji, Tianyue Luo, and Yanjun Wu.
\newblock A wolf in sheep's clothing: practical black-box adversarial attacks for evading learning-based windows malware detection in the wild.
\newblock In {\em Proceedings of the 33rd USENIX Conference on Security Symposium}, SEC '24, USA, 2024. USENIX Association.

\bibitem{app12157546}
Faming Lu, Zhaoyang Cai, Zedong Lin, Yunxia Bao, and Mengfan Tang.
\newblock Research on the construction of malware variant datasets and their detection method.
\newblock {\em Applied Sciences}, 12(15), 2022.
\newblock URL: \url{https://www.mdpi.com/2076-3417/12/15/7546}, \href {https://doi.org/10.3390/app12157546} {\path{doi:10.3390/app12157546}}.

\bibitem{Lucas23}
Keane Lucas, Samruddhi Pai, Weiran Lin, Lujo Bauer, Michael~K. Reiter, and Mahmood Sharif.
\newblock Adversarial training for {Raw-Binary} malware classifiers.
\newblock In {\em 32nd USENIX Security Symposium (USENIX Security 23)}, pages 1163--1180, Anaheim, CA, August 2023. USENIX Association.
\newblock URL: \url{https://www.usenix.org/conference/usenixsecurity23/presentation/lucas}.

\bibitem{lucas2021malware}
Keane Lucas, Mahmood Sharif, Lujo Bauer, Michael~K. Reiter, and Saurabh Shintre.
\newblock Malware makeover: Breaking ml-based static analysis by modifying executable bytes.
\newblock In {\em Proceedings of the 2021 ACM Asia Conference on Computer and Communications Security}, ASIA CCS '21, page 744–758, New York, NY, USA, 2021. Association for Computing Machinery.
\newblock \href {https://doi.org/10.1145/3433210.3453086} {\path{doi:10.1145/3433210.3453086}}.

\bibitem{SAFE}
Luca Massarelli, Giuseppe~Antonio Di~Luna, Fabio Petroni, Leonardo Querzoni, and Roberto Baldoni.
\newblock Function representations for binary similarity.
\newblock {\em IEEE Transactions on Dependable and Secure Computing}, 19(4):2259--2273, 2022.
\newblock \href {https://doi.org/10.1109/TDSC.2021.3051852} {\path{doi:10.1109/TDSC.2021.3051852}}.

\bibitem{microsoft2024mddr}
{Microsoft}.
\newblock Advanced technologies at the core of {Microsoft Defender Antivirus}.
\newblock \url{https://learn.microsoft.com/en-us/defender-endpoint/adv-tech-of-mdav}, 2025.
\newblock [Last Accessed: April 23, 2026].

\bibitem{microsoftcallingconventions}
{Microsoft}.
\newblock Calling conventions, 2025.
\newblock [Last Accessed: Apr 15, 2026].
\newblock URL: \url{https://learn.microsoft.com/en-us/cpp/cpp/calling-conventions?view=msvc-170}.

\bibitem{pespecifications}
Microsoft.
\newblock Pe format, 2025.
\newblock [Last Accessed: Dec 9, 2025].
\newblock URL: \url{https://learn.microsoft.com/en-us/windows/win32/debug/pe-format}.

\bibitem{mikolov2013efficient}
Tomas Mikolov, Kai Chen, Greg Corrado, and Jeffrey Dean.
\newblock Efficient estimation of word representations in vector space.
\newblock {\em arXiv preprint arXiv:1301.3781}, 2013.

\bibitem{ming2017impeding}
Jiang Ming, Zhi Xin, Pengwei Lan, Dinghao Wu, Peng Liu, and Bing Mao.
\newblock Impeding behavior-based malware analysis via replacement attacks to malware specifications.
\newblock {\em Journal of Computer Virology and Hacking Techniques}, 13:193--207, 2017.

\bibitem{mwangi2024adversarial}
Maina~Bernard Mwangi and Shin-Ming Cheng.
\newblock An adversarial attack on ml-based iot malware detection using binary diversification techniques.
\newblock {\em IEEE Access}, 12:185172--185186, 2024.
\newblock \href {https://doi.org/10.1109/ACCESS.2024.3513713} {\path{doi:10.1109/ACCESS.2024.3513713}}.

\bibitem{peng2025evading}
Hao Peng, Zehao Yu, Dandan Zhao, Zhiguo Ding, Jieshuai Yang, Bo~Zhang, Jianming Han, Xuhong Zhang, Shouling Ji, and Ming Zhong.
\newblock Evading control flow graph based gnn malware detectors via active opcode insertion method with maliciousness preserving.
\newblock {\em Scientific Reports}, 15(1):9174, 2025.

\bibitem{QIAO2022102762}
Yanchen Qiao, Weizhe Zhang, Zhicheng Tian, Laurence~T. Yang, Yang Liu, and Mamoun Alazab.
\newblock Adversarial malware sample generation method based on the prototype of deep learning detector.
\newblock {\em Computers \& Security}, 119:102762, 2022.
\newblock URL: \url{https://www.sciencedirect.com/science/article/pii/S0167404822001572}, \href {https://doi.org/10.1016/j.cose.2022.102762} {\path{doi:10.1016/j.cose.2022.102762}}.

\bibitem{raff2018malware}
Edward Raff, Jon Barker, Jared Sylvester, Robert Brandon, Bryan Catanzaro, and Charles~K. Nicholas.
\newblock Malware detection by eating a whole exe.
\newblock In {\em AAAI Workshops}, 2017.
\newblock URL: \url{https://api.semanticscholar.org/CorpusID:33641567}.

\bibitem{raff2021classifying}
Edward Raff, William Fleshman, Richard Zak, Hyrum~S Anderson, Bobby Filar, and Mark McLean.
\newblock Classifying sequences of extreme length with constant memory applied to malware detection.
\newblock In {\em Proceedings of the AAAI Conference on Artificial Intelligence}, volume~35, pages 9386--9394, 2021.

\bibitem{idapro}
Hex Rays.
\newblock Ida pro - a powerful disassembler, decompiler and a versatile debugger. in one tool., 2025.
\newblock [Last Accessed: Dec 31, 2025].
\newblock URL: \url{https://hex-rays.com/ida-pro}.

\bibitem{rigaki2023power}
Maria Rigaki and Sebastian Garcia.
\newblock The power of meme: Adversarial malware creation with model-based reinforcement learning.
\newblock In {\em European Symposium on Research in Computer Security}, pages 44--64. Springer, 2023.

\bibitem{sechen2003timberwolf}
C.~Sechen and A.~Sangiovanni-Vincentelli.
\newblock The timberwolf placement and routing package.
\newblock {\em IEEE Journal of Solid-State Circuits}, 20(2):510--522, 1985.
\newblock \href {https://doi.org/10.1109/JSSC.1985.1052337} {\path{doi:10.1109/JSSC.1985.1052337}}.

\bibitem{someya2023fcgat}
Minami Someya, Yuhei Otsubo, and Akira Otsuka.
\newblock Fcgat: Interpretable malware classification method using function call graph and attention mechanism.
\newblock In {\em Proceedings of Network and Distributed Systems Security (NDSS) Symposium}, volume~1, 2023.

\bibitem{someya2023graph}
Minami Someya, Yuhei Otsubo, and Akira Otsuka.
\newblock Graph neural network based function call graph embedding for malware classification.
\newblock {\em Journal of Surveillance, Security and Safety}, 4(2):47--61, 2023.

\bibitem{song2025fcghunter}
Shiwen Song, Xiaofei Xie, Ruitao Feng, Qi~Guo, and Sen Chen.
\newblock Fcghunter: Towards evaluating robustness of graph-based android malware detection.
\newblock {\em IEEE Transactions on Software Engineering}, 52(2):428--448, 2026.
\newblock \href {https://doi.org/10.1109/TSE.2025.3626788} {\path{doi:10.1109/TSE.2025.3626788}}.

\bibitem{sophos2025ransomware}
{Sophos}.
\newblock The state of ransomware 2025.
\newblock \url{https://www.sophos.com/en-us/content/state-of-ransomware}, 2025.
\newblock Median ransom payment \$1M in 2025 (down from \$2M in 2024); average recovery cost \$1.53M.

\bibitem{suciu2019exploring}
Octavian Suciu, Scott~E Coull, and Jeffrey Johns.
\newblock Exploring adversarial examples in malware detection.
\newblock In {\em 2019 IEEE Security and Privacy Workshops (SPW)}, pages 8--14. IEEE, 2019.

\bibitem{surfshark2025malware}
{Surfshark}.
\newblock Windows users, beware: 7× more malware than macos.
\newblock \url{https://surfshark.com/research/chart/malware-cases-windows-macOS}, 2025.
\newblock [Last Accessed: April 23, 2026].

\bibitem{11575074}
Kai Tan, Dongyang Zhan, Haining Yu, Hongli Zhang, Bei Zhao, and Xiaojun Jia.
\newblock { A Practical Black-box Adversarial Attack on Graph-learning-based Models for Malware Detection }.
\newblock {\em IEEE Transactions on Dependable and Secure Computing}, (01):1--18, June 2026.
\newblock URL: \url{https://doi.ieeecomputersociety.org/10.1109/TDSC.2026.3706542}, \href {https://doi.org/10.1109/TDSC.2026.3706542} {\path{doi:10.1109/TDSC.2026.3706542}}.

\bibitem{enigma}
Enigma Protector~Developers Team.
\newblock Enigma: A professional system for executable files licensing and protection., 2025.
\newblock [Last Accessed: Dec 10, 2025].
\newblock URL: \url{https://enigmaprotector.com/}.

\bibitem{upx}
The~UPX Team.
\newblock Upx: The ultimate packer for executables., 2025.
\newblock [Last Accessed: Dec 10, 2025].
\newblock URL: \url{https://upx.github.io/}.

\bibitem{tian2024functionality}
Buwei Tian, Junyong Jiang, Zichen He, Xin Yuan, Lu~Dong, and Changyin Sun.
\newblock Functionality-verification attack framework based on reinforcement learning against static malware detectors.
\newblock {\em IEEE Transactions on Information Forensics and Security}, 19:8500--8514, 2024.
\newblock \href {https://doi.org/10.1109/TIFS.2024.3453047} {\path{doi:10.1109/TIFS.2024.3453047}}.

\bibitem{robust_malware_detection_challenge}
Sicco Verwer, Azqa Nadeem, Christian Hammerschmidt, Laurens Bliek, Abdullah Al-Dujaili, and Una-May O'Reilly.
\newblock The robust malware detection challenge and greedy random accelerated multi-bit search.
\newblock In {\em Proceedings of the 13th ACM Workshop on Artificial Intelligence and Security}, AISec'20, page 61–70, New York, NY, USA, 2020. Association for Computing Machinery.
\newblock \href {https://doi.org/10.1145/3411508.3421374} {\path{doi:10.1145/3411508.3421374}}.

\bibitem{wang2020mdea}
Xiruo Wang and Risto Miikkulainen.
\newblock Mdea: Malware detection with evolutionary adversarial learning.
\newblock In {\em 2020 IEEE Congress on Evolutionary Computation (CEC)}, pages 1--8, 2020.
\newblock \href {https://doi.org/10.1109/CEC48606.2020.9185810} {\path{doi:10.1109/CEC48606.2020.9185810}}.

\bibitem{wang2025optilock}
Zeng Wang, Lilas Alrahis, Animesh Basak~Chowdhury, Dominik Germek, Ramesh Karri, and Ozgur Sinanoglu.
\newblock Optilock: Automated optimization of learning-resilient logic locking.
\newblock {\em IEEE Access}, 13:166649--166669, 2025.
\newblock \href {https://doi.org/10.1109/ACCESS.2025.3612444} {\path{doi:10.1109/ACCESS.2025.3612444}}.

\bibitem{wu2021mcbg}
Bolun Wu, Yuanhang Xu, and Futai Zou.
\newblock Malware classification by learning semantic and structural features of control flow graphs.
\newblock In {\em 20th {IEEE} International Conference on Trust, Security and Privacy in Computing and Communications, TrustCom 2021, Shenyang, China, October 20-22, 2021}, pages 540--547. {IEEE}, 2021.
\newblock URL: \url{https://doi.org/10.1109/TrustCom53373.2021.00084}, \href {https://doi.org/10.1109/TRUSTCOM53373.2021.00084} {\path{doi:10.1109/TRUSTCOM53373.2021.00084}}.

\bibitem{https://doi.org/10.1049/ise2.12082}
Yafei Wu, Jian Shi, Peicheng Wang, Dongrui Zeng, and Cong Sun.
\newblock Deepcatra: Learning flow- and graph-based behaviours for android malware detection.
\newblock {\em IET Information Security}, 17(1):118--130, 2023.
\newblock URL: \url{https://ietresearch.onlinelibrary.wiley.com/doi/abs/10.1049/ise2.12082}, \href {https://arxiv.org/abs/https://ietresearch.onlinelibrary.wiley.com/doi/pdf/10.1049/ise2.12082} {\path{arXiv:https://ietresearch.onlinelibrary.wiley.com/doi/pdf/10.1049/ise2.12082}}, \href {https://doi.org/10.1049/ise2.12082} {\path{doi:10.1049/ise2.12082}}.

\bibitem{xie2025chain}
Peng Xie, Yequan Bie, Jianda Mao, Yangqiu Song, Yang Wang, Hao Chen, and Kani Chen.
\newblock Chain of attack: On the robustness of vision-language models against transfer-based adversarial attacks.
\newblock In {\em 2025 IEEE/CVF Conference on Computer Vision and Pattern Recognition (CVPR)}, pages 14679--14689, 2025.
\newblock \href {https://doi.org/10.1109/CVPR52734.2025.01368} {\path{doi:10.1109/CVPR52734.2025.01368}}.

\bibitem{yan2019classifying}
Jiaqi Yan, Guanhua Yan, and Dong Jin.
\newblock Classifying malware represented as control flow graphs using deep graph convolutional neural network.
\newblock In {\em 2019 49th Annual IEEE/IFIP International Conference on Dependable Systems and Networks (DSN)}, pages 52--63, 2019.
\newblock \href {https://doi.org/10.1109/DSN.2019.00020} {\path{doi:10.1109/DSN.2019.00020}}.

\bibitem{Yuan2020BlackBoxAA}
Junkun Yuan, Shaofang Zhou, Lanfen Lin, Feng Wang, and Jia Cui.
\newblock Black-box adversarial attacks against deep learning based malware binaries detection with gan.
\newblock In {\em European Conference on Artificial Intelligence}, 2020.
\newblock URL: \url{https://api.semanticscholar.org/CorpusID:221714626}.

\bibitem{zapzalka2024semantics}
Dylan Zapzalka, Saeed Salem, and David Mohaisen.
\newblock Semantics-preserving node injection attacks against gnn-based acfg malware classifiers.
\newblock {\em IEEE Transactions on Dependable and Secure Computing}, 22(1):549--560, 2025.
\newblock \href {https://doi.org/10.1109/TDSC.2024.3409410} {\path{doi:10.1109/TDSC.2024.3409410}}.

\bibitem{zhang2022semantics}
Lan Zhang, Peng Liu, Yoon-Ho Choi, and Ping Chen.
\newblock Semantics-preserving reinforcement learning attack against graph neural networks for malware detection.
\newblock {\em IEEE Transactions on Dependable and Secure Computing}, 20(2):1390--1402, 2023.
\newblock \href {https://doi.org/10.1109/TDSC.2022.3153844} {\path{doi:10.1109/TDSC.2022.3153844}}.

\bibitem{zhao2021structural}
Kaifa Zhao, Hao Zhou, Yulin Zhu, Xian Zhan, Kai Zhou, Jianfeng Li, Le~Yu, Wei Yuan, and Xiapu Luo.
\newblock Structural attack against graph based android malware detection.
\newblock In {\em Proceedings of the 2021 ACM SIGSAC Conference on Computer and Communications Security}, CCS '21, page 3218–3235, New York, NY, USA, 2021. Association for Computing Machinery.
\newblock \href {https://doi.org/10.1145/3460120.3485387} {\path{doi:10.1145/3460120.3485387}}.

\bibitem{zhu2024learning}
Rongyi Zhu, Zeliang Zhang, Zhuo Liu, Chenliang Xu, and Susan Liang.
\newblock Learning to transform dynamically for better adversarial transferability.
\newblock In {\em 2024 IEEE/CVF Conference on Computer Vision and Pattern Recognition (CVPR)}, pages 24273--24283, 2024.
\newblock \href {https://doi.org/10.1109/CVPR52733.2024.02291} {\path{doi:10.1109/CVPR52733.2024.02291}}.

\bibitem{zou2023universal}
Andy Zou, Zifan Wang, J.~Zico Kolter, and Matt Fredrikson.
\newblock Universal and transferable adversarial attacks on aligned language models.
\newblock {\em ArXiv}, abs/2307.15043, 2023.
\newblock URL: \url{https://api.semanticscholar.org/CorpusID:260202961}.

\end{thebibliography}

\appendix
\section{Appendix}\label{sec:appendix}

\section*{Ethics Considerations}\label{sec:ethics}
All experiments were conducted on known malware samples from MalwareBazaar under a static-only pipeline. No adversarial variants were executed outside isolated environments, distributed, or publicly released. The generated adversarial variants are not publicly released, and we will share only cryptographic hashes of the original samples to enable reproducibility. This research does not involve human subjects and does not require IRB approval.





\subsection{SAFE+GNN: Architecture and Training Details}
\label{app:SAFE+GNN}
\paragraph{Original Work.}
Someya et al.~\cite{someya2023graph} proposed SAFE+GNN, a malware \emph{family} classification method that combines Function Call Graph structure with learned function embeddings. The original model was evaluated on two datasets (MalwareBazaar: 3{,}927 samples, 6 families; BIG-2015: 10{,}810 samples, 9 families) using ten-fold cross-validation, achieving F1-scores of 98.27\% and 98.31\%, respectively. The original implementation uses Ghidra for FCG extraction and the SAFE neural network~\cite{SAFE}, a bidirectional GRU with self-attention, for function feature extraction.

\paragraph{Our adaptation.}
We adapt the SAFE+GNN architecture for \emph{binary} malware detection (malware vs.\ benign) with the following modifications:

\begin{enumerate}
  \item \textbf{Task}: Binary classification (2 output classes) instead of multi-class family classification.
  \item \textbf{Feature extraction}: The original SAFE+GNN uses the SAFE neural encoder~\cite{SAFE}, a pretrained bidirectional GRU with self-attention and the only publicly available pretrained model was trained exclusively on AMD64 (64-bit) instruction sequences, while our evaluation corpus consists of x86 (32-bit) PE binaries. The 32-bit and 64-bit x86 ISAs differ in register naming, addressing modes, and instruction encoding, so applying the 64-bit model to 32-bit disassembly maps the majority of instructions to unknown tokens. We therefore replace SAFE with a Word2Vec~\cite{mikolov2013efficient} CBOW model (vector size 100, window 2, 100 training epochs) trained on tokenized 32-bit disassembly from training corpus for SAFE+GNN, mean-pooling per-instruction embeddings to obtain 100-dimensional function-level feature vectors. We follow the same Word2Vec CBOW embedding approach used by FCGAT~\cite{someya2023fcgat}, which employs identical parameters (vector size 100, window 2) for function feature creation in FCG-based classification. This also ensures computational efficiency during attack simulation, where function embeddings must be recomputed at each classifier query.
  \item \textbf{FCG extraction}: We use IDA Home~\cite{idapro} instead of Ghidra for consistency with our MalConv and MalGraph evaluation pipeline.
  \item \textbf{Dataset}: The model was trained and evaluated on 20{,}000 samples from a 2024-collected corpus (Section~\ref{app:SAFE+GNN_dataset}).
\end{enumerate}

\noindent
All hyperparameters were kept the same as the original implementation: 3$\times$ SAGEConv layers (mean aggregation, hidden dimension 64), LeakyReLU activation (negative slope 0.01), dropout rates 0.2/0.3/0.4 (input/conv/readout), global mean pool readout, and a Linear (64$\to$2) classifier head (29{,}506 total parameters). Training uses AdamW (lr\,=\,0.005, weight decay\,=\,0.01), cross-entropy loss, batch size 128, and early stopping on validation F1-macro with patience 50. We selected the best checkpoint at epoch 100 and evaluated it on 2{,}000 held-out test samples (1{,}000 malware, 1{,}000 benign). Table~\ref{tab:safegnn_test_metrics} reports the full test-set performance.

\paragraph{Word2Vec embedding.}
Following FCGAT~\cite{someya2023fcgat}, we represented each assembly instruction as one Word2Vec token. We removed commas and comments following semicolons, replaced hexadecimal and decimal numeric constants with \texttt{N}, and joined each opcode and its operands with underscores (e.g., \texttt{push\_eax}, \texttt{mov\_ebp\_esp}, and \texttt{call\_N}). We trained a Gensim CBOW model with an embedding dimension of 100, a context window of 2, and 100 epochs on tokenized disassembly from the 16{,}000 training files. Validation, test, and FPR-calibration samples were excluded. The minimum token count was set to 5 to prune rare address-specific tokens, yielding a vocabulary of 4{,}817{,}392 distinct instruction tokens.

\begin{table}[h]
\centering
\caption{SAFE+GNN test-set performance (epoch 100, 2{,}000 samples).}
\label{tab:safegnn_test_metrics}
\footnotesize
\setlength{\tabcolsep}{4pt}
\begin{tabular}{@{}lcccc@{}}
\toprule
\textbf{TP} & \textbf{TN} & \textbf{FP} & \textbf{FN} & \textbf{Accuracy} \\
\midrule
988 & 917 & 83 & 12 & 95.25\% \\
\bottomrule
\end{tabular}
\\[4pt]
\begin{tabular}{@{}lccccc@{}}
\toprule
\textbf{Precision} & \textbf{Recall (TPR)} & \textbf{F1} & \textbf{TNR} & \textbf{FPR} & \textbf{FNR} \\
\midrule
0.9225 & 0.9880 & 0.9541 & 0.9170 & 0.0830 & 0.0120 \\
\bottomrule
\end{tabular}
\end{table}

\subsection{\textsc{PatchAtCall} Implementation Details}\label{app:patch-at-call}

We describe the implementation details of the \textsc{PatchAtCall} procedure (Algorithm~\ref{alg:patch-at-call}) for patching at a call site $c$ with semantic NOPs. We allocate a contiguous region in a newly injected PE section at a given Relative Virtual Address (RVA). The trampoline's Virtual Address (VA) is computed as $\texttt{imagebase} + \text{RVA}$, and the $k$ dummy function bodies are placed sequentially immediately after the trampoline. Each dummy function's VA is determined by accumulating the sizes of the preceding bodies from the trampoline's end. All \texttt{call} and \texttt{jmp} operands are encoded as 5-byte near-relative instructions, and the original 5-byte \texttt{call} at the hijacked site is overwritten with a same-sized \texttt{jmp} to the trampoline VA.

\subsection{Call-Site Ranking Strategies}\label{app:structural-filtering}

The site-selection strategies rank candidate call sites using signals derived from the malware binary before the NOP search phase.
We describe two ranking strategies: assembly-level instruction-density ranking and structural CFG centrality ranking.

\subsubsection{Instruction-Density Ranking}

Assembly code-based instruction-density features have been used to identify malware-informative regions in prior studies~\cite{downing2021deepreflect, huang2025malfocus}. We rank call sites by these signals to prioritize perturbation at classifier-sensitive locations.

For each candidate call site $c$, we extract a fixed-size window of assembly instructions centered at $c$ (using $w_{size}$ instructions before and after). Windows are collected via IDA Home. For each window $w$, we compute a scalar score $s_w$ based on the density of the following instruction categories, following the feature set of DeepReflect~\cite{downing2021deepreflect}:
\begin{itemize}
  \item \textbf{Arithmetic and Logic Instructions}: counts of basic math, bit-shifting, and logical operations (e.g., \texttt{xor}, \texttt{add}, \texttt{sub}, \texttt{shl}).
  \item \textbf{Transfer Instructions}: counts of data-movement operations (e.g., \texttt{mov}, \texttt{push}, \texttt{pop}) reflecting memory manipulation and argument passing.
\end{itemize}

Windows with fewer than \texttt{min\_inst} instructions are discarded.
Remaining windows are sorted in descending order of $s_w$, after which a one-dimensional IoU-based non-maximum suppression step removes windows with $>50\%$ overlap, preventing the ranked list from clustering around a single code region when adjacent call sites produce overlapping instruction windows. The top-$K_w$ non-overlapping call-site addresses form the ranked candidate set $\mathcal{C}_{asm}$.
The full procedure is in Algorithm~\ref{alg:window-ranking}.

\begin{algorithm}[h]
\fontsize{7}{8}\selectfont
\caption{Assembly-Based Call Site Ranking}
\label{alg:window-ranking}
\begin{algorithmic}[1]
\Require Set of call-site windows $\mathcal{W}$, minimum instruction threshold \texttt{min\_inst}, IoU threshold $\tau_{iou}$, desired number of windows $K_w$.
\Ensure Ranked set of top-$K_w$ call sites $\mathcal{C}_{asm}$.

\For{each window $w \in \mathcal{W}$}
    \State Extract instruction features (counts and ratios)
    \If{number of instructions in $w$ $< \texttt{min\_inst}$} Discard $w$
    \Else
        \State Store $w$ and its raw features
    \EndIf
\EndFor

\State Compute min-max statistics for every feature across remaining windows

\For{each window $w$}
    \State Normalize feature values using min-max statistics
    \State Compute score $s_w$ from instruction-density features by summing normalized feature counts
\EndFor

\State Sort all windows in descending order of $s_w$

\State Initialize empty set $\mathcal{R}$
\For{each window $w$ in sorted order}
    \If{$\mathrm{IoU}(w, w') < \tau_{iou}$ for all $w' \in \mathcal{R}$}
        \State Add $w$ to $\mathcal{R}$ \Comment{1-D IoU-based NMS}
    \EndIf
\EndFor

\State Let $\mathcal{W}_{top}$ be the top-$K_w$ windows in $\mathcal{R}$ (or all if $|\mathcal{R}| < K_w$)
\State \Return $\mathcal{C}_{asm} \gets \{c_w : w \in \mathcal{W}_{top}\}$, where $c_w$ is the call-site address centered in $w$

\end{algorithmic}
\end{algorithm}

\subsubsection{Structural CFG Centrality Ranking}

For the structural direction, we investigate whether a call site's topological position within the function's CFG influences how effectively a perturbation at that site disrupts the classifier's decision.
This is particularly relevant for graph-based classifiers such as MalGraph that operate directly on CFG and FCG representations.
We adapt graph-theoretic centrality measures from prior work on adversarial malware generation~\cite{zapzalka2024semantics} and compute them from CFGs. 

For each candidate call site $c$, we identify the basic block $BB_c$ containing $c$ within the CFG of its enclosing function and compute a centrality score $\mathcal{K}(BB_c)$. We compute centrality at the CFG basic-block level rather than the FCG level to distinguish among call sites within the same function.
Table~\ref{tab:centrality_measures} summarizes the measures considered.

\begin{table}[h]
\centering
\footnotesize
\caption{Structural centrality measures computed per call site from IDA-extracted CFGs.}
\label{tab:centrality_measures}
\begin{tabular}{lp{6.2cm}}
\toprule
\textbf{Measure} & \textbf{Description} \\
\midrule
Raw Degree       & Sum of in and out-edges of $BB_c$. Captures local connectivity. \\
Betweenness      & Fraction of all-pairs shortest paths passing through $BB_c$. Reflects its role as a structural bridge. \\
Eigenvector      & Recursive importance proportional to neighbor importance. Captures influence propagation. \\
\bottomrule
\end{tabular}
\end{table}

Call sites are ranked by $\mathcal{K}(BB_c)$ and the top-$K_s$ are selected.
The sorting direction (ascending or descending) is treated as a hyperparameter. The complete procedure is in Algorithm~\ref{alg:structural-ranking}.

\begin{algorithm}[h]
\fontsize{7}{8}\selectfont
\caption{Structural CFG-Based Call Site Ranking}
\label{alg:structural-ranking}
\begin{algorithmic}[1]
\Require Set of candidate call sites $\mathcal{C}$, centrality measure $\mathcal{K} \in \{\text{raw\_degree}, \text{betweenness}, \text{eigenvector}\}$, sorting direction $d \in \{\text{ascending}, \text{descending}\}$, desired number of sites $K_s$.
\Ensure Ranked set of top-$K_s$ call sites $\mathcal{C}_{top}$.

\For{each call site $c \in \mathcal{C}$}
    \State Identify function $F_c$ containing $c$ and its CFG $G_{F_c}$
    \State Identify basic block $BB_c \in G_{F_c}$ containing $c$
    \State Compute centrality score $\mathcal{K}(BB_c)$ on $G_{F_c}$
\EndFor

\State Sort $\mathcal{C}$ by $\mathcal{K}(BB_c)$ according to direction $d$

\State \Return the top-$K_s$ call sites $\mathcal{C}_{top}$ (or all if $|\mathcal{C}| < K_s$)

\end{algorithmic}
\end{algorithm}

\subsection{Greedy Search Algorithms}\label{app:greedy-outer}

Algorithm~\ref{alg:greedy-sem-nop} details the outer greedy loop over candidate call sites.
It follows the same structure as the SA outer loop (Algorithm~\ref{alg:sa_sem_nop}) without temperature tracking or probabilistic acceptance.
The inner per-site procedure is in Algorithm~\ref{alg:local-greedy}.

\begin{algorithm}[h]
\fontsize{7}{8}\selectfont
\caption{Greedy Semantic NOP Search over Call Sites}
\label{alg:greedy-sem-nop}
\begin{algorithmic}[1]
\Require Malware $m_0$, classifier $\mathcal{P}(\cdot)$ with threshold $\tau^{clsf}$, call sites $\mathcal{C}$, NOP library $\mathcal{N}$, size budget $B$, max iterations $I_{\max}$.
\Ensure $(m^*, p^*, \texttt{bypassed})$: best binary, its score, and evasion flag.
\State $m^* \gets m_0$,\ $p^* \gets \mathcal{P}(m_0)$
\State $\texttt{no\_improve} \gets 0$,\ $\texttt{bypassed} \gets \texttt{false}$,\ $i \gets 0$
\While{$i < I_{\max}$ \textbf{and} $\mathcal{C} \neq \emptyset$ \textbf{and} $\texttt{no\_improve} < \texttt{patience}$}
    \If{$p^* < \tau^{clsf}$} $\texttt{bypassed} \gets \texttt{true}$;\ \textbf{break} \EndIf
    \State Select call site $c \in \mathcal{C}$
    \State $(m_{\text{site}}, p_{\text{site}}, b_{\text{in}}) \gets \textsc{LocalGreedy}(c, m_0, m^*, p^*, \mathcal{N}, B,  J_{\max}, \tau^{clsf})$
    \If{$b_{\text{in}}$} $m^* \gets m_{\text{site}}$,\ $p^* \gets p_{\text{site}}$,\ $\texttt{bypassed} \gets \texttt{true}$; \textbf{break}
    \EndIf
    \If{$p_{\text{site}} < p^*$} $m^* \gets m_{\text{site}}$,\ $p^* \gets p_{\text{site}}$,\ $\texttt{no\_improve} \gets 0$ \Comment{Update best}
    \Else
        \State $\texttt{no\_improve} \gets \texttt{no\_improve} + 1$
    \EndIf
    \State \textsc{UpdateCallSiteList}$(\mathcal{C}, c, m^*)$ \Comment{Append new calls from displacement}
    \State $i \gets i + 1$
\EndWhile
\State \Return $(m^*, p^*, \texttt{bypassed})$
\end{algorithmic}
\end{algorithm}

Algorithm~\ref{alg:local-greedy} details the inner greedy loop invoked at each call site by the outer greedy procedure (Algorithm~\ref{alg:greedy-sem-nop}).
For each candidate semantic NOP sampled from $\mathcal{N}$, the inner loop patches the current global state via \textsc{PatchAtCall} (Algorithm~\ref{alg:patch-at-call}), queries the classifier, and retains the lowest-scoring candidate as the site-level best.

\begin{algorithm}[h]
\fontsize{7}{8}\selectfont
\caption{\textsc{LocalGreedy}: Inner Greedy Loop at a Fixed Call Site}
\label{alg:local-greedy}
\begin{algorithmic}[1]
\Require Call site $c$, original malware $m_0$, global state $m^*$ with score $p^*$, NOP library $\mathcal{N}$, size budget $B$, max inner iterations $J_{\max}$, classifier threshold $\tau^{clsf}$.
\Ensure $(m_{\text{site}}, p_{\text{site}}, \texttt{bypassed})$: best state for this call site, its score, and evasion flag.
\State $m_{\text{site}} \gets m^*$,\ $p_{\text{site}} \gets p^*$
\State $\texttt{bypassed} \gets \texttt{false}$
\For{$j = 1$ to $J_{\max}$}
    \State $n_j \sim \textsc{SelectSemNops}(\mathcal{N})$
    \State $\hat{m} \gets \textsc{PatchAtCall}(m^*, c, n_j)$ \Comment{From global $m^*$}
    \If{\textsc{SizeIncrease}$(\hat{m}, m_0) > B$} \textbf{continue} \EndIf
    \State $p_{\text{cand}} \gets \mathcal{P}(\hat{m})$
    \If{$p_{\text{cand}} < \tau^{clsf}$} $m_{\text{site}} \gets \hat{m}$,\ $p_{\text{site}} \gets p_{\text{cand}}$,\ $\texttt{bypassed} \gets \texttt{true}$
        \State \textbf{return} $(m_{\text{site}}, p_{\text{site}}, \texttt{bypassed})$
    \EndIf
    \If{$p_{\text{cand}} < p_{\text{site}}$} $m_{\text{site}} \gets \hat{m}$,\ $p_{\text{site}} \gets p_{\text{cand}}$ \Comment{Strict greedy improvement}
    \EndIf
\EndFor
\State \Return $(m_{\text{site}}, p_{\text{site}}, \texttt{bypassed})$
\end{algorithmic}
\end{algorithm}

\subsection{Default Hyperparameters}\label{app:hyperparameters}
For our experiments, we set the maximum number of outer iterations
(call sites) to $I_{\max}=20$ for MalConv and MalGraph and
$I_{\max}=45$ for SAFE+GNN. The maximum number of inner iterations
(semantic-NOP candidates per call site) is $J_{\max}=20$ for MalConv
and MalGraph, $J_{\max}=1$ for SAFE+GNN Greedy, and $J_{\max}=3$ for
SAFE+GNN SA and Adaptive SA. The size-increase budget is $B=5\%$ for
MalConv and MalGraph and $B=30\%$ for SAFE+GNN. Per-sample timeouts are 15 minutes for MalConv and MalGraph, 25 minutes
for SAFE+GNN Greedy, and 35 minutes for SAFE+GNN SA and Adaptive SA. For SA, we initialize the temperature at $T_0=0.9$, apply a cooling factor of $\alpha=0.9$ after each outer iteration, and use a minimum temperature of $T_{\min}=0.01$ across classifiers. We also implement an adaptive timeout mechanism based on attack progress. For semantic NOPs, the number of distinct types sampled per iteration $s$ is 1 for MalConv and 3 for MalGraph, and the maximum repetition count per type $r$ is 300 for MalConv and 100 for MalGraph (see Appendix~\ref{app:nop-sensitivity}). For SAFE+GNN, semantic-NOP-based trampoline gap filling and body NOPs are disabled ($r=0$ and body-NOP count zero) and function-body content is instead provided by the alternative function body pool. This configuration reflects the finding that semantic NOP body content yields diminishing returns for this classifier (Section ~\ref{sec:ablation}) 85\% vs.\ 17\% ASR, so the search budget is allocated toward outer call-site iterations instead. Injected dummy functions use
$k_{\min}=1$ and $k_{\max}=3$ for MalConv and MalGraph, and
$k_{\min}=2{,}500$ and $k_{\max}=3{,}500$ for SAFE+GNN. MalConv and
MalGraph use 3 NOP sequences per injected function body. For instruction-density ranking, we use a window of
$w_{size}=50$ instructions on either side of the call site, a minimum instruction threshold of \texttt{min\_inst}=30, an IoU threshold of $\tau_{iou}=0.5$, and retain the top $K_w=20$ windows. For structural centrality ranking, we select the top $K_s=20$ call sites by centrality score. Baseline comparisons use the same hyperparameters where applicable.

The SAFE+GNN minimal function body pool consists of five
approximately 5-byte push-wrapped patterns, sampled with weights:
\texttt{push ecx; xor ecx,ecx; pop ecx; ret} (0.40),
\texttt{push ecx; test ecx,ecx; pop ecx; ret} (0.30),
\texttt{push ebx; xor ebx,ebx; pop ebx; ret} (0.15),
\texttt{push ebx; test ebx,ebx; pop ebx; ret} (0.10), and
\texttt{push eax; test eax,eax; pop eax; ret} (0.05). The
\texttt{push}/\texttt{pop} wrapping preserves register state, including
the return-value register \texttt{eax}. All patterns are in the SAFE+GNN
Word2Vec vocabulary.

\textbf{Adversarial finetuning hyperparameters.}
We finetune each pretrained model with a low learning rate for a small number of epochs: MalConv at lr\,=\,$5{\times}10^{-5}$ for 1 epoch, MalGraph at lr\,=\,$1{\times}10^{-4}$ for 2 epochs, and SAFE+GNN at lr\,=\,$5{\times}10^{-4}$ for 2 epochs. These were selected after tuning over learning rates and epoch counts. More aggressive configurations caused training loss to diverge for MalConv and MalGraph, likely due to distribution mismatch between the unavailable original training data and our independently collected finetuning set. For SAFE+GNN, which we trained from scratch (Appendix ~\ref{app:SAFE+GNN}), the adversarial finetuning samples were generated from the same 2024 corpus, ensuring no distribution mismatch. We recalibrate the finetuned models at 1\% FPR on the same benign calibration sets used for the pretrained models (Section~\ref{app:dataset-details}). We evaluate at this FPR because it corresponds to the harder
attack threshold (lower ASR across all methods in
Table~\ref{tab:full_comparison}). Recalibration shifts each pretrained model's threshold, so ASR values in Section~\ref{sec:rq5} are not directly comparable to the main results in Table~\ref{tab:full_comparison}. To ensure a fair comparison, we compute ASR only on the intersection of samples that both models detect: 1{,}123 for MalConv, 1{,}936 for MalGraph, and 1{,}000 for SAFE+GNN (where all samples are detected by both models).

\subsection{Dataset Construction Details}\label{app:dataset-details}\label{app:SAFE+GNN_dataset}

\textbf{Size filtering rationale.}
We collected 20{,}456 x86 Windows PE malware samples from MalwareBazaar~\cite{MalwareBazaar} 2025 submissions and partitioned them into three classifier-specific attack pools with the following size constraints:

\begin{compactitem}
\item \textbf{MalConv:} $\leq 1.9$\,MB. MalConv accepts inputs of at most 2\, MB, and our attack appends a new PE section to each sample, so we apply a tighter cutoff to ensure the injected section remains within the input window.
\item \textbf{MalGraph:} $\leq {\sim}8.4$\,MB. No inherent size constraint, but to control IDA analysis cost and memory usage during CFG/FCG extraction, we removed large outliers. The 95th-percentile sizes of the threshold-filtered MalGraph samples were 8.3707\, MB and 14.7546\, MB at 0.1\% and 1\% FPR, respectively. We applied the smaller cutoff (8.3707\, MB) uniformly across both FPRs.
\item \textbf{SAFE+GNN:} 10\,KB-10\,MB. The larger per-site injection volume ($k_{\min}/k_{\max}=2{,}500/3{,}500$) increases both the injected section size and the IDA reanalysis time per query, so we select a computationally tractable range.
\end{compactitem}

\noindent These size constraints are imposed by the target classifiers and IDA resource usage, not by the attack itself, which imposes no inherent size limit. After size filtering, we applied the following shared pipeline across all three pools: \wcircle{1}~Removed packed binaries using Detect-It-Easy (DIE)~\cite{die}, \wcircle{2}~Filtered out samples unparsable by the Python \texttt{pefile} library, \wcircle{3}~Retained only samples with at least one identifiable internal \texttt{call} instruction (the minimum requirement for our attack surface), and \wcircle{4}~Scored each pool against the respective classifier at 0.1\% and 1\% FPR operating points, retaining only samples above the classifier threshold. The final attack pools are 559/3{,}662 (MalConv), 2{,}605/3{,}234 (MalGraph), and 1{,}000/1{,}000 (SAFE+GNN) at 0.1\%/1\% FPR. SAFE+GNN uses 1{,}000 samples per threshold as the higher dummy node injection volume per sample increases per-attack computation cost.

\textbf{SAFE+GNN training corpus.}
Since MalConv and MalGraph use pretrained models from~\cite{ling2024wolf} (trained on 210{,}251 Windows executables~\cite{ling2022malgraph}), only SAFE+GNN required a dedicated training corpus. We collected 14{,}000 x86 Windows PE malware samples from MalwareBazaar for 2024, filtered packed binaries using DIE~\cite{die}, and retained only samples for which IDA successfully extracted both an FCG and per-function tokenized disassembly, yielding 10{,}000 usable samples. We sourced benign samples (17{,}886, used 10,000) from the Practical Security Analytics PE dataset~\cite{lesterpsadataset}.

\begin{table}[h]
\centering
\caption{SAFE+GNN dataset partitions. All partitions have zero pairwise SHA-256 overlap. Attack pools are drawn from a separate 2025 corpus. MB stands for MalwareBazaar and PSA for Practical Security Analytics.}
\label{tab:SAFE+GNN_dataset}
\footnotesize
\setlength{\tabcolsep}{3pt}
\begin{tabular}{@{}lrrrl@{}}
\toprule
\textbf{Partition} & \textbf{Malware} & \textbf{Benign} & \textbf{Total} & \textbf{Source} \\
\midrule
Training       & 8{,}000 & 8{,}000 & 16{,}000 & MB 2024 + PSA \\
Validation     & 1{,}000 & 1{,}000 & 2{,}000  & MB 2024 + PSA \\
Test           & 1{,}000 & 1{,}000 & 2{,}000  & MB 2024 + PSA \\
FPR calibration& ---     & 7{,}886 & 7{,}886  & PSA (held-out) \\
\midrule
Attack (0.1\% FPR) & 1{,}000 & --- & 1{,}000 & MB 2025 \\
Attack (1\% FPR)   & 1{,}000 & --- & 1{,}000 & MB 2025 \\
\bottomrule
\end{tabular}
\end{table}

\begin{figure}[h]
\centering
\includegraphics[width=0.75\columnwidth]{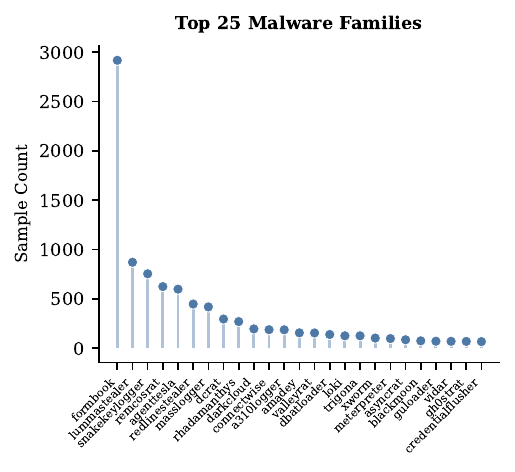}
\caption{Distribution of the 25 most prevalent malware families across the deduplicated union of all six attack/test pools.}
\label{fig:family_dist}
\end{figure}

\textbf{Attack Dataset Malware Families.}
We also report the malware family counts of our attack pool for each configuration. The family names were collected from the same MalwareBazaar~\cite{MalwareBazaar} API from their generated reports for each sample. Among all the unique malware samples across six pools (three classifiers at two FPR thresholds), we identify 217 unique malware families. An additional 8-15\% of samples per pool carry no family label in MalwareBazaar and are excluded from all family-level statistics. The largest pool by family diversity is MalGraph at 1\% FPR with 163 families. The composition of each pool is listed in Table~\ref{tab:family_counts}. Figure~\ref{fig:family_dist} shows the distribution of the 25 most prevalent families across the deduplicated union of all pools. 

\begin{table}[h]
\centering
\footnotesize
\caption{Attack pool malware family composition per configuration. Family counts exclude unlabeled samples.}
\label{tab:family_counts}
\setlength{\tabcolsep}{3.5pt}
\begin{tabular}{lcccccc}
\toprule
& \multicolumn{2}{c}{\textbf{MalConv}} & \multicolumn{2}{c}{\textbf{MalGraph}} & \multicolumn{2}{c}{\textbf{SAFE+GNN}} \\
\cmidrule(lr){2-3} \cmidrule(lr){4-5} \cmidrule(lr){6-7}
& 0.1\% & 1\% & 0.1\% & 1\% & 0.1\% & 1\% \\
\midrule
Samples          & 559   & 3{,}662 & 2{,}605 & 3{,}234 & 1{,}000 & 1{,}000 \\
Known-family     & 465   & 3{,}304 & 2{,}399 & 2{,}761 & 857     & 855     \\
Unique families  & 88    & 153     & 120     & 163     & 66      & 65      \\
\bottomrule
\end{tabular}
\end{table}

\textbf{SRL baseline and defense pools.}
For the SRL baseline and defense experiments, we assembled two additional pools from MalwareBazaar samples collected between March and November 2024. We randomly selected 2{,}000 samples from March to August 2024 to serve as the SRL baseline training set. To construct the defense evaluation pool (Section~\ref{sec:rq5}), we generated adversarial samples using our method on the 2024 dataset, applied the same DIE-based packer filter, and retained 300 samples for adversarial fine-tuning. We additionally collected 300 random benign Windows PE binaries for mixed-label adversarial fine-tuning of each target detector.

\textbf{Design justifications.}
Since the MalGuise training dataset, including the benign samples used to calibrate the pretrained classifiers, is not publicly available (we contacted the authors but did not receive the data), both the fine-tuning benign set and the threshold-calibration pools are drawn from our own independently collected benign set. The choice of a small number of fine-tuning samples and a disjoint 2024 adversarial pool is justified for two reasons: \wcircle{1}~A larger fine-tuning set would risk degrading the pretrained models' discriminative ability, since the distribution mismatch between the original (unavailable) training data and our fine-tuning set could hurt model performance, and \wcircle{2}~Using adversarial samples drawn from the same 2025 test pool and then evaluating on that pool would bias the ASR estimate, so the 2024 pool provides a cleanly disjoint fine-tuning set.
To recalibrate detection thresholds at controlled false-positive rates, we use dedicated benign calibration sets for each classifier: 2{,}762 samples for MalConv and 1{,}683 for MalGraph, both independently collected with no overlap with the fine-tuning benigns, and the same 7{,}886 held-out PSA benign set used for the pretrained SAFE+GNN model, since the 300 benign finetuning samples were drawn from the training split and do not overlap with the calibration set.

\subsection{Traditional Obfuscation Tools}
\label{app:upx_enigma}

We use two popular Windows executable obfuscation tools - UPX~\cite{upx}, a general-purpose executable packer that compresses binaries and decompresses them at runtime and Enigma~\cite{enigma}, a commercial Windows executable protector that applies various obfuscation mechanisms, like API simulations, code virtualizations, etc.~\cite{ling2024wolf} against MalConv and Malgraph classifiers. Each tool is applied directly to the original malware samples, and the packed/protected binaries are evaluated under the same detector configurations as our attack.

\begin{table}[h]
\centering
\caption{ASR (\%) of traditional obfuscation tools.}
\label{tab:upx_enigma}
\footnotesize
\setlength{\tabcolsep}{4pt}
\begin{tabular}{@{}lcccc@{}}
\toprule
 & \multicolumn{2}{c}{\textbf{MalConv}} & \multicolumn{2}{c}{\textbf{MalGraph}} \\
\cmidrule(lr){2-3} \cmidrule(lr){4-5}
 & 0.1\% & 1\% & 0.1\% & 1\% \\
\midrule
UPX & 3.76 & 0.33 & 2.73 & 1.24 \\
Enigma & 0.54 & 0.00 & 1.38 & 0.28 \\
\bottomrule
\end{tabular}
\end{table}

Both tools achieve negligible ASR ($\leq$3.76\%) across
all configurations (Table ~\ref{tab:upx_enigma}). For MalGraph, packing is
counterproductive as it destroys function boundaries,
produces degenerate graphs, and yields binaries flagged
as maximally suspicious. \system{}'s
structure-preserving perturbations yield fully analyzable
binaries that the classifier processes normally.

\subsection{High-Confidence Subsample Analysis}
\label{app:high_confidence}

To verify that \system{}'s effectiveness is not driven by borderline classifications, we filter the Table~\ref{tab:full_comparison} attack outcomes to
samples where the pre-attack classifier confidence
exceeds increasingly strict thresholds. For each
classifier and FPR, we retain only samples with
P(malware) $\geq \tau$ and recompute ASR.
Table~\ref{tab:high_confidence} reports the results.

\begin{table}[h]
\centering
\caption{ASR (\%) restricted to high-confidence samples.
$N$ is the number of qualifying samples at each cutoff.
ASR remains above 81\% at all cutoffs including
$\geq$0.999.}
\label{tab:high_confidence}
\footnotesize
\setlength{\tabcolsep}{2.5pt}
\renewcommand{\arraystretch}{1.05}
\begin{tabular}{@{}ll rr rr rr@{}}
\toprule
& & \multicolumn{2}{c}{$\geq$\textbf{0.95}}
& \multicolumn{2}{c}{$\geq$\textbf{0.99}}
& \multicolumn{2}{c}{$\geq$\textbf{0.999}} \\
\cmidrule(lr){3-4} \cmidrule(lr){5-6} \cmidrule(lr){7-8}
\textbf{Classifier} & \textbf{FPR}
& $N$ & ASR & $N$ & ASR & $N$ & ASR \\
\midrule
MalConv & 0.1\% & 559 & 100.0 & 223 & 100.0 & 40 & 100.0 \\
MalConv & 1\% & 595 & 98.5 & 213 & 99.1 & 39 & 97.4 \\
\midrule
MalGraph & 0.1\% & 2431 & 95.7 & 862 & 89.1 & 574 & 86.4 \\
MalGraph & 1\% & 1806 & 91.3 & 632 & 85.0 & 399 & 81.0 \\
\midrule
SAFE+GNN & 0.1\% & 1000 & 97.4 & 1000 & 97.4 & 930 & 98.0 \\
SAFE+GNN & 1\% & 963 & 84.5 & 899 & 84.0 & 822 & 83.6 \\
\bottomrule
\end{tabular}
\end{table}

ASR reduces as the confidence threshold
increases but remains above 81\% for all classifiers
at all cutoffs. MalConv ASR is effectively invariant
to filtering (97-100\%). MalGraph shows moderate
reduction at $\geq$0.999, but remains above 80\% even on the most confident predictions. SAFE+GNN is nearly
flat across all thresholds (85.0\% $\to$ 83.6\% at
1\% FPR), with 82-100\% of the test pool already
exceeding P(malware) = 0.95, indicating that the
classifier assigns high confidence to nearly all
samples before any perturbation is applied.

The number of qualifying samples at each cutoff
varies across classifiers because different
architectures produce different score distributions.
SAFE+GNN concentrates scores near 1.0 so most
samples pass even the strictest cutoff, while
MalConv and MalGraph spread scores more broadly at
1\% FPR. In all cases, every attack set sample was
correctly classified as malware above the respective
FPR threshold before any adversarial modification.

\subsection{Malware Family Overlap Analysis}
\label{app:family_overlap}

A potential concern with temporal evaluation gaps is
that the attack set may contain malware families not
represented in the classifier's training data, allowing
concept drift to inflate ASR. We analyze the family
overlap between SAFE+GNN's training corpus (2024) and
the attack test set (2025) to assess whether this is
the case.

Using MalwareBazaar family labels, we compute two
overlap metrics between the SAFE+GNN training set
(2024) and each FPR attack pool (2025). We define
\emph{family overlap} as the fraction of labeled
attack test-set families that also appear in the training
corpus, and \emph{sample coverage} as the fraction of
labeled attack test samples whose family appears in training.
Samples without a family label (14\% of each pool)
are excluded from both metrics.

\begin{table}[h]
\centering
\caption{SAFE+GNN family overlap between training
(2024) and attack test (2025) corpora. Known Fam.\
= families with labels in the test pool. Unseen =
families present only in the test set.}
\label{tab:family_overlap}
\footnotesize
\setlength{\tabcolsep}{3.5pt}
\renewcommand{\arraystretch}{1.05}
\begin{tabular}{@{}lccccc@{}}
\toprule
\textbf{FPR} & \textbf{Known} & \textbf{Overlap}
& \textbf{Unseen} & \textbf{Fam.\ Overlap}
& \textbf{Sample Cov.} \\
 & \textbf{Fam.} & \textbf{Fam.}
& \textbf{Fam.} & & \\
\midrule
0.1\% & 66 & 49 & 17 & 74.2\% & 97.1\% \\
1\%   & 65 & 47 & 18 & 72.3\% & 95.0\% \\
\bottomrule
\end{tabular}
\end{table}

Table~\ref{tab:family_overlap} shows that 72-74\%
of test families overlap with training, and these
overlapping families account for 95-97\% of labeled
test samples. The evaluated attack set is therefore
not dominated by unseen families, and unseen-family
novelty is unlikely to explain the observed ASR. Together with the
high-confidence analysis discussed in Appendix ~\ref{app:high_confidence}, these results indicate that the attack succeeds on a
test set whose family composition is well represented
in the classifier's training distribution.

For MalConv and MalGraph, the original training data
was not disclosed by the authors of~\cite{ling2024wolf}
despite repeated requests. However, these models
correctly classify all test samples above the
respective FPR thresholds, and ASR remains 84.5-100\%
even when restricted to samples with
P(malware) $\geq$ 0.95
(Table~\ref{tab:high_confidence}), indicating that
the classifiers are functional on the 2025 corpus.


\subsection{Search Strategy Convergence Efficiency Additional Details}
\label{app:search_efficiency}
\begin{figure}[tb]
\centering
\includegraphics[width=\columnwidth]{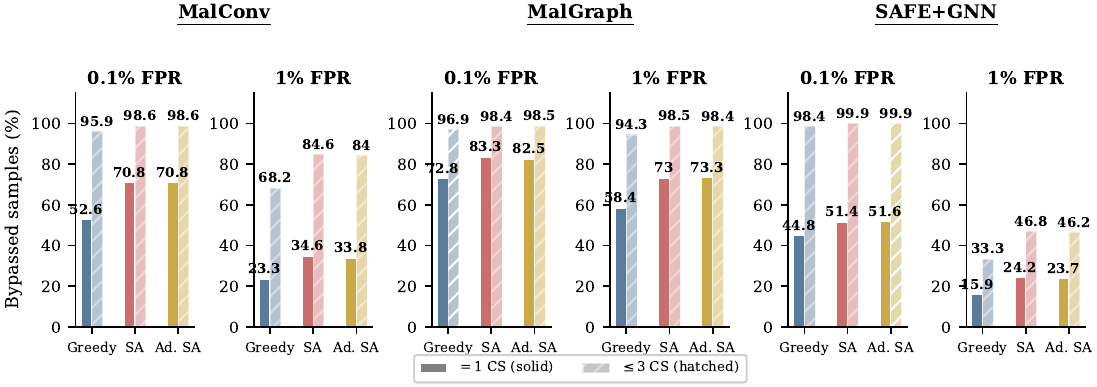}
\caption{Search strategy convergence efficiency: \% of bypassed samples requiring 1 (solid) or $\leq$\,3 (hatched) call site perturbations to cross the decision boundary.}
\label{fig:search_efficiency}
\end{figure}

We provide the details of how the choice of search strategy affects attack convergence efficiency in this section. Figure~\ref{fig:search_efficiency} compares the \% of bypassed samples that achieve evasion within a given perturbation budget. Across all three classifiers, SA and Adaptive SA consistently require fewer call site modifications than Greedy, showing that SA's ability to probabilistically accept worse semantic NOPs yields more targeted solutions and faster convergence. For MalConv and Malgraph, this gap is 10-18\, pp higher for SA variants across all configurations and most pronounced under MalConv at 1\% FPR, where Greedy achieves single-perturbation evasion for only 23.3\% of samples compared to 34.6\% for SA, and it widens further at the $\leq$\,3  threshold. For SAFE+GNN, the pattern also holds with SA achieving 24.2\% single-perturbation evasion vs 15.9\% at 1\% FPR, and 51.4\% vs 44.8\% at 0.1\% FPR.



\subsection{Call-Site Ranking Results}\label{app:structural-filtering-results}

\begin{figure}[tb]
\centering
\includegraphics[width=0.78\columnwidth]{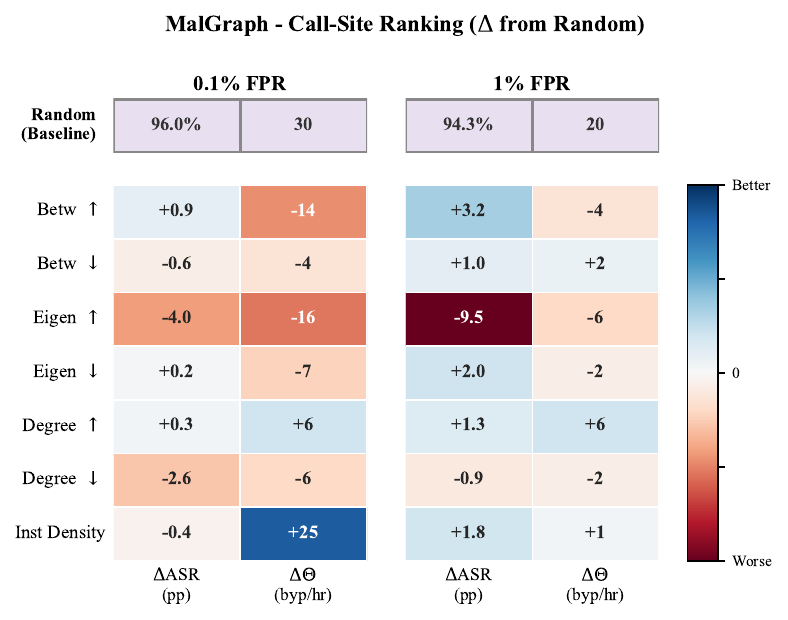}
\caption{Effect of structural call site ranking on MalGraph. Each cell shows the delta ($\Delta$) from random selection. Absolute random-baseline values appear in the shaded row at top. Color encodes direction: blue = improvement, red = degradation.}
\label{fig:structural_mg}
\end{figure}

Figure~\ref{fig:structural_mg} shows that for Malgraph, Betweenness$\uparrow$ yields the largest ASR gain (+3.2\, pp at 1\% FPR), as such basic blocks act as topological bridges, and displacing call sites at such nodes creates perturbations that influence more downstream paths in the graph. This comes at a throughput cost ($-$4\,byp/hr), as the search requires more iterations at these constrained sites. Degree$\uparrow$ offers the best throughput gain (+6\,byp/hr) with a modest ASR improvement (+1.28\,pp). Instruction density delivers the strongest throughput gain at 0.1\% FPR (+25\,byp/hr), suggesting assembly-level feature density correlates with classifier sensitivity. At the other extreme, Eigenvector$\uparrow$ degrades ASR by $-$9.51\,pp at 1\% FPR, as selecting call sites in nodes embedded in already well-connected subgraphs produces perturbations that are structurally marginal relative to the neighborhood's existing connectivity, contributing little to shifting the learned representation. Degree$\downarrow$ also underperforms random selection across configurations.

\begin{figure}[tb]
\centering
\includegraphics[width=0.78\columnwidth]{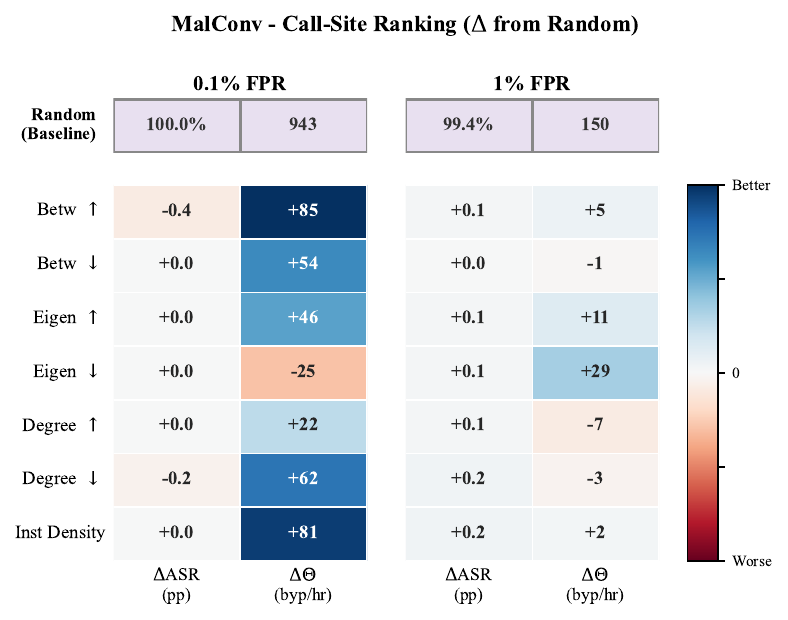}
\caption{Effect of structural call-site ranking on MalConv. Each cell shows the delta ($\Delta$) from random selection. Absolute random-baseline values appear in the shaded row at top. Color encodes direction: blue = improvement, red = degradation.}
\label{fig:structural_mc}
\end{figure}

Figure~\ref{fig:structural_mc} shows the effect of structural call-site ranking
on MalConv. ASR is largely insensitive to ranking strategy ($\pm$0.25\,pp). Throughput, however, shows more variation. The largest throughput gain comes from Eigenvector$\downarrow$ (+29\,byp/hr at 1\% FPR) rather than instruction
density, which produces comparatively modest improvements at that threshold. Throughput gains are also more pronounced at the 0.1\% FPR operating point, where fewer perturbation iterations are needed per sample.

\subsection{DAG-Constrained Injection}\label{app:dag-construction}

The FCG complexity ablation in Section~\ref{par:fcg_complexity} replaces the isolated leaf functions of Algorithm~\ref{alg:patch-at-call} with a small random directed acyclic graph (DAG) of inter-function calls.
Algorithm~\ref{alg:dag-construction} describes the DAG construction procedure.

\begin{algorithm}[h]
\fontsize{7.5}{8.5}\selectfont
\caption{\textsc{BuildRandomDAG}: DAG Construction for Dummy Function Chained Injection}
\label{alg:dag-construction}
\begin{algorithmic}[1]
\Require Number of dummy functions $k$, edge probability $p_{edge} \in [0,1]$.
\Ensure DAG adjacency $G$ over nodes $\{0, \dots, k{-}1\}$.

\State $G[i] \gets \emptyset$ for all $i \in \{0, \dots, k{-}1\}$
\State $\pi \gets \textsc{RandomShuffle}(\{0, \dots, k{-}1\})$

\For{each $u$ in $\pi$}
    \State $A_u \gets \textsc{Ancestors}(u, G)$ \Comment{Transitive parents via BFS}
    \For{each $v \in \{0, \dots, k{-}1\} \setminus (\{u\} \cup A_u)$}
        \If{$\textsc{Uniform}(0,1) < p_{edge}$}
            \State $G[u] \gets G[u] \cup \{v\}$ \Comment{Directed edge $u \to v$}
        \EndIf
    \EndFor
\EndFor

\State \textbf{assert} $G$ is acyclic \Comment{DFS cycle validation}
\State \Return $G$
\end{algorithmic}
\end{algorithm}

Given the DAG $G$, each dummy function body $\texttt{body}_i$ from Algorithm~\ref{alg:patch-at-call} is augmented with explicit \texttt{call} instructions to its children $G[i]$, interleaved among the semantic NOPs. The trampoline structure remains identical to Algorithm~\ref{alg:patch-at-call}. 

\subsection{SA Temperature Sensitivity}\label{app:temp-sensitivity}

The initial temperature $T_0$ controls the geometric decay rate of the acceptance probability in the SA algorithms. To test its effect on attack performance, we vary $T_0 \in \{0.3, 0.6, 0.9, 1.2\}$ and measure ASR and $\Theta$ at 0.1\% FPR for both classifiers. We use $T_0{=}0.9$ as the default value in all other experiments.

\begin{figure}[h]
\centering
\includegraphics[width=\columnwidth]{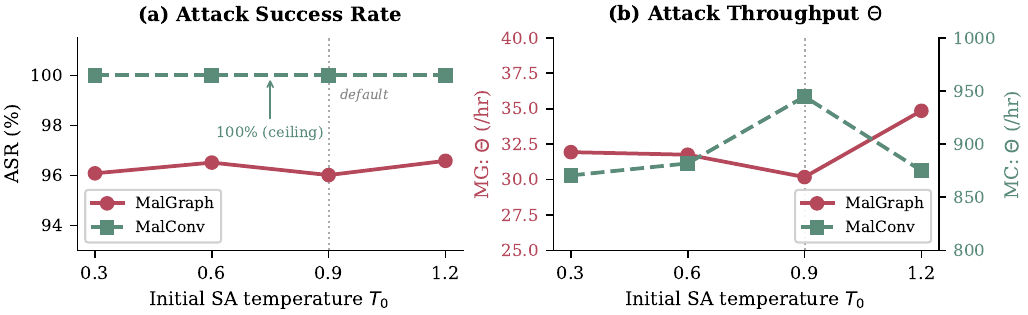}
\caption{Sensitivity of Adaptive SA to initial temperature $T_0$ at 0.1\% FPR. Dotted line marks the default ($T_0{=}0.9$). (a)~ASR varies by $<$0.6\,pp for MalGraph. MalConv is at 100\% ceiling. (b)~Attack Throughput $\Theta$ (successful bypasses/hr) on independent axes (MalGraph left, MalConv right).}
\label{fig:temp_sensitivity}
\end{figure}

Figure~\ref{fig:temp_sensitivity} shows that the ASR is largely insensitive to this hyperparameter. MalGraph ASR varies by less than 0.6\, pp across the full range, while MalConv remains unchanged. Throughput remains consistent up to 0.9 but increases by ~5 bypasses/hour for the 1.2 temperature setting for MalGraph, suggesting that higher temperature settings may be beneficial for faster malware generation. Interestingly for MalConv, we observe the highest throughput at the default value of 0.9 and slightly lower throughput for other values.

\subsection{Semantic NOP Variety Sensitivity}\label{app:nop-sensitivity}

At each injection step, the attack draws $s$ NOP types uniformly from the pool of 25 templates and repeats each maximum up to $r$ times via random sampling between (0,$r$), with $s \times r \approx 300$ held approximately constant to control the total perturbation budget. The defaults are $s{=}3, r{=}100$ for MalGraph and $s{=}1, r{=}300$ for MalConv for all of our experiments. We vary $s$ and $r$ within $\{(1, 300), (3, 100), (5, 60), (7, 43)\}$ and measure ASR and throughput at 0.1\% FPR.

\begin{figure}[h]
\centering
\includegraphics[width=\columnwidth]{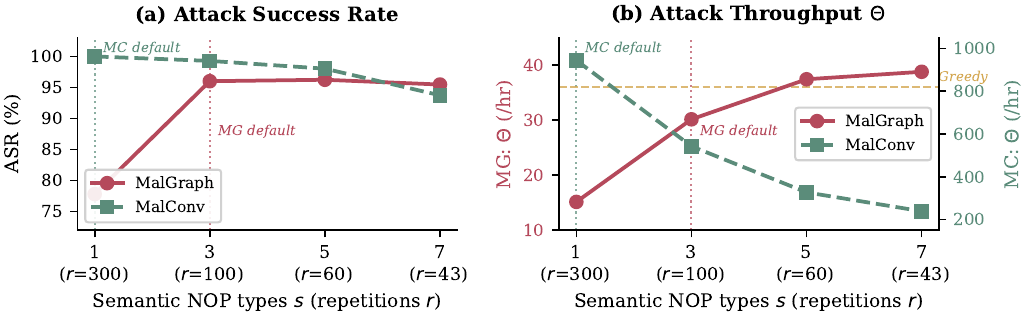}
\caption{Sensitivity to semantic NOP variety at 0.1\% FPR. The $x$-axis shows the number of NOP types $s$ drawn per injection step, with the maximum repetitions $r$ adjusted inversely ($s \times r \approx 300$). Defaults differ per classifier (MG: $s{=}3$; MC: $s{=}1$), marked with annotations. (b)~Dashed line marks Greedy throughput from Table~\ref{tab:full_comparison} for MG.}
\label{fig:semnop_sensitivity}
\end{figure}

Figure~\ref{fig:semnop_sensitivity} reveals classifier-dependent sensitivity. For MalGraph, restricting to a single NOP type ($s{=}1$) causes an 18.2,pp ASR drop, but increasing variety beyond the default has negligible effect; throughput improves with variety, reaching 39,byp/hr at $s{=}7$ versus 15,byp/hr at $s{=}1$. For MalConv, the opposite holds: the single-type default achieves 100\% ASR, and increasing variety degrades both ASR ($-$6.26,pp at $s{=}7$) and throughput (239 vs.\ 945,byp/hr). This asymmetry reflects MalGraph's GNN benefiting from structurally distinct perturbations, while MalConv's byte-level convolutions favor concentrated repetition.

The preceding analysis varies how many types are drawn per
injection step from a diverse pool. We additionally test whether pool diversity itself matters by replacing the full pool ($\sim$180 sequences) with a single instruction type (\texttt{mov reg, reg} with 16 general-purpose register variants) at 0.1\% FPR. Table~\ref{tab:nop_pool_ablation} shows pool diversity is critical for MalGraph (ASR collapses from 96.27\% to 5.34\%), as identical instruction patterns produce indistinguishable embeddings that the GNN learns to disregard. MalConv ASR is preserved (99.5\%), though throughput drops from 945 to 612 byp/hr due to reduced byte-level variation.

\begin{table}[h]
\centering
\footnotesize
\caption{Diverse NOP pool vs.\ single-type pool
(\texttt{mov reg, reg} only) at 0.1\% FPR using Adaptive SA.}
\label{tab:nop_pool_ablation}
\setlength{\tabcolsep}{4pt}
\begin{tabular}{@{}llcc@{}}
\toprule
\textbf{Classifier} & \textbf{Pool} & \textbf{ASR (\%)} & $\boldsymbol{\Theta}$ \textbf{(byp/hr)} \\
\midrule
MalGraph & Diverse (default) & 96.27 & 30.31 \\
MalGraph & Single-type       &  5.34 & 23 \\
\midrule
MalConv  & Diverse (default) & 100.0 & 945 \\
MalConv  & Single-type       & 99.46 & 612 \\
\bottomrule
\end{tabular}
\end{table}

\subsection{Dummy Function Count Sensitivity}\label{app:k-sensitivity}

The number of dummy functions injected per call site, $k \sim \mathrm{Uniform}(k_{\min}, k_{\max})$, controls the volume of structural perturbation. In all main experiments we use $k_{\min}{=}1$, $k_{\max}{=}3$ for MalConv and MalGraph. We vary $k_{\min}/k_{\max} \in \{1/1,\; 5/5,\; 7/7\}$ and compare against the default $1/3$ at 0.1\% FPR using Adaptive SA. SAFE+GNN requires a substantially larger injection volume ($k_{\min}/k_{\max}{=}2500/3500$) due to its global mean pool aggregation; the effect of varying $k$ for SAFE+GNN is analyzed separately in Section~\ref{sec:ablation} as part of the $k$-range sensitivity analysis.

\begin{figure}[h]
\centering
\includegraphics[width=\columnwidth]{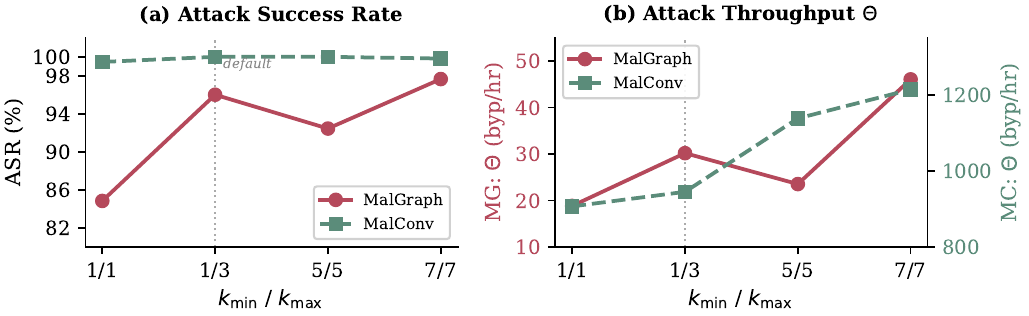}
\caption{Sensitivity to $k_{\min}/k_{\max}$ at 0.1\% FPR. Dotted line marks the default ($1/3$). (a)~ASR. $k{=}1/1$ drops MalGraph ASR by 11.5\,pp; $k \geq 3$ is stable. MalConv is invariant. (b)~Throughput on independent axes (MalGraph left, MalConv right).}
\label{fig:k_sensitivity_mc_mg}
\end{figure}

Figure~\ref{fig:k_sensitivity_mc_mg} shows that $k{=}1/1$ causes an 11.5\,pp MalGraph ASR drop, as a single injected node per call site is insufficient to shift the GNN's representation. For $k \geq 3$, ASR stabilizes: $k{=}7/7$ is only 1.7\,pp above the default. MalConv is flat across all values ($~$99.5\%). While higher $k$ increases throughput for both classifiers, we retain $k_{\min}/k_{\max}{=}1/3$ as the default since ASR is within 1.7,pp of the best configuration and has minimal structural footprint.

\subsection{Semantic Preservation Failure Analysis}\label{app:semantic-failures}

This section provides detailed failure analysis for the semantic preservation evaluation in Section~\ref{sec:rq4}. Of the 1{,}706 samples evaluated across all six (classifier, FPR) configurations, 147 (8.6\%) fail the SPR test. Table~\ref{tab:spr_failure_breakdown} provides per-configuration breakdowns.

\begin{table}[h]
\centering
\caption{Per-configuration SPR failure breakdown. \emph{Diverged}: adversarial produced $\geq 20$ API calls but sequence distance exceeded $\text{dist}_\Delta$. \emph{Short-trace}: adversarial produced $< 20$ API calls despite both original runs exceeding this threshold.}
\label{tab:spr_failure_breakdown}
\footnotesize
\setlength{\tabcolsep}{2pt}
\renewcommand{\arraystretch}{1.05}
\begin{tabular}{@{}llccccc@{}}
\toprule
\textbf{Classifier} & \textbf{FPR} & $N$ & \textbf{Failures} & \textbf{Diverged} & \textbf{Short-trace} & \textbf{SPR (\%)} \\
\midrule
MalConv  & 0.1\% & 211 & 29 & 15 & 14 & 86.26 \\
MalConv  & 1\%   & 300 & 39 & 26 & 13 & 87.00 \\
MalGraph & 0.1\% & 300 & 30 &  3 & 27 & 90.00 \\
MalGraph & 1\%   & 300 & 21 & 10 & 11 & 93.00 \\
SAFE+GNN  & 0.1\% & 300 & 19 &  8 & 11 & 93.67 \\
SAFE+GNN  & 1\%   & 295 &  9 &  2 &  7 & 96.95 \\
\midrule
\textbf{Total} & & \textbf{1{,}706} & \textbf{147} & \textbf{64} & \textbf{83} & \\
\bottomrule
\end{tabular}
\end{table}

\paragraph{\textbf{Overlay analysis (MalConv and MalGraph)}.} The dominant cause of failure for MalConv and MalGraph is overlay-dependent malware. Of the 54 diverged failures, 52 (96\%) carry a PE overlay (data appended after the last PE section, commonly used for encrypted payloads or second-stage code). The overlay content is identical in original and adversarial binaries (verified by SHA-256), but the absolute file offset shifts forward by at least one memory page because the inserted \texttt{.htext} section sits between the last original section and the overlay. Entry points and all original section layouts are otherwise unchanged, verified on all 119 failed and a 200-sample control drawn from successful adversarial samples. Of the 54 diverged adversarial samples, 51 (94\%) statically import the Win32 file-I/O chain typically used for self-reading (\texttt{GetModuleFileName}, \texttt{CreateFile}, \texttt{GetFileSize}, \texttt{SetFilePointer}, \texttt{ReadFile}). The co-occurrence of overlay presence (96\%), offset displacement with preserved content, and self-reading API imports suggests that these samples fail because they read their own image at absolute offsets that are no longer valid in the adversarial binary. MalConv configurations account for 41 of the 54 diverged failures, reflecting a higher proportion of overlay-bearing samples in its bypass pools. Of the 22 short-trace samples that produced 1-19 API calls, only 7 import overlay-related APIs, suggesting a different failure mechanism for this category.

\paragraph{\textbf{Overlay analysis (SAFE+GNN)}.}
SAFE+GNN shows the lowest failure rate across all configurations (28 out of 595, vs.\ 119 out of 1{,}111 for MC/MG). The failure mechanism differs from MalConv and MalGraph. Short-trace failures are overlay-dominated - at 1\% FPR, all 7 short-trace failures (100\%) carry PE overlays, whereas only 22\% of successfully preserved samples in the same configurations do. This is consistent with the overlay-offset-shift pattern observed in MalConv and MalGraph. At 0.1\% FPR, 5 of 11 short-trace failures (45.5\%) carry overlays vs.\ 24\% baseline. However, diverged failures are \emph{not} overlay-dominated, as only 2 of 10 diverged samples across both FPRs carry overlays. We attribute these to the larger structural footprint of the SAFE+GNN attack (thousands of injected call sites per sample), which may alter execution timing or dispatch order in samples with path-dependent behavior. The higher overall SPR for SAFE+GNN is consistent with the alternative function bodies being a structurally minimal perturbation ($\sim$5 bytes per injected function) that does not alter the original code's instruction stream beyond the call-site displacement itself.

\subsection{Analysis with Antivirus Engines}
\label{app:av_analysis}

Both \system{} and its primary baseline, MalGuise, are structural perturbation attacks as they perturb the CFG and/or FCG topology and inject semantic NOP sequences to evade ML-based classifiers. Neither attack modifies, removes, or overwrites any of the original malware's code, data, import tables, string constants, or embedded signatures. The adversarial binary is a strict superset of the original, with all original bytes preserved at their relative positions within existing sections. To evaluate whether these perturbations transfer to production antivirus engines that rely on detection features orthogonal to the targeted ML classifiers, we scan a stratified random sample of $N{=}5{,}029$ adversarial binaries from \system{} (drawn from the full pool of 10{,}060 samples across Malgraph and Malconv and both FPR thresholds, seed=42) and an independently drawn sample of the same size from MalGuise's adversarial output pool. SAFE+GNN adversarial samples share the same PE-level injection mechanism but use minimal function bodies, so we expect AV transfer rates to be comparable to those observed for MalConv and MalGraph.

All antivirus engines were freshly installed with default configurations on April~24,~2026, with signature databases current as of installation. Bitdefender\footnote{\url{https://www.bitdefender.com/}}~(v10.3.1), Norton\footnote{\url{https://us.norton.com/}}~(v26.3.1), and ClamAV\footnote{\url{https://docs.clamav.net/}}~(v1.5.2, signatures~27980) were installed on macOS and used for static file scanning. We performed Microsoft Defender scans on a Windows~10 virtual machine using \texttt{MpCmdRun.exe} in custom-scan mode with local signature and heuristic detection. Since all engines perform static signature and heuristic analysis on PE file content, the host operating system does not affect detection outcomes for Windows PE malware. Table~\ref{tab:av_evasion} reports the results for both attacks.

\begin{table}[tb]
\centering
\caption{ASR (\%) against four antivirus engines for adversarial samples produced by \system{} and MalGuise ($N{=}5{,}029$ per attack). Both structural perturbation attacks show comparable ASR against commercial engines (14-18\%)}
\label{tab:av_evasion}
\setlength{\tabcolsep}{6pt}
\renewcommand{\arraystretch}{1.1}
\scriptsize
\begin{tabular}{@{}lcc@{}}
\toprule
\textbf{AV Engine} & \textbf{\system{} ASR (\%)} & \textbf{MalGuise ASR (\%)} \\
\midrule
Bitdefender        & 14.28 & 15.51 \\
Microsoft Defender & 18.34 & 15.83 \\
Norton             & 17.12 & 15.34 \\
\midrule
ClamAV             & 74.31 & 74.78 \\
\bottomrule
\end{tabular}
\end{table}
The results reveal a consistent pattern across both attacks. Against the three commercial engines, \system{} achieves 14-18\% ASR, and MalGuise achieves 15-16\%, falling within 1-3 percentage points on every engine. Commercial engines that combine signature matching, heuristics, behavioral analysis, and cloud-based ML layers detect the sample through orthogonal features (import tables, string constants, API call sequences, behavioral signatures) that structural perturbations leave untouched by design. The near-identical ASR across both attacks (MalGuise's CFG-only redividing vs.\ \system{}'s dual CFG+FCG node injection) shows that bounded AV transfer is a fundamental property of this attack class. The attack surface that \system{} characterizes is the vulnerability of graph-based ML classifiers to topological perturbation, not the overall robustness of the full AV detection pipeline.

ClamAV, a signature-only engine without behavioral or ML layers, diverges sharply at 74-75\% ASR for both attacks, as overwritten call-site instructions and injected sections break byte-pattern signatures whose matched span overlaps the modified regions in the samples. The 4-5$\times$ gap quantifies the defensive value of multi-layered detection.


\subsection{Structural Heuristic Defense: Extended Analysis}
\label{app:defense_sweep}

\noindent\textbf{Original malware size distribution.}
The size threshold used for H\textsubscript{size} is derived from the original (unmodified) malware population. The malware samples have a high coefficient of variation (0.97) and a heavy right skew (2.74), indicating extreme natural size variance, which is why any single-threshold size heuristic produces high false positive rates.


\smallskip
\noindent\textbf{Threshold sweep.}
We sweep size thresholds by varying the base percentile of the original malware size distribution (P50, P75, P90) and the overhead multiplier applied to it (5--30\%). Table~\ref{tab:sweep} confirms a monotonic tradeoff: reducing FPR below 8\% pushes TPR below 10\%, regardless of the base percentile.

\begin{table}[h]
\centering
\caption{H\textsubscript{size} TPR/FPR across base
  percentiles and overhead multipliers.  Leaf and DAG
  pooled across all classifiers and operating points.}
\label{tab:sweep}
\footnotesize
\begin{tabular}{llrrr}
\toprule
\textbf{Anchor} & \textbf{Overhead} & \textbf{TPR (Leaf)} & \textbf{TPR (DAG)} & \textbf{FPR} \\
\midrule
P50 & $\times$1.05  & 53.3\% & 52.0\% & 13.4\% \\
P50 & $\times$1.16  & 32.1\% & 30.9\% & 12.6\% \\
P50 & $\times$1.30  & 25.0\% & 24.1\% & 11.3\% \\
\midrule
P75 & $\times$1.05  & 25.5\% & 24.5\% & 11.4\% \\
P75 & $\times$1.30  & 12.3\% & 12.3\% & 9.2\%  \\
\midrule
P90 & $\times$1.05  & 7.5\%  & 7.6\%  & 6.1\%  \\
P90 & $\times$1.30  & 5.6\%  & 5.7\%  & 5.1\%  \\
\bottomrule
\end{tabular}
\end{table}

\smallskip
\noindent\textbf{DAG variant and H\textsubscript{leaf}.}
H\textsubscript{leaf} flags samples whose injected section contains only leaf functions (no outgoing calls). The DAG variant (Table~\ref{tab:fcg_chaining}) adds inter-dummy function call edges, so most injected functions are no longer leaves. For SAFE+GNN ($k \approx 3{,}000$ functions per site), even a low per-edge probability produces enough edges that every function acquires at least one call, reducing H\textsubscript{leaf} to 0\%. For MalConv/MalGraph ($k=1$-$3$ per site), some functions remain leaves by chance because fewer potential callees exist.

\subsection{Attack Mechanism Analysis and Feature Space Visualization}\label{app:feature-space}
\textbf{Attack Mechanism Analysis.}
MalGraph's hierarchical GNN embeds each function's CFG and aggregates via global pooling over the FCG, so injected dummy functions directly contribute to the program embedding and shift the pooled representation. For MalConv, the injected section and displaced call sites perturb the raw-byte feature space directly. For SAFE+GNN, global mean pooling averages all function node embeddings with equal weight, so injecting dummy functions with distinct Word2Vec embeddings shifts the averaged representation past the decision boundary at sufficient volume (Section~\ref{sec:ablation} shows that removing injection collapses ASR from 85\% to 0.1\%).

\begin{figure*}[t]
\centering
\includegraphics[width=0.8\textwidth]{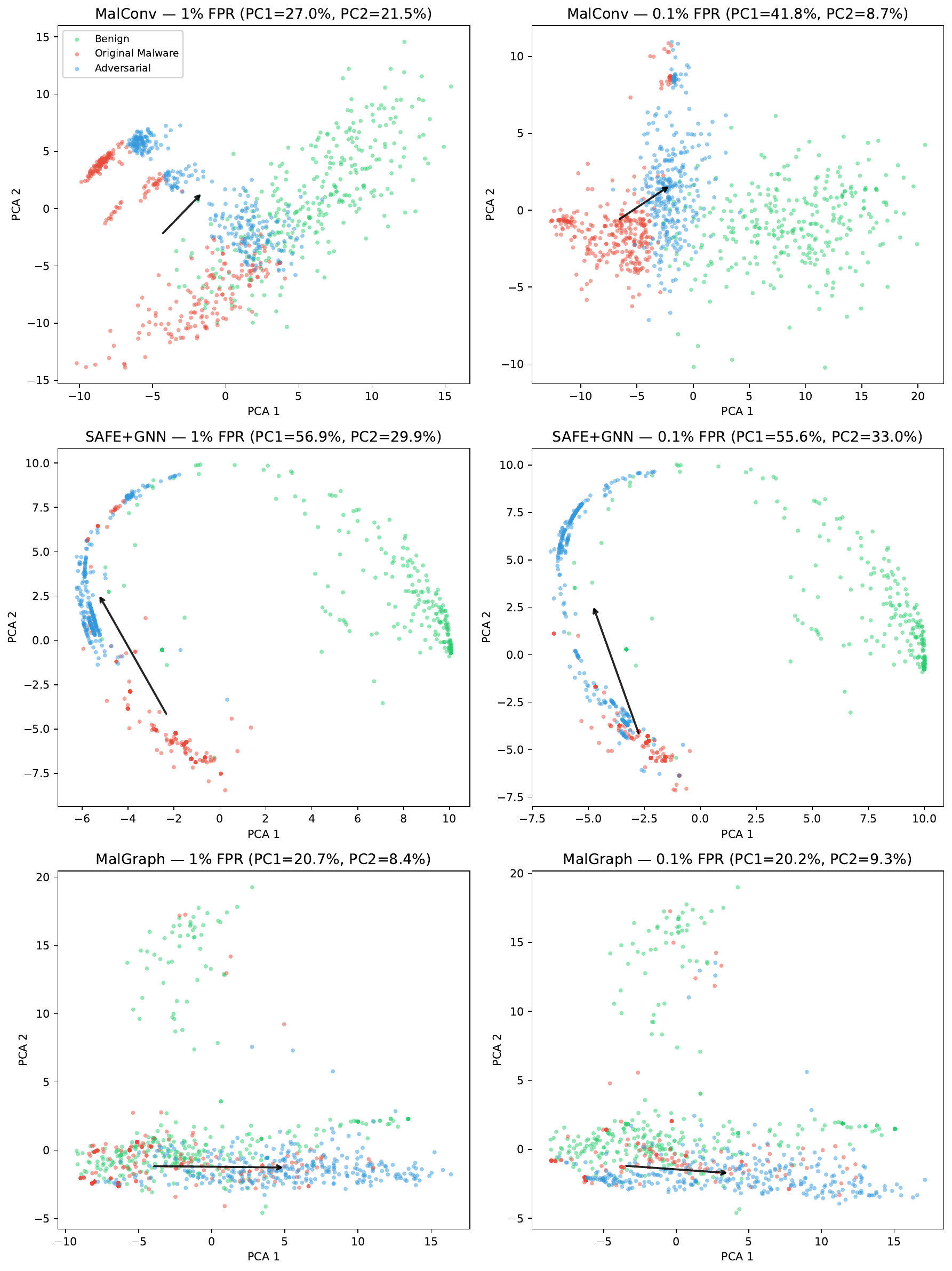}
\caption{PCA projection of penultimate-layer embeddings (L2-normalized, standardized) for benign (green), original malware (red), and adversarial (blue) samples. Arrows point from the original malware average to the adversarial average. MalConv (128-d): adversarial samples partially overlap with the benign region, particularly at 0.1\% FPR. SAFE+GNN (64-d, 87-89\% variance captured): adversarial samples shift away from the original malware cluster. MalGraph (200-d, 29\% variance captured): adversarial is displaced from the original cluster into a distinct region.}
\label{fig:feature_space}
\end{figure*}

\textbf{Feature Space Visualization.}
To visualize how \system{}'s perturbations affect each classifier's learned representation, we extract the penultimate-layer embedding (the embedding just before the classification head) for 300 benign, 300 original malware, and 300 adversarial (bypassed) samples per configuration. Embeddings are L2-normalized to remove magnitude differences and projected to 2D via PCA. Figure~\ref{fig:feature_space} shows the results across all three classifiers at both FPR thresholds. Arrows point from the average position of original malware samples to the average position of adversarial samples, showing the direction the attack moves the representation.

The three classifiers exhibit distinct patterns. MalConv adversarial representations partially overlap the benign region and shift away from the malware region, particularly at 0.1\% FPR. SAFE+GNN's large amount of node injection dilutes the graph-level representation through mean pooling (mean L2 norm drops from 4.4 to 0.4), shifting samples away from the malware cluster. MalGraph injection inflates the embedding through max pooling (mean norm increases from 81 to 191), displacing samples into a distinct high-magnitude region. L2 normalization reduces the displacement, but adversarial samples remain farther from benign samples than the originals. These opposite effects (collapse vs.\ inflation) arise from the same injection mechanism interacting with different architectures.


\end{document}